\documentclass[reprint,
runinaddress,
nofootinbib,
amsmath,amssymb,
aps,
prd,
floatfix
]{revtex4-2}

\def\aap{Astron. Astrophys.}
\def\apj{Astrophys. J.}
\def\aapr{Astron. Astrophys. Rev.}
\def\apjl{Astrophys. J. Lett.}
\def\apjs{Astrophys. J. Suppl. Ser.}
\def\mnras{Mon. Not. R. Astron. Soc.}
\def\araa{Annu. Rev. Astron. Astrophys.}
\def\aj{Astron. J.}

\def\na{New Astronomy}

\def\nat{Nature}

\def\prd{Phys. Rev. D}
\def\pasp{Publ. Astron. Soc. Pac.}

\def\ssr{Space Sci. Rev.}

\usepackage{xspace}
\usepackage{subcaption}
\usepackage{multirow}
\usepackage{graphicx}
\usepackage{dcolumn}
\usepackage{bm}
\usepackage{orcidlink}
\usepackage{hyperref}
\hypersetup{
	colorlinks=true,        
	linkcolor=red,         
	citecolor=blue,        
	urlcolor=blue,           
}
\usepackage{dsfont}
\usepackage{comment}
\usepackage{color}
\usepackage[normalem]{ulem}
\usepackage{soul}
\usepackage{float}

\begin{document}

\preprint{APS/123-QED}

\title{A Mass-Independent Damping Timescale in Black Hole Accretion Systems}

\author{Haoyang Zhang\,\orcidlink{0000-0003-3392-320X}$^{1,2}$}
\author{Shenbang Yang$^{3}$}%
\author{Li Zhang\,\orcidlink{0000-0002-5880-8497}$^{1,2}$}
\author{Benzhong Dai\,\orcidlink{0000-0001-7908-4996}$^{1,2}$}
\email{bzhdai@ynu.edu.cn}

\affiliation{%
	$^{1}$Department of Astronomy, Yunnan University, Kunming 650091, China\\
	$^{2}$Key Laboratory of Astroparticle Physics of Yunnan Province, Yunnan University, Kunming 650091, China\\
	$^{3}$Faculty of Science, Kunming University of Science and Technology, Kunming 650500, China
}%

\date{\today}

\begin{abstract}
The scaling laws reveal the underlying structural similarities shared by astrophysical systems across vastly different scales. In black hole accretion systems, the scaling relations between the characteristic damping timescales (CDTs) of light curves and black hole mass offer valuable insights into the underlying physical structure of accretion disks. Here, we investigate the long-term hard X-ray CDTs of 106 black hole and neutron star accretion systems using light curves from the \textit{Swift} Burst Alert Telescope 157-month catalog. Unexpectedly, for the first time, we discover a mass-independent CDT in these black hole accretion systems, in contrast to well-established scaling laws. This puzzling phenomenon can be attributed to conductive timescales arising from disk–corona interactions, instead of the intrinsic accretion disk processes characterized by scaling laws, and it may further modulate jet emission in blazars. This result demonstrates thermal conduction as a key mechanism driving hard X-ray variability and offers new observational evidence for the disk–corona–jet connection. 
\end{abstract}

\maketitle

\section{Introduction}
Stellar-mass black holes (SBHs; 5--15 $M_{\odot}$) and neutron stars that form at the end of stellar evolution may accrete matter from their companion stars to form X-ray binary (XRB) systems \citep{2006ARA&A..44...49R,2007A&ARv..15....1D}. Similarly, at galaxy scales, supermassive black holes (SMBHs; $10^{6}-10^{10} M_{\odot}$) may accrete gas from their host galaxy to form active galactic nuclei (AGNs) \citep{1995PASP..107..803U,2019ApJ...875L...1E}. Apart from their physical structures, these two types of black hole accretion systems also share many observational similarities, such as similar broadband spectral shapes \citep{2017SSRv..207....5R}, variability ranging from minutes to years \citep{2006ARA&A..44...49R,2015A&A...574A..33R,2015ApJS..218...18D,2017A&ARv..25....2P}, quasiperiodic oscillations (QPOs) \citep{2006ARA&A..44...49R,2008Natur.455..369G}, and radio and $\gamma$-ray emission from relativistic jets \citep{1999ARA&A..37..409M,2018Natur.562...82A,2019ARA&A..57..467B,lhaasocollaboration2024ultrahighenergygammarayemissionassociated}. Although significant advancements in understanding have been made using observations, many mysteries remain regarding the underlying physical mechanisms of these phenomena.

Due to the significant mass difference between these two accretion systems, the study of mass-dependent scaling laws has garnered widespread attention. Specifically, the well-known relationship between the mass of the balck hole and galaxy velocity dispersion \cite{2000ApJ...539L..13G} and the X-ray and radio luminosities scale with the mass of the black hole, known as the fundamental plane of black hole activity \cite{2003MNRAS.345.1057M,2006A&A...456..439K,2016MNRAS.455.2551N,2017SSRv..207....5R}. In terms of variability, the power spectral density (PSD) of light curves (LCs) of these accretion systems can be modeled as a bending power-law (BPL) spectrum (i.e.,  $P(f)=[Af^{-\alpha_{\text{low}}}/(1+(f/f_{\text{bend}})^{\alpha_{\text{high}}-\alpha_{\text{low}}})]+C$, where $A$, $f_{\text{bend}}$, $\alpha_{\text{high}}$, $\alpha_{\text{low}}$, and $C$ are the normalization, break frequency, high-frequency slope, low-frequency slope, and Poisson noise, respectively.) with a characteristic damping timescale (CDT; $\tau_{\text{damping}}$) \cite{2013MNRAS.433..907E}. \cite{2006Natur.444..730M} and \cite{2015SciA....1E0686S} established several scaling relationships for the CDT: $\tau_{\text{damping}} \sim (M_{\text{BH}}, L_{\text{bol}})$ in X-rays, and $\tau_{\text{damping}} \sim (M_{\text{BH}}, \dot{M}, R)$ in the optical band, respectively. Furthermore, direct monotonic relationships ($\tau_{\text{damping}}\propto M_{\text{BH}}$) have been reported with similar slopes ($\sim0.5$) in nonjetted and jetted accretion systems \cite{2021Sci...373..789B,2024ApJ...967L..18Z}. In addition, the scaling relationship $T_{\text{QPO}} \propto M_{\text{BH}}$ in X-rays has been reported by \cite{2015ApJ...798L...5Z}. \cite{2021ApJ...906...92S} and \cite{2025MNRAS.538.2161Z} further discussed the applicability of this relationship in different physical scenarios.

A variety of physical processes has been suggested to explain the BPL features in the PSDs. In the accretion disk, the dynamical timescale, thermal equilibrium timescale, cooling and heating front propagation timescales, viscous timescale, and Compton cooling timescale can explain the observational CDTs, which have magnitudes ranging from several months to years  \citep{2006ASPC..360..265C,2012A&A...540L...2I,2021ApJ...907...96S,2021Sci...373..789B}. Recently, \citet{2024ApJ...966....8Z} demonstrated that the corona-heated accretion disk reprocessing model can naturally generate LCs with CDTs that match those measured observationally. In addition, all of the above models can effectively explain the scaling laws between CDTs and black hole masses \citep{2012A&A...540L...2I,2021Sci...373..789B}. The rotational energy of black holes and angular momentum of accretion flows can be extracted from a large-scale magnetic field \citep{1977MNRAS.179..433B,1982MNRAS.199..883B}, providing energy support for the formation of jets; this mechanism is known as the disk--jet connection. Therefore, the CDT of jet variability may originated from accretion disks \citep{2024ApJ...967L..18Z}. Within the jet, shocks and magnetic reconnection can also produce CDT variability on scales of several hours to days \citep{2014ApJ...791...21F}. 

A recent study investigated the long-term ($\sim15$ yr) $\gamma$-ray  variability of black hole accretion systems, finding two microquasars (a subclass of XRBs with relativistic jets) \citep{1999ARA&A..37..409M}) that exhibited variability properties remarkably similar to those of blazars. Their result suggests that the CDTs of the long-term $\gamma$-ray LCs of blazars and microquasars do not follow the expected scaling relation with black hole mass \citep{2025PhRvD.111h3049S}. The $\gamma$-ray CDT of one of the microquasars (LS~I~+61 303) can be explained by the extreme, superorbital period of an XRB \citep{2024ApJ...972...80Z}, while the origin of the CDT ($\sim69$ days in the observer frame) of the other source (Cygnus~X-3) remains unclear. This discovery poses a challenge to the well-established scaling law of black hole accretion systems. Given the rarity of XRBs with detectable $\gamma$-ray emission, it becomes crucial to examine the variability of more common XRBs across other wavelengths with statistically significant samples.

In this work, we apply, for the first time, a Gaussian process method that incorporates the damped random walk (DRW) process for modeling LCs from the \textit{Swift} 157-month catalog of AGNs and XRBs in hard X-rays (14--195 keV). Our aim is to detect the CDT in the long-term LCs and reveal the underlying physical processes behind the phenomenon. The rest of this paper is structured as follows. We briefly introduce the data and DRW method in Sec. \ref{sec:d_m}. In Sec. \ref{sec:r_d}, we show the modeling results and discuss the possible physical origins of the CDTs. We present our conclusions in Sec. \ref{sec:conclusion}.

\begin{figure*}[ht]
	\includegraphics[width=1.\textwidth]{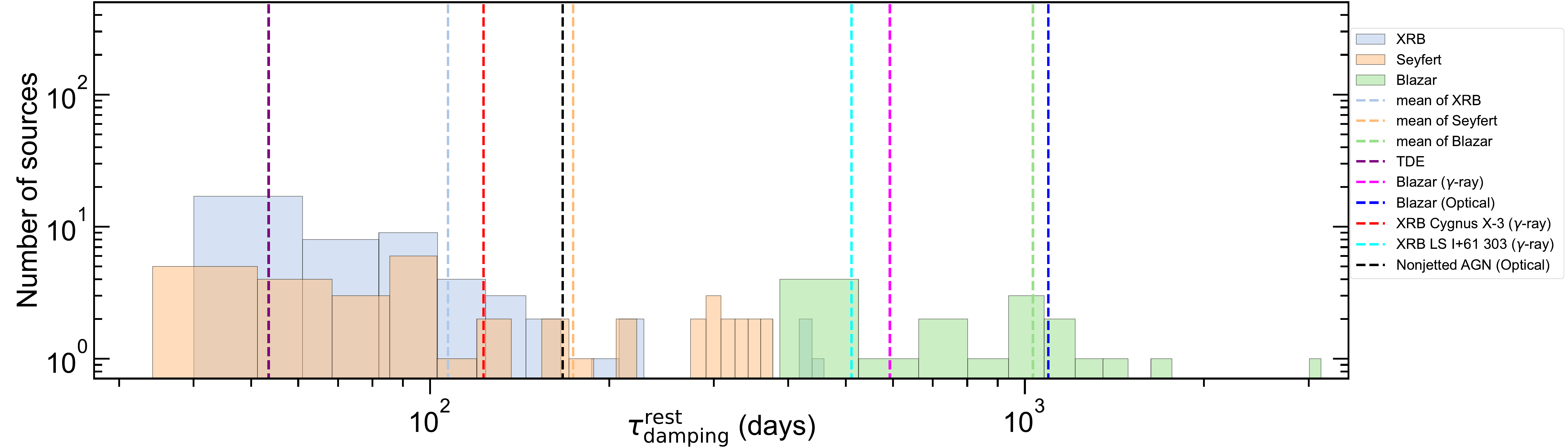}
	\caption{Comparison of the CDT distributions in our results and previous studies. Light blue, light yellow, and light green indicate the distributions of XRBs, Seyfert galaxies, and blazars, respectively. The blue dashed line indicates the peak location of the CDT distribution of blazars in the optical \citep{2012ApJ...760...51R}. Two microquasars observed in $\gamma$ rays by \citet{2025PhRvD.111h3049S} are shown. The CDT of a TDE is shown as the purple dashed line. The other dashed lines correspond to the means of our three samples and previous work. \label{fig:dis}}
\end{figure*}

\begin{table*}
	\renewcommand{\arraystretch}{1.3}
	\setlength{\tabcolsep}{3pt}
	\centering
	\caption{Statistical information of our results compared with previous work.}
	\begin{tabular}{cccccc}
		\hline
		Statistical Information of $\tau_{\text{damping}}^{\text{rest}}$ (days) & XRBs (X-ray) & Seyferts (X-ray) & Blazars (X-ray) & Blazars ($\gamma$-ray) & Nonjetted AGN (optical) \\
		\hline
		$\tau_{\text{min}}^{\text{rest}}$        & 40.04  & 34.13  & 387.15  & 79.04   & 2.00    \\
		$\tau_{\text{max}}^{\text{rest}}$        & 459.44 & 376.79 & 3146.43 & 3463.38 & 398.11  \\
		$\tau_{\text{median}}^{\text{rest}}$     & 80.64  & 127.07 & 954.56  & 385.91  & 158.49  \\
		$\tau_{\text{mean}}^{\text{rest}}$       & 107.17 & 173.98 & 1031.60 & 592.20  & 167.15  \\
		\hline
	\end{tabular}
	{Notes---The statistical information of the blazars ($\gamma$-ray) and nonjetted AGN (optical) is derived from the combined results of previous works \citep{2024MNRAS.527.2672S,2024ApJ...967L..18Z,2025PhRvD.111h3049S,2021Sci...373..789B}.}
	\label{tab:statistical}
\end{table*}

\begin{figure*}
	\includegraphics[width=\textwidth]{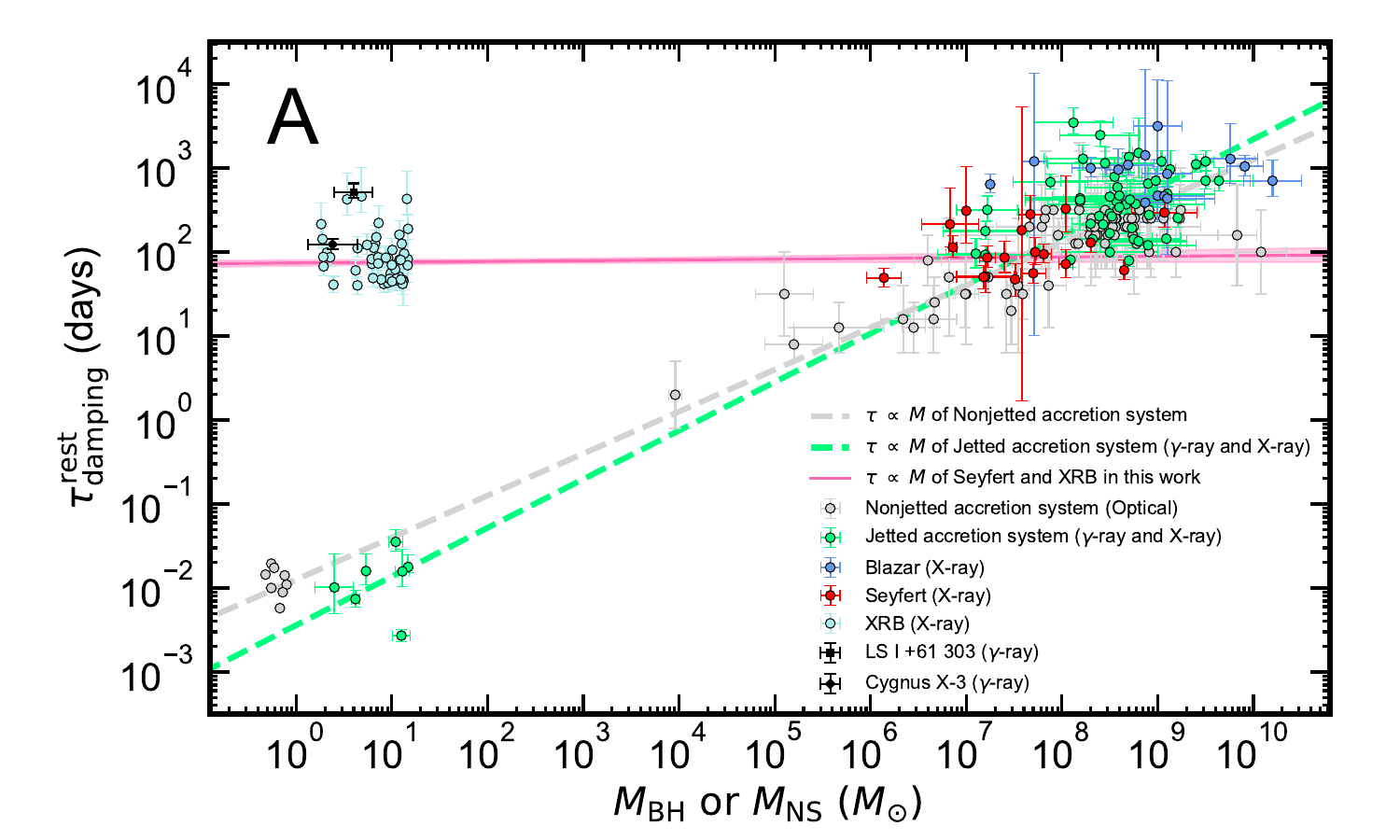}
	\includegraphics[width=\textwidth]{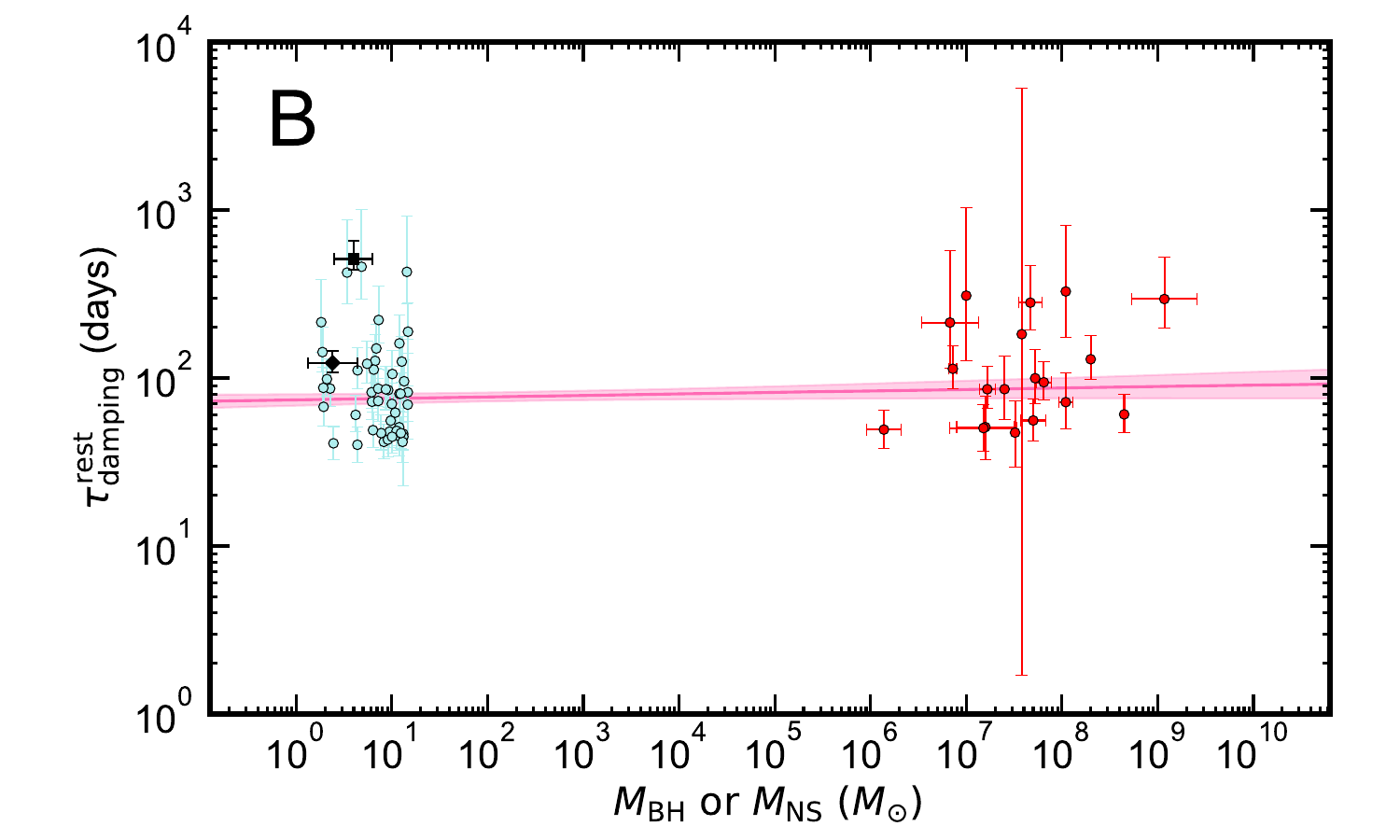}
	\caption{The $\tau_{\text{damping}}^{\text{rest}}\propto M_{\text{BH}}$ relations of our results and previous work. In panel (A), the gray and spring-green dashed lines indicate the best linear fit for nonjetted accretion systems, combined with previous work of blazars in $\gamma$ rays. The hot-pink dashed lines and shaded region respectively represent the best fit and 1$\sigma$ confidence interval of XRBs and Seyfert galaxies of this work. Panel (B) shows a zoomed-in view of the linear relation of our work. \label{fig:lin_rel}}
\end{figure*}

\section{Data and Method}\label{sec:d_m}
Mounted on board \textit{Swift} is the Burst Alert Telescope (BAT), which is dedicated to the detection and continuous observation of $\gamma$-ray bursts \citep{2005SSRv..120..143B}. The \textit{Swift}/BAT 157-month survey
catalog includes 1888 valid sources \citep{2025ApJ...989..161L}. Among these sources, AGNs account for the majority, with Seyfert galaxies comprising approximately 50\% and blazars about 10\%. XRBs make up around 13\%. We selected all low-mass XRBs that host accretion disks \citep{2006csxs.book..623T} and all AGN sources from the \textit{Swift}/BAT 157-month catalog.\footnote{\url{https://swift.gsfc.nasa.gov/bat_survey/bs157mon/data/BAT_157m_catalog_20250525.txt}} The data reduction for each source in the eight energy bands (14--20, 20--24, 24--35, 35--50, 50--75, 75--100, 100--150, and 150--195 keV) has been correctly carried out \citep{2025ApJ...989..161L}. The detection threshold of telescope for each source is $4.8\sigma$ to ensure the high signal-to-noise ratio of the LCs. Calibrated using the Crab spectrum, a weighted combination of eight energy bands is obtained, with the count rate expressed in units of “Crab” \citep{2025ApJ...989..161L}. The duration and cadence of the LCs are $\sim$157 months and $\sim$1 month, respectively. The 14--195 keV “Crab-calibrated” LC data products used in this work can be downloaded from the official website.\footnote{\url{https://swift.gsfc.nasa.gov/results/bs157mon/}} We excluded LC data points with exposure times shorter than one day, as this procedure effectively filters out those with very large error bars \citep{2023MNRAS.526.4040M}.

The LCs of AGNs and XRBs have been well modeled as a DRW process in many studies \citep{2009ApJ...698..895K,2010ApJ...721.1014M,2012ApJ...760...51R,2013ApJ...765..106Z,2017A&A...597A.128K,2021ApJ...907...96S,2021Sci...373..789B,2022MNRAS.514..164S,2022ApJ...930..157Z,2023ApJ...944..103Z,2024MNRAS.527.2672S,2024ApJ...967L..18Z,2025PhRvD.111h3049S}. The DRW model is a specific form of the continuous time autoregressive moving average process \citep{2009ApJ...698..895K,2014ApJ...788...33K}, where $p=1$ and $q=0$; in some studies, this model is referred to as the Ornstein--Uhlenbeck process \citep{1996AmJPh..64..225G}. The PSD of this model has a BPL shape, but with $\alpha_{\text{high}}\sim2$, $\alpha_{\text{low}}\sim0$ and thus can be used to search for the CDT of each considered LC. The stochastic differential equation of this process can be derived \citep{2014ApJ...788...33K} as

\begin{equation}
	\Bigg[\frac{d}{dt}+\frac{1}{\tau_{\text{DRW}}}\Bigg]\text{y}(t)=\sigma_{\text{DRW}}\epsilon(t),
\end{equation}
where $\tau_{\text{DRW}}$ is the CDT, $\sigma_{\text{DRW}}$ is the amplitude, and $\epsilon(t)$ is white noise. 

A more complex, alternative model, i.e., a damped harmonic oscillator, which has more parameters, has also been used to describe the LCs of AGNs \citep{2022ApJ...936..132Y,2025ApJ...984...45X} as well as searching for QPOs in LCs \citep{2021ApJ...907..105Y,2023ApJ...946...52Z}. However, the parameters of this model often fail to be constrained during the modeling process \citep{2022ApJ...936..132Y,2024ApJ...966....8Z}. Moreover, when this model is used to determine the CDT, it generally produces results comparable to those obtained with DRW \citep{2024ApJ...966....8Z}.

In our method, we use the fast and scalable Gaussian process modeling tool \texttt{celerite}\footnote{\url{https://celerite.readthedocs.io/en/stable/}} deveploped by \citet{2017AJ....154..220F}. This tool not only enables fast operation of DRW modeling of LCs, but also avoids the aliasing and red noise leakage that are common in Fourier techniques due to limited observational duration and unevenly sampling. The covariance function (kernal function) in \texttt{celerite} is defined as
\begin{equation}
	k(t_{nm})=2\sigma_{\text{DRW}}^{2}\cdot\text{exp}(-t_{nm}/\tau_{\text{DRW}}),
\end{equation}
where $t_{nm}$ is the time lag between two measurements \citep{2017AJ....154..220F}. 
Next, the PSD of this process is
\begin{equation}
	S(\omega)=\sqrt{\frac{8}{\pi}}\sigma^{2}_{\text{DRW}}\tau_{\text{DRW}}\frac{1}{1+(\omega\tau_{\text{DRW}})^{2}}.
\end{equation}
Corresponding to the BPL model, the bend timescale ($f_{\text{bend}}$) can be obtained by $\tau_{\text{DRW}}=1/(2\pi f_{\text{bend}})$.  

In our modeling, we perform maximum likelihood estimation using the \texttt{L-BFGS-B} optimizer to find the initial parameters for subsequent modeling with the Markov Chain Monte Carlo (MCMC) technique. Then, we use the MCMC sampler \texttt{emcee}\footnote{\url{https://pypi.org/project/emcee/}} to get the posterior distributions of the parameters. Specifically, we conduct $10^{5}$ steps of sampling, and discard the first 30000 steps for burn-in. 

The CDTs obtained from DRW modeling must be filtered by two criteria: the maximum value of the CDT should less than one-tenth of the baseline and the minimum CDT should be larger than the average cadence of the LCs \citep{2022ApJ...930..157Z}. Otherwise, the CDTs will deviate from the true physical timescales responsible for the observed emission \citep{2017A&A...597A.128K,2021Sci...373..789B}. In addition, a good fit implies the standardized residuals (SRs) are well represented by white noise. The Kolmogorov–Smirnov test was applied to compare the distribution between the SRs and white noise, and the fits with $P<0.05$ were filtered out. Furthermore, the autocorrelation functions (ACFs) of the SRs and squared SRs should both fall within the 95\% confidence interval of white noise \citep{2014ApJ...788...33K,2021ApJ...907..105Y,2023ApJ...946...52Z}. The fitting results of a source that pass all the above criteria are considered valid.

\section{Results and Discussion}\label{sec:r_d}
We obtained the CDTs for 39 Seyfert galaxies (19 with estimated black hole masses), 17 blazars (15 with estimated black hole masses), 49 XRBs, and one tidal disruption event (TDE) as our valid sample, all of which passed the selection criteria and showed good fits. In Appendix \ref{App:A}, we show two examples of the fitting results from our valid sample. The complete fitting results for all 106 valid sources are provided in Appendix \ref{App:B} and the Supplementary Materials \cite{zhang_2025_17830360}. 

Subsequently, we transformed the CDT in the observer frame to the rest frame by the following relation:
\begin{equation}
	\tau_{\text{damping}}^{\text{rest}}=\frac{\tau^{\text{obs}}_{\text{damping}}}{1+z}\delta_{\text{D}},
\end{equation}
where $z$ is the redshift (for XRBs, $z\sim0$), and $\delta_{\text{D}}$ is the Doppler factor. As the hard X-ray emission of Seyfert galaxies is produced by inverse-Compton scattering within the corona \citep{2025FrASS..1130392L}, we set $\delta_{\text{D}}\sim 1$ for Seyfert galaxies. For blazars, hard X-ray emission can originate from both the jet and the corona via inverse-Compton scattering. However, both observations and theoretical models suggest that the jet component is dominant \citep{2006ARA&A..44..463H,2019ARA&A..57..467B,2025A&A...698L..19L}. Therefore, we adopt an average Doppler factor of $\delta_{\text{D}} = 10$ \citep{2022ApJ...930..157Z}. In XRBs, only a few sources have had their transient jets observed in the hard state \citep{1999ARA&A..37..409M,2006ARA&A..44...49R}. Even in the hard state, it is still not clear whether the hard X-ray radiation comes from the jet or the corona \citep{1997ApJ...482..448N,2022Sci...378..650K}, and some studies have shown that the Doppler beaming effect of the jets in XRBs is very weak \citep{2017ApJ...851..144L,2018Natur.562...82A}. Thus, for the XRBs in this work, $\delta_{\text{D}}=1$. 

Our analysis reveals clear distinctions in CDT distributions across different sources and bands. The statistical information of our results is presented in Table \ref{tab:statistical}. The distributions of our results and previous work are shown in Figure \ref{fig:dis}. The CDTs of the blazars are larger than those of the nonjetted AGNs (include Seyfert galaxies) in all wave bands. More specifically, in blazars, the mean value of our results is close to the peak position of the distribution in the optical band (see the light-green dashed lines and blue dashed lines in Figure \ref{fig:dis}), while the results at $\gamma$ rays (pink dashed lines) are smaller than at hard X-rays and optical. For the Seyfert galaxies, the results closely match those of the nonjetted AGNs (optical), as shown in Table \ref{tab:statistical} and Figure \ref{fig:dis}. The XRBs display a similar distribution shape to those of the Seyfert galaxies.

Subsequently, we utilized the linear regression tool \texttt{linmix},\footnote{\url{https://github.com/jmeyers314/linmix}} which considers data errors, to find the scaling relationship at hard X-rays. Since the hard X-ray Seyfert galaxies and XRBs both originate from coronae, we explored the scaling relationship between these two. Given that only a small fraction of the sources in our sample have estimated black hole masses, we adopt random values drawn from a uniform distribution in the range 1.4--15 $M_{\odot}$, covering the typical mass range from neutron stars to SBHs. Because the black hole masses in XRBs span a relatively narrow range (approximately 1 order of magnitude), this approach does not significantly affect the determination of the scaling law from XRBs to AGNs. The $\tau_{\text{damping}}^{\text{rest}}\propto M_{\text{BH}}$ relation for XRBs and Seyfert is

\begin{equation}
	\tau_{\text{damping}}^{\text{rest}}=87.21^{+11.47}_{-12.91} \ \text{days} \left( \frac{M_{\text{BH}}}{10^{8}M_{\odot}} \right)^{0.01^{+0.01}_{-0.01}}. \label{equ:1}
\end{equation}

The Pearson correlation coefficient of Equation \ref{equ:1} is $\sim0.15$, indicating that the CDTs have no correlation with black hole mass. A visualization of $\tau_{\text{damping}}^{\text{rest}}\propto M_{\text{BH}}$ is shown as the hot-pink lines in panel (A) of Figure \ref{fig:lin_rel}. For comparison, we also show the $\tau_{\text{damping}}^{\text{rest}}\propto M_{\text{BH}}$ relation for the combined results of jetted accretion systems in $\gamma$ rays (spring-green dashed lines) and nonjetted accretion systems in the optical (gray dashed lines) from previous work \citep{2024MNRAS.527.2672S,2021Sci...373..789B,2024ApJ...967L..18Z,2025PhRvD.111h3049S}. The Pearson correlation coefficients of the jetted and nonjetted systems are $\sim0.96$ and $\sim0.98$, respectively. The slopes of two relations are $\sim0.5$ and $\sim0.58$, respectively, which differ markedly from our results. Panel (B) of Figure \ref{fig:lin_rel} shows a close-up view of the linear relation of this work.

In previous studies, CDTs with scale of months to years in nonjetted accretion systems can be explained by some physical timescales, such as the dynamical timescale, thermal equilibrium timescale, cooling and heating front propagation timescale, viscous timescale, and Compton cooling timescale \citep{2006ASPC..360..265C,2012A&A...540L...2I,2021ApJ...907...96S,2021Sci...373..789B}. According to above paradigm, all of these physical timescales can be scaled with black hole mass:
\begin{equation}
    \begin{aligned}
	& t_{\text{th}}=1680 \left( \frac{\alpha}{0.01} \right)^{-1} \times \left( \frac{M_{\text{BH}}}{10^{8}M_{\odot}} \right)  \left( \frac{R}{100R_{\text{S}}} \right)^{3/2} \text{days} \\ 
	& \text{and} \quad t_{\text{C}}\propto \frac{M_{\text{BH}}^2}{\dot{M}},  \label{equ:2}
    \end{aligned}
\end{equation}
where $t_{\text{th}}\sim\alpha^{-1}t_{\text{dyn}}\sim(H/R)t_{\text{front}}\sim(H/R)^{2}t_{\text{visc}}$, $\alpha$ is the dimensionless viscosity parameter, which usually ranges from 0.01--0.5 in various accretion states \citep{1998tbha.conf..148N,2010ApJ...713...52D,2012A&A...545A.115K,2019NewA...70....7M}, $R_{\text{S}}$ is the Schwarzschild
radius, $R$ is the radial extent of accretion disk, and $t_{\text{C}}$ is the Compton cooling timescale. 

\begin{figure*}[t]
	\includegraphics[width=\textwidth]{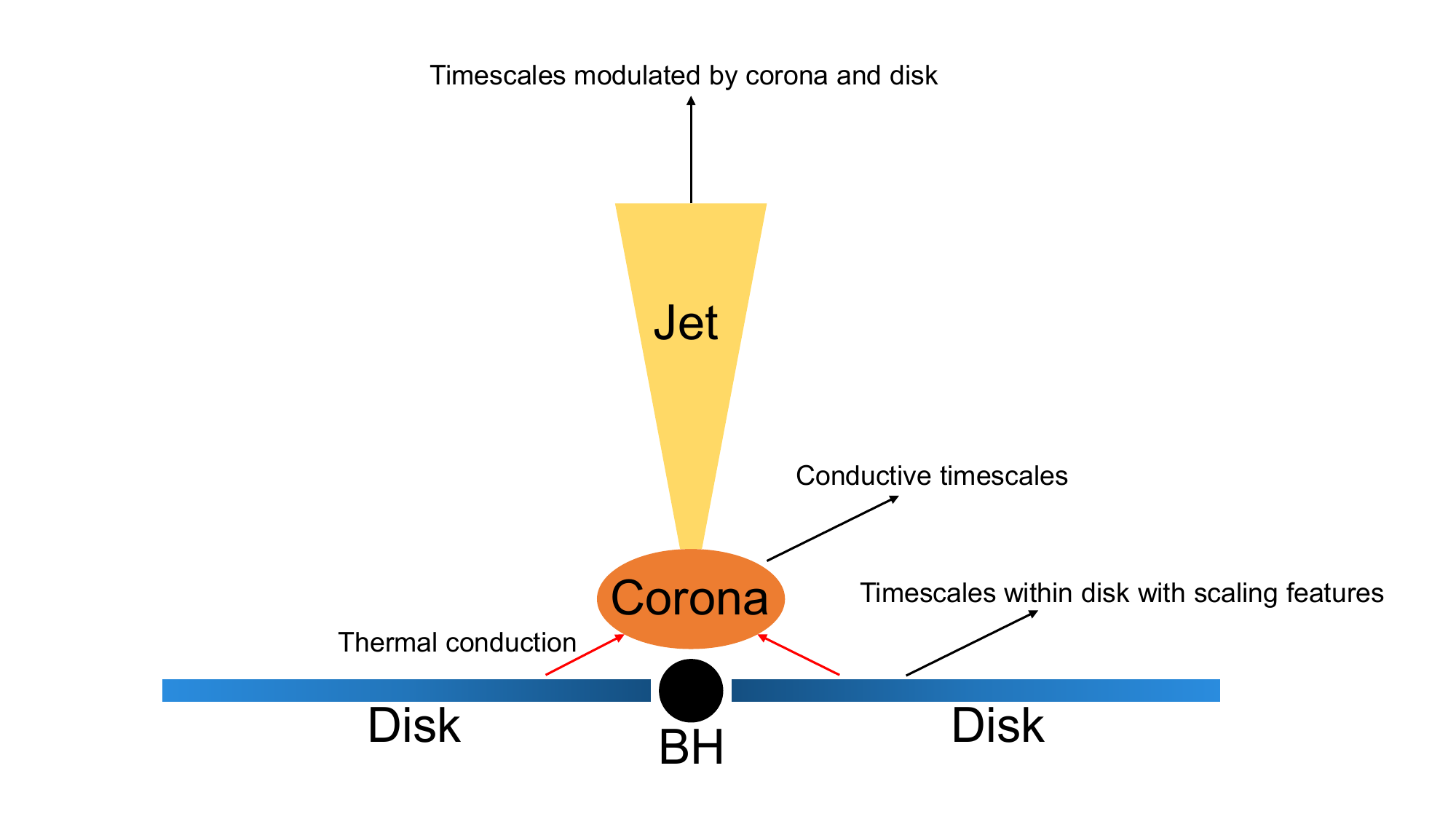}
	\caption{Schematic of CDTs originate from different positions of a black hole accretion system. The CDT, which originates from the accretion disk, follows a scaling relation. Meanwhile, thermal conduction between the disk and the corona gives rise to a conductive timescale. These two processes act together to modulate the timescale manifested in the jet emission.} \label{fig:car_fig}
\end{figure*}

Because of the disk--jet connection \citep{1977MNRAS.179..433B,1982MNRAS.199..883B,2003ApJ...593..667M,2003ApJ...593..184L,2004ApJ...615L...9W,2017ApJ...840...46I,2021ApJ...910L...3W}, CDTs in jets with approximately similar scales as disks should originate from the disks \citep{2019ApJ...885...12R,2024ApJ...967L..18Z}. In addition, within the jet, shocks and magnetic reconnection can affect the variability originating from the accretion disk. Specifically, disturbances/perturbations will cause the system to take a longer time to reach stability \citep{2025MNRAS.537.2380Z}. In result, the CDTs in jetted are expected to be larger than those in nonjetted accretion systems. Therefore, according to Equation \ref{equ:2}, the CDT points (spring-green points) and correlations found for the jetted accretion systems (represented spring-green dashed lines in Figure \ref{fig:lin_rel}, respectively) closely match the theoretical expectations \citep{2021Sci...373..789B,2024ApJ...967L..18Z}. 

In this work, since the hard X-ray CDTs of Seyfert galaxies are very similar to those of nonjetted AGNs in the optical band (see Table \ref{tab:statistical}), the hard X-ray CDTs from the corona are likely to originate in the accretion disk. The physical timescales in the accretion disk can provide reasonable alternative explanations (Equation \ref{equ:2}). Same as previous studies in other bands, the hard X-ray CDTs in blazars are larger than those in Seyfert galaxies (see Table \ref{tab:statistical}). The TDE, as a transient SMBH accretion system, exhibits a smaller CDT than seen for the AGNs (see in Figure \ref{fig:dis}). In the future, larger samples are needed to determine the statistical properties of TDEs.

For XRBs, the above mechanisms operate over timescales ranging from minutes to hours. These can be responsible for the data point in the lower-left corner of panel (A) of Figure \ref{fig:lin_rel}. However, we detected that the CDTs of XRBs have magnitudes similar to those of Seyfert galaxies and nonjetted accretion systems (see Equation \ref{equ:1} and the pale turquoise points in Figure \ref{fig:lin_rel}). It is necessary to seek alternative physical mechanisms to explain this puzzling phenomenon. Hard X-ray emission of XRBs is believed to originate from the inverse-Compton scattering process within the corona \citep{2014ARA&A..52..529Y}. Thermal conduction can occur between the cooler ($\sim10^{4}$ K) accretion disk and the hotter ($\sim10^{8}$ K) corona \citep{1999MNRAS.308..751R,2012ApJ...754...81L}. Based on this, \citet{2025ApJ...988..206Z} demonstrated that for corona emission from AGNs, by invoking reasonable parameter ranges, the conductive timescale can be responsible for CDTs ranging from months to years. According to \citet{2025ApJ...988..206Z}, the conductive timescale can be expressed as
\begin{equation}
	\tau_{\text{cond}} \sim \frac{u}{q/L} \sim \frac{3n_{e}L^{2}k_{\text{B}}}{2f\kappa_{\text{Sp}}} \: \text{s}, \label{equ:3}
\end{equation}
where $u$, $q$, $L$, $n_{e}$, $k_{\text{B}}$, $f$, and $\kappa_{\text{Sp}}$ are the thermal energy density, heat flux, corona size, electron number density, Boltzmann constant, magnetic field suppression factor, and Spitzer thermal conductivity, respectively. For XRBs and AGNs, the main differences in the above parameters lie in the electron number density and corona size. Assuming that the coronae of AGNs and XRBs have comparable sizes (i.e., a few Schwarzschild radii) \citep{2025FrASS..1130392L}. Due to the difference in mass scale of black holes, the physical sizes of the coronae of AGNs are 5--7 orders of magnitude larger than those of XRBs. The electron number density of XRBs and AGNs is $\sim10^{18}-10^{23}$ $\text{cm}^{-3}$ and $\sim10^{10}-10^{12}$ $\text{cm}^{-3}$, respectively \citep{2020AAS...23534604G,2022MNRAS.512..728C}. Thus, the $n_{e}L^{2}$ term in Equation \ref{equ:3} can be at the same order of magnitude for AGNs and XRBs. Therefore, this physical scenario naturally explains the mass-independent hard X-ray variability from XRBs to AGNs. 

In Figure \ref{fig:lin_rel}, similar to our XRB results, the two microquasars have long-term, $\gamma$-ray CDTs spanning months to years, as represented by the black points \citep{2025PhRvD.111h3049S}. Since the physical processes in coronae may modulate the jet emission (i.e., the disk--corona--jet connection) \citep{2020MNRAS.496..245Z,2021ApJ...910L...3W}, we suggest that the $\gamma$-ray CDTs of these two types of sources may originate from their coronae. Like microquasars, the CDTs of the blazars in this work may also originate from coronae. In order to present a clearer explanation of our physical framework, we plotted a schematic of CDTs originate from different positions of a black hole accretion system in Figure \ref{fig:car_fig}.

Consequently, the CDTs of AGNs with magnitudes ranging from months to years can be explained by a series of timescales originating from the accretion disk (Equation \ref{equ:2}), which have a scaling property. The conductive timescale resulting from the interaction between the corona and accretion disk can explain the mass-independent variability from XRBs to AGNs. In AGNs, it is not possible with current data to distinguish whether the month-to-year timescales originate from the accretion disk or corona.
\section{Conclusion}\label{sec:conclusion}
We modeled the long-term hard X-ray variability of XRBs and AGNs from the \textit{Swift}/BAT 157-month LC catalog using the DRW model. We obtained 106 valid CDT measurements. We discovered a new puzzling pattern of variability, namely that the long-term hard X-ray CDTs from XRBs to AGNs are independent of the mass of the accreting object. This clearly differs from the scaling relations with black hole mass reported in previous studies. By adopting reasonable estimates for the electron number density and corona size, we explain that this variability arises from the conductive timescale within the corona. This physical process can further modulates the jet radiation in blazars. 

This work demonstrates thermal conduction as a key mechanism driving hard X-ray variability. Furthermore, this work offers new observational evidence for the disk--corona--jet connection. To unravel the underlying physical processes of this phenomenon, comprehensive global general relativistic magnetohydrodynamic numerical simulations and high-cadence, long-term, multiwavelength observations are essential.
\section*{Acknowledgements}
This work was partly supported by the National Science Foundation of China (grant nos. 12263007 and 12233006), and the High-level Talent Support Program of Yunnan Province.
\section*{Data available}
The detailed fitting results of 106 sources and the original data used for the figures in this paper can be found in the Supplementary Materials \citep{zhang_2025_17830360}.
\bibliography{ref}

\providecommand{\noopsort}[1]{}
\begin{thebibliography}{81}%
\makeatletter
\providecommand \@ifxundefined [1]{%
 \@ifx{#1\undefined}
}%
\providecommand \@ifnum [1]{%
 \ifnum #1\expandafter \@firstoftwo
 \else \expandafter \@secondoftwo
 \fi
}%
\providecommand \@ifx [1]{%
 \ifx #1\expandafter \@firstoftwo
 \else \expandafter \@secondoftwo
 \fi
}%
\providecommand \natexlab [1]{#1}%
\providecommand \enquote  [1]{``#1''}%
\providecommand \bibnamefont  [1]{#1}%
\providecommand \bibfnamefont [1]{#1}%
\providecommand \citenamefont [1]{#1}%
\providecommand \href@noop [0]{\@secondoftwo}%
\providecommand \href [0]{\begingroup \@sanitize@url \@href}%
\providecommand \@href[1]{\@@startlink{#1}\@@href}%
\providecommand \@@href[1]{\endgroup#1\@@endlink}%
\providecommand \@sanitize@url [0]{\catcode `\\12\catcode `\$12\catcode
  `\&12\catcode `\#12\catcode `\^12\catcode `\_12\catcode `\%12\relax}%
\providecommand \@@startlink[1]{}%
\providecommand \@@endlink[0]{}%
\providecommand \url  [0]{\begingroup\@sanitize@url \@url }%
\providecommand \@url [1]{\endgroup\@href {#1}{\urlprefix }}%
\providecommand \urlprefix  [0]{URL }%
\providecommand \Eprint [0]{\href }%
\providecommand \doibase [0]{https://doi.org/}%
\providecommand \selectlanguage [0]{\@gobble}%
\providecommand \bibinfo  [0]{\@secondoftwo}%
\providecommand \bibfield  [0]{\@secondoftwo}%
\providecommand \translation [1]{[#1]}%
\providecommand \BibitemOpen [0]{}%
\providecommand \bibitemStop [0]{}%
\providecommand \bibitemNoStop [0]{.\EOS\space}%
\providecommand \EOS [0]{\spacefactor3000\relax}%
\providecommand \BibitemShut  [1]{\csname bibitem#1\endcsname}%
\let\auto@bib@innerbib\@empty
\bibitem [{\citenamefont {{Remillard}}\ and\ \citenamefont
  {{McClintock}}(2006)}]{2006ARA&A..44...49R}%
  \BibitemOpen
  \bibfield  {author} {\bibinfo {author} {\bibfnamefont {R.~A.}\ \bibnamefont
  {{Remillard}}}\ and\ \bibinfo {author} {\bibfnamefont {J.~E.}\ \bibnamefont
  {{McClintock}}},\ }\bibfield  {title} {\bibinfo {title} {{X-Ray Properties of
  Black-Hole Binaries}},\ }\href
  {https://doi.org/10.1146/annurev.astro.44.051905.092532} {\bibfield
  {journal} {\bibinfo  {journal} {\araa}\ }\textbf {\bibinfo {volume} {44}},\
  \bibinfo {pages} {49} (\bibinfo {year} {2006})},\ \Eprint
  {https://arxiv.org/abs/astro-ph/0606352} {arXiv:astro-ph/0606352 [astro-ph]}
  \BibitemShut {NoStop}%
\bibitem [{\citenamefont {{Done}}\ \emph {et~al.}(2007)\citenamefont {{Done}},
  \citenamefont {{Gierli{\'n}ski}},\ and\ \citenamefont
  {{Kubota}}}]{2007A&ARv..15....1D}%
  \BibitemOpen
  \bibfield  {author} {\bibinfo {author} {\bibfnamefont {C.}~\bibnamefont
  {{Done}}}, \bibinfo {author} {\bibfnamefont {M.}~\bibnamefont
  {{Gierli{\'n}ski}}},\ and\ \bibinfo {author} {\bibfnamefont {A.}~\bibnamefont
  {{Kubota}}},\ }\bibfield  {title} {\bibinfo {title} {{Modelling the behaviour
  of accretion flows in X-ray binaries. Everything you always wanted to know
  about accretion but were afraid to ask}},\ }\href
  {https://doi.org/10.1007/s00159-007-0006-1} {\bibfield  {journal} {\bibinfo
  {journal} {\aapr}\ }\textbf {\bibinfo {volume} {15}},\ \bibinfo {pages} {1}
  (\bibinfo {year} {2007})},\ \Eprint {https://arxiv.org/abs/0708.0148}
  {arXiv:0708.0148 [astro-ph]} \BibitemShut {NoStop}%
\bibitem [{\citenamefont {{Urry}}\ and\ \citenamefont
  {{Padovani}}(1995)}]{1995PASP..107..803U}%
  \BibitemOpen
  \bibfield  {author} {\bibinfo {author} {\bibfnamefont {C.~M.}\ \bibnamefont
  {{Urry}}}\ and\ \bibinfo {author} {\bibfnamefont {P.}~\bibnamefont
  {{Padovani}}},\ }\bibfield  {title} {\bibinfo {title} {{Unified Schemes for
  Radio-Loud Active Galactic Nuclei}},\ }\href {https://doi.org/10.1086/133630}
  {\bibfield  {journal} {\bibinfo  {journal} {\pasp}\ }\textbf {\bibinfo
  {volume} {107}},\ \bibinfo {pages} {803} (\bibinfo {year} {1995})},\ \Eprint
  {https://arxiv.org/abs/astro-ph/9506063} {arXiv:astro-ph/9506063 [astro-ph]}
  \BibitemShut {NoStop}%
\bibitem [{\citenamefont {{Event Horizon Telescope Collaboration}}\ \emph
  {et~al.}(2019)\citenamefont {{Event Horizon Telescope Collaboration}},
  \citenamefont {{Akiyama}}, \citenamefont {{Alberdi}}, \citenamefont {{Alef}},
  \citenamefont {{Asada}}, \citenamefont {{Azulay}}, \citenamefont {{Baczko}},
  \citenamefont {{Ball}}, \citenamefont {{Balokovi{\'c}}}, \citenamefont
  {{Barrett}}, \citenamefont {{Bintley}}, \citenamefont {{Blackburn}},
  \citenamefont {{Boland}}, \citenamefont {{Bouman}}, \citenamefont {{Bower}},
  \citenamefont {{Bremer}}, \citenamefont {{Brinkerink}}, \citenamefont
  {{Brissenden}}, \citenamefont {{Britzen}}, \citenamefont {{Broderick}},
  \citenamefont {{Broguiere}}, \citenamefont {{Bronzwaer}}, \citenamefont
  {{Byun}}, \citenamefont {{Carlstrom}}, \citenamefont {{Chael}}, \citenamefont
  {{Chan}}, \citenamefont {{Chatterjee}}, \citenamefont {{Chatterjee}},
  \citenamefont {{Chen}}, \citenamefont {{Chen}}, \citenamefont {{Cho}},
  \citenamefont {{Christian}}, \citenamefont {{Conway}}, \citenamefont
  {{Cordes}}, \citenamefont {{Crew}}, \citenamefont {{Cui}}, \citenamefont
  {{Davelaar}}, \citenamefont {{De Laurentis}}, \citenamefont {{Deane}},
  \citenamefont {{Dempsey}}, \citenamefont {{Desvignes}}, \citenamefont
  {{Dexter}}, \citenamefont {{Doeleman}}, \citenamefont {{Eatough}},
  \citenamefont {{Falcke}}, \citenamefont {{Fish}}, \citenamefont {{Fomalont}},
  \citenamefont {{Fraga-Encinas}}, \citenamefont {{Freeman}}, \citenamefont
  {{Friberg}}, \citenamefont {{Fromm}}, \citenamefont {{G{\'o}mez}},
  \citenamefont {{Galison}}, \citenamefont {{Gammie}}, \citenamefont
  {{Garc{\'\i}a}}, \citenamefont {{Gentaz}}, \citenamefont {{Georgiev}},
  \citenamefont {{Goddi}}, \citenamefont {{Gold}}, \citenamefont {{Gu}},
  \citenamefont {{Gurwell}}, \citenamefont {{Hada}}, \citenamefont {{Hecht}},
  \citenamefont {{Hesper}}, \citenamefont {{Ho}}, \citenamefont {{Ho}},
  \citenamefont {{Honma}}, \citenamefont {{Huang}}, \citenamefont {{Huang}},
  \citenamefont {{Hughes}}, \citenamefont {{Ikeda}}, \citenamefont {{Inoue}},
  \citenamefont {{Issaoun}}, \citenamefont {{James}}, \citenamefont
  {{Jannuzi}}, \citenamefont {{Janssen}}, \citenamefont {{Jeter}},
  \citenamefont {{Jiang}}, \citenamefont {{Johnson}}, \citenamefont
  {{Jorstad}}, \citenamefont {{Jung}}, \citenamefont {{Karami}}, \citenamefont
  {{Karuppusamy}}, \citenamefont {{Kawashima}}, \citenamefont {{Keating}},
  \citenamefont {{Kettenis}}, \citenamefont {{Kim}}, \citenamefont {{Kim}},
  \citenamefont {{Kim}}, \citenamefont {{Kino}}, \citenamefont {{Koay}},
  \citenamefont {{Koch}}, \citenamefont {{Koyama}}, \citenamefont {{Kramer}},
  \citenamefont {{Kramer}}, \citenamefont {{Krichbaum}}, \citenamefont {{Kuo}},
  \citenamefont {{Lauer}}, \citenamefont {{Lee}}, \citenamefont {{Li}},
  \citenamefont {{Li}}, \citenamefont {{Lindqvist}}, \citenamefont {{Liu}},
  \citenamefont {{Liuzzo}}, \citenamefont {{Lo}}, \citenamefont {{Lobanov}},
  \citenamefont {{Loinard}}, \citenamefont {{Lonsdale}}, \citenamefont {{Lu}},
  \citenamefont {{MacDonald}}, \citenamefont {{Mao}}, \citenamefont
  {{Markoff}}, \citenamefont {{Marrone}}, \citenamefont {{Marscher}},
  \citenamefont {{Mart{\'\i}-Vidal}}, \citenamefont {{Matsushita}},
  \citenamefont {{Matthews}}, \citenamefont {{Medeiros}}, \citenamefont
  {{Menten}}, \citenamefont {{Mizuno}}, \citenamefont {{Mizuno}}, \citenamefont
  {{Moran}}, \citenamefont {{Moriyama}}, \citenamefont {{Moscibrodzka}},
  \citenamefont {{M{\"u}ller}}, \citenamefont {{Nagai}}, \citenamefont
  {{Nagar}}, \citenamefont {{Nakamura}}, \citenamefont {{Narayan}},
  \citenamefont {{Narayanan}}, \citenamefont {{Natarajan}}, \citenamefont
  {{Neri}}, \citenamefont {{Ni}}, \citenamefont {{Noutsos}}, \citenamefont
  {{Okino}}, \citenamefont {{Olivares}}, \citenamefont {{Ortiz-Le{\'o}n}},
  \citenamefont {{Oyama}}, \citenamefont {{{\"O}zel}}, \citenamefont
  {{Palumbo}}, \citenamefont {{Patel}}, \citenamefont {{Pen}}, \citenamefont
  {{Pesce}}, \citenamefont {{Pi{\'e}tu}}, \citenamefont {{Plambeck}},
  \citenamefont {{PopStefanija}}, \citenamefont {{Porth}}, \citenamefont
  {{Prather}}, \citenamefont {{Preciado-L{\'o}pez}}, \citenamefont {{Psaltis}},
  \citenamefont {{Pu}}, \citenamefont {{Ramakrishnan}}, \citenamefont {{Rao}},
  \citenamefont {{Rawlings}}, \citenamefont {{Raymond}}, \citenamefont
  {{Rezzolla}}, \citenamefont {{Ripperda}}, \citenamefont {{Roelofs}},
  \citenamefont {{Rogers}}, \citenamefont {{Ros}}, \citenamefont {{Rose}},
  \citenamefont {{Roshanineshat}}, \citenamefont {{Rottmann}}, \citenamefont
  {{Roy}}, \citenamefont {{Ruszczyk}}, \citenamefont {{Ryan}}, \citenamefont
  {{Rygl}}, \citenamefont {{S{\'a}nchez}}, \citenamefont
  {{S{\'a}nchez-Arguelles}}, \citenamefont {{Sasada}}, \citenamefont
  {{Savolainen}}, \citenamefont {{Schloerb}}, \citenamefont {{Schuster}},
  \citenamefont {{Shao}}, \citenamefont {{Shen}}, \citenamefont {{Small}},
  \citenamefont {{Sohn}}, \citenamefont {{SooHoo}}, \citenamefont {{Tazaki}},
  \citenamefont {{Tiede}}, \citenamefont {{Tilanus}}, \citenamefont {{Titus}},
  \citenamefont {{Toma}}, \citenamefont {{Torne}}, \citenamefont {{Trent}},
  \citenamefont {{Trippe}}, \citenamefont {{Tsuda}}, \citenamefont {{van
  Bemmel}}, \citenamefont {{van Langevelde}}, \citenamefont {{van Rossum}},
  \citenamefont {{Wagner}}, \citenamefont {{Wardle}}, \citenamefont
  {{Weintroub}}, \citenamefont {{Wex}}, \citenamefont {{Wharton}},
  \citenamefont {{Wielgus}}, \citenamefont {{Wong}}, \citenamefont {{Wu}},
  \citenamefont {{Young}},\ and\ \citenamefont
  {{Young}}}]{2019ApJ...875L...1E}%
  \BibitemOpen
  \bibfield  {author} {\bibinfo {author} {\bibnamefont {{Event Horizon
  Telescope Collaboration}}}, \bibinfo {author} {\bibfnamefont
  {K.}~\bibnamefont {{Akiyama}}}, \bibinfo {author} {\bibfnamefont
  {A.}~\bibnamefont {{Alberdi}}}, \bibinfo {author} {\bibfnamefont
  {W.}~\bibnamefont {{Alef}}}, \bibinfo {author} {\bibfnamefont
  {K.}~\bibnamefont {{Asada}}}, \bibinfo {author} {\bibfnamefont
  {R.}~\bibnamefont {{Azulay}}}, \bibinfo {author} {\bibfnamefont {A.-K.}\
  \bibnamefont {{Baczko}}}, \bibinfo {author} {\bibfnamefont {D.}~\bibnamefont
  {{Ball}}}, \bibinfo {author} {\bibfnamefont {M.}~\bibnamefont
  {{Balokovi{\'c}}}}, \bibinfo {author} {\bibfnamefont {J.}~\bibnamefont
  {{Barrett}}}, \bibinfo {author} {\bibfnamefont {D.}~\bibnamefont
  {{Bintley}}}, \bibinfo {author} {\bibfnamefont {L.}~\bibnamefont
  {{Blackburn}}}, \bibinfo {author} {\bibfnamefont {W.}~\bibnamefont
  {{Boland}}}, \bibinfo {author} {\bibfnamefont {K.~L.}\ \bibnamefont
  {{Bouman}}}, \bibinfo {author} {\bibfnamefont {G.~C.}\ \bibnamefont
  {{Bower}}}, \bibinfo {author} {\bibfnamefont {M.}~\bibnamefont {{Bremer}}},
  \bibinfo {author} {\bibfnamefont {C.~D.}\ \bibnamefont {{Brinkerink}}},
  \bibinfo {author} {\bibfnamefont {R.}~\bibnamefont {{Brissenden}}}, \bibinfo
  {author} {\bibfnamefont {S.}~\bibnamefont {{Britzen}}}, \bibinfo {author}
  {\bibfnamefont {A.~E.}\ \bibnamefont {{Broderick}}}, \bibinfo {author}
  {\bibfnamefont {D.}~\bibnamefont {{Broguiere}}}, \bibinfo {author}
  {\bibfnamefont {T.}~\bibnamefont {{Bronzwaer}}}, \bibinfo {author}
  {\bibfnamefont {D.-Y.}\ \bibnamefont {{Byun}}}, \bibinfo {author}
  {\bibfnamefont {J.~E.}\ \bibnamefont {{Carlstrom}}}, \bibinfo {author}
  {\bibfnamefont {A.}~\bibnamefont {{Chael}}}, \bibinfo {author} {\bibfnamefont
  {C.-k.}\ \bibnamefont {{Chan}}}, \bibinfo {author} {\bibfnamefont
  {S.}~\bibnamefont {{Chatterjee}}}, \bibinfo {author} {\bibfnamefont
  {K.}~\bibnamefont {{Chatterjee}}}, \bibinfo {author} {\bibfnamefont {M.-T.}\
  \bibnamefont {{Chen}}}, \bibinfo {author} {\bibfnamefont {Y.}~\bibnamefont
  {{Chen}}}, \bibinfo {author} {\bibfnamefont {I.}~\bibnamefont {{Cho}}},
  \bibinfo {author} {\bibfnamefont {P.}~\bibnamefont {{Christian}}}, \bibinfo
  {author} {\bibfnamefont {J.~E.}\ \bibnamefont {{Conway}}}, \bibinfo {author}
  {\bibfnamefont {J.~M.}\ \bibnamefont {{Cordes}}}, \bibinfo {author}
  {\bibfnamefont {G.~B.}\ \bibnamefont {{Crew}}}, \bibinfo {author}
  {\bibfnamefont {Y.}~\bibnamefont {{Cui}}}, \bibinfo {author} {\bibfnamefont
  {J.}~\bibnamefont {{Davelaar}}}, \bibinfo {author} {\bibfnamefont
  {M.}~\bibnamefont {{De Laurentis}}}, \bibinfo {author} {\bibfnamefont
  {R.}~\bibnamefont {{Deane}}}, \bibinfo {author} {\bibfnamefont
  {J.}~\bibnamefont {{Dempsey}}}, \bibinfo {author} {\bibfnamefont
  {G.}~\bibnamefont {{Desvignes}}}, \bibinfo {author} {\bibfnamefont
  {J.}~\bibnamefont {{Dexter}}}, \bibinfo {author} {\bibfnamefont {S.~S.}\
  \bibnamefont {{Doeleman}}}, \bibinfo {author} {\bibfnamefont {R.~P.}\
  \bibnamefont {{Eatough}}}, \bibinfo {author} {\bibfnamefont {H.}~\bibnamefont
  {{Falcke}}}, \bibinfo {author} {\bibfnamefont {V.~L.}\ \bibnamefont
  {{Fish}}}, \bibinfo {author} {\bibfnamefont {E.}~\bibnamefont {{Fomalont}}},
  \bibinfo {author} {\bibfnamefont {R.}~\bibnamefont {{Fraga-Encinas}}},
  \bibinfo {author} {\bibfnamefont {W.~T.}\ \bibnamefont {{Freeman}}}, \bibinfo
  {author} {\bibfnamefont {P.}~\bibnamefont {{Friberg}}}, \bibinfo {author}
  {\bibfnamefont {C.~M.}\ \bibnamefont {{Fromm}}}, \bibinfo {author}
  {\bibfnamefont {J.~L.}\ \bibnamefont {{G{\'o}mez}}}, \bibinfo {author}
  {\bibfnamefont {P.}~\bibnamefont {{Galison}}}, \bibinfo {author}
  {\bibfnamefont {C.~F.}\ \bibnamefont {{Gammie}}}, \bibinfo {author}
  {\bibfnamefont {R.}~\bibnamefont {{Garc{\'\i}a}}}, \bibinfo {author}
  {\bibfnamefont {O.}~\bibnamefont {{Gentaz}}}, \bibinfo {author}
  {\bibfnamefont {B.}~\bibnamefont {{Georgiev}}}, \bibinfo {author}
  {\bibfnamefont {C.}~\bibnamefont {{Goddi}}}, \bibinfo {author} {\bibfnamefont
  {R.}~\bibnamefont {{Gold}}}, \bibinfo {author} {\bibfnamefont
  {M.}~\bibnamefont {{Gu}}}, \bibinfo {author} {\bibfnamefont {M.}~\bibnamefont
  {{Gurwell}}}, \bibinfo {author} {\bibfnamefont {K.}~\bibnamefont {{Hada}}},
  \bibinfo {author} {\bibfnamefont {M.~H.}\ \bibnamefont {{Hecht}}}, \bibinfo
  {author} {\bibfnamefont {R.}~\bibnamefont {{Hesper}}}, \bibinfo {author}
  {\bibfnamefont {L.~C.}\ \bibnamefont {{Ho}}}, \bibinfo {author}
  {\bibfnamefont {P.}~\bibnamefont {{Ho}}}, \bibinfo {author} {\bibfnamefont
  {M.}~\bibnamefont {{Honma}}}, \bibinfo {author} {\bibfnamefont {C.-W.~L.}\
  \bibnamefont {{Huang}}}, \bibinfo {author} {\bibfnamefont {L.}~\bibnamefont
  {{Huang}}}, \bibinfo {author} {\bibfnamefont {D.~H.}\ \bibnamefont
  {{Hughes}}}, \bibinfo {author} {\bibfnamefont {S.}~\bibnamefont {{Ikeda}}},
  \bibinfo {author} {\bibfnamefont {M.}~\bibnamefont {{Inoue}}}, \bibinfo
  {author} {\bibfnamefont {S.}~\bibnamefont {{Issaoun}}}, \bibinfo {author}
  {\bibfnamefont {D.~J.}\ \bibnamefont {{James}}}, \bibinfo {author}
  {\bibfnamefont {B.~T.}\ \bibnamefont {{Jannuzi}}}, \bibinfo {author}
  {\bibfnamefont {M.}~\bibnamefont {{Janssen}}}, \bibinfo {author}
  {\bibfnamefont {B.}~\bibnamefont {{Jeter}}}, \bibinfo {author} {\bibfnamefont
  {W.}~\bibnamefont {{Jiang}}}, \bibinfo {author} {\bibfnamefont {M.~D.}\
  \bibnamefont {{Johnson}}}, \bibinfo {author} {\bibfnamefont {S.}~\bibnamefont
  {{Jorstad}}}, \bibinfo {author} {\bibfnamefont {T.}~\bibnamefont {{Jung}}},
  \bibinfo {author} {\bibfnamefont {M.}~\bibnamefont {{Karami}}}, \bibinfo
  {author} {\bibfnamefont {R.}~\bibnamefont {{Karuppusamy}}}, \bibinfo {author}
  {\bibfnamefont {T.}~\bibnamefont {{Kawashima}}}, \bibinfo {author}
  {\bibfnamefont {G.~K.}\ \bibnamefont {{Keating}}}, \bibinfo {author}
  {\bibfnamefont {M.}~\bibnamefont {{Kettenis}}}, \bibinfo {author}
  {\bibfnamefont {J.-Y.}\ \bibnamefont {{Kim}}}, \bibinfo {author}
  {\bibfnamefont {J.}~\bibnamefont {{Kim}}}, \bibinfo {author} {\bibfnamefont
  {J.}~\bibnamefont {{Kim}}}, \bibinfo {author} {\bibfnamefont
  {M.}~\bibnamefont {{Kino}}}, \bibinfo {author} {\bibfnamefont {J.~Y.}\
  \bibnamefont {{Koay}}}, \bibinfo {author} {\bibfnamefont {P.~M.}\
  \bibnamefont {{Koch}}}, \bibinfo {author} {\bibfnamefont {S.}~\bibnamefont
  {{Koyama}}}, \bibinfo {author} {\bibfnamefont {M.}~\bibnamefont {{Kramer}}},
  \bibinfo {author} {\bibfnamefont {C.}~\bibnamefont {{Kramer}}}, \bibinfo
  {author} {\bibfnamefont {T.~P.}\ \bibnamefont {{Krichbaum}}}, \bibinfo
  {author} {\bibfnamefont {C.-Y.}\ \bibnamefont {{Kuo}}}, \bibinfo {author}
  {\bibfnamefont {T.~R.}\ \bibnamefont {{Lauer}}}, \bibinfo {author}
  {\bibfnamefont {S.-S.}\ \bibnamefont {{Lee}}}, \bibinfo {author}
  {\bibfnamefont {Y.-R.}\ \bibnamefont {{Li}}}, \bibinfo {author}
  {\bibfnamefont {Z.}~\bibnamefont {{Li}}}, \bibinfo {author} {\bibfnamefont
  {M.}~\bibnamefont {{Lindqvist}}}, \bibinfo {author} {\bibfnamefont
  {K.}~\bibnamefont {{Liu}}}, \bibinfo {author} {\bibfnamefont
  {E.}~\bibnamefont {{Liuzzo}}}, \bibinfo {author} {\bibfnamefont {W.-P.}\
  \bibnamefont {{Lo}}}, \bibinfo {author} {\bibfnamefont {A.~P.}\ \bibnamefont
  {{Lobanov}}}, \bibinfo {author} {\bibfnamefont {L.}~\bibnamefont
  {{Loinard}}}, \bibinfo {author} {\bibfnamefont {C.}~\bibnamefont
  {{Lonsdale}}}, \bibinfo {author} {\bibfnamefont {R.-S.}\ \bibnamefont
  {{Lu}}}, \bibinfo {author} {\bibfnamefont {N.~R.}\ \bibnamefont
  {{MacDonald}}}, \bibinfo {author} {\bibfnamefont {J.}~\bibnamefont {{Mao}}},
  \bibinfo {author} {\bibfnamefont {S.}~\bibnamefont {{Markoff}}}, \bibinfo
  {author} {\bibfnamefont {D.~P.}\ \bibnamefont {{Marrone}}}, \bibinfo {author}
  {\bibfnamefont {A.~P.}\ \bibnamefont {{Marscher}}}, \bibinfo {author}
  {\bibfnamefont {I.}~\bibnamefont {{Mart{\'\i}-Vidal}}}, \bibinfo {author}
  {\bibfnamefont {S.}~\bibnamefont {{Matsushita}}}, \bibinfo {author}
  {\bibfnamefont {L.~D.}\ \bibnamefont {{Matthews}}}, \bibinfo {author}
  {\bibfnamefont {L.}~\bibnamefont {{Medeiros}}}, \bibinfo {author}
  {\bibfnamefont {K.~M.}\ \bibnamefont {{Menten}}}, \bibinfo {author}
  {\bibfnamefont {Y.}~\bibnamefont {{Mizuno}}}, \bibinfo {author}
  {\bibfnamefont {I.}~\bibnamefont {{Mizuno}}}, \bibinfo {author}
  {\bibfnamefont {J.~M.}\ \bibnamefont {{Moran}}}, \bibinfo {author}
  {\bibfnamefont {K.}~\bibnamefont {{Moriyama}}}, \bibinfo {author}
  {\bibfnamefont {M.}~\bibnamefont {{Moscibrodzka}}}, \bibinfo {author}
  {\bibfnamefont {C.}~\bibnamefont {{M{\"u}ller}}}, \bibinfo {author}
  {\bibfnamefont {H.}~\bibnamefont {{Nagai}}}, \bibinfo {author} {\bibfnamefont
  {N.~M.}\ \bibnamefont {{Nagar}}}, \bibinfo {author} {\bibfnamefont
  {M.}~\bibnamefont {{Nakamura}}}, \bibinfo {author} {\bibfnamefont
  {R.}~\bibnamefont {{Narayan}}}, \bibinfo {author} {\bibfnamefont
  {G.}~\bibnamefont {{Narayanan}}}, \bibinfo {author} {\bibfnamefont
  {I.}~\bibnamefont {{Natarajan}}}, \bibinfo {author} {\bibfnamefont
  {R.}~\bibnamefont {{Neri}}}, \bibinfo {author} {\bibfnamefont
  {C.}~\bibnamefont {{Ni}}}, \bibinfo {author} {\bibfnamefont {A.}~\bibnamefont
  {{Noutsos}}}, \bibinfo {author} {\bibfnamefont {H.}~\bibnamefont {{Okino}}},
  \bibinfo {author} {\bibfnamefont {H.}~\bibnamefont {{Olivares}}}, \bibinfo
  {author} {\bibfnamefont {G.~N.}\ \bibnamefont {{Ortiz-Le{\'o}n}}}, \bibinfo
  {author} {\bibfnamefont {T.}~\bibnamefont {{Oyama}}}, \bibinfo {author}
  {\bibfnamefont {F.}~\bibnamefont {{{\"O}zel}}}, \bibinfo {author}
  {\bibfnamefont {D.~C.~M.}\ \bibnamefont {{Palumbo}}}, \bibinfo {author}
  {\bibfnamefont {N.}~\bibnamefont {{Patel}}}, \bibinfo {author} {\bibfnamefont
  {U.-L.}\ \bibnamefont {{Pen}}}, \bibinfo {author} {\bibfnamefont {D.~W.}\
  \bibnamefont {{Pesce}}}, \bibinfo {author} {\bibfnamefont {V.}~\bibnamefont
  {{Pi{\'e}tu}}}, \bibinfo {author} {\bibfnamefont {R.}~\bibnamefont
  {{Plambeck}}}, \bibinfo {author} {\bibfnamefont {A.}~\bibnamefont
  {{PopStefanija}}}, \bibinfo {author} {\bibfnamefont {O.}~\bibnamefont
  {{Porth}}}, \bibinfo {author} {\bibfnamefont {B.}~\bibnamefont {{Prather}}},
  \bibinfo {author} {\bibfnamefont {J.~A.}\ \bibnamefont
  {{Preciado-L{\'o}pez}}}, \bibinfo {author} {\bibfnamefont {D.}~\bibnamefont
  {{Psaltis}}}, \bibinfo {author} {\bibfnamefont {H.-Y.}\ \bibnamefont {{Pu}}},
  \bibinfo {author} {\bibfnamefont {V.}~\bibnamefont {{Ramakrishnan}}},
  \bibinfo {author} {\bibfnamefont {R.}~\bibnamefont {{Rao}}}, \bibinfo
  {author} {\bibfnamefont {M.~G.}\ \bibnamefont {{Rawlings}}}, \bibinfo
  {author} {\bibfnamefont {A.~W.}\ \bibnamefont {{Raymond}}}, \bibinfo {author}
  {\bibfnamefont {L.}~\bibnamefont {{Rezzolla}}}, \bibinfo {author}
  {\bibfnamefont {B.}~\bibnamefont {{Ripperda}}}, \bibinfo {author}
  {\bibfnamefont {F.}~\bibnamefont {{Roelofs}}}, \bibinfo {author}
  {\bibfnamefont {A.}~\bibnamefont {{Rogers}}}, \bibinfo {author}
  {\bibfnamefont {E.}~\bibnamefont {{Ros}}}, \bibinfo {author} {\bibfnamefont
  {M.}~\bibnamefont {{Rose}}}, \bibinfo {author} {\bibfnamefont
  {A.}~\bibnamefont {{Roshanineshat}}}, \bibinfo {author} {\bibfnamefont
  {H.}~\bibnamefont {{Rottmann}}}, \bibinfo {author} {\bibfnamefont {A.~L.}\
  \bibnamefont {{Roy}}}, \bibinfo {author} {\bibfnamefont {C.}~\bibnamefont
  {{Ruszczyk}}}, \bibinfo {author} {\bibfnamefont {B.~R.}\ \bibnamefont
  {{Ryan}}}, \bibinfo {author} {\bibfnamefont {K.~L.~J.}\ \bibnamefont
  {{Rygl}}}, \bibinfo {author} {\bibfnamefont {S.}~\bibnamefont
  {{S{\'a}nchez}}}, \bibinfo {author} {\bibfnamefont {D.}~\bibnamefont
  {{S{\'a}nchez-Arguelles}}}, \bibinfo {author} {\bibfnamefont
  {M.}~\bibnamefont {{Sasada}}}, \bibinfo {author} {\bibfnamefont
  {T.}~\bibnamefont {{Savolainen}}}, \bibinfo {author} {\bibfnamefont {F.~P.}\
  \bibnamefont {{Schloerb}}}, \bibinfo {author} {\bibfnamefont {K.-F.}\
  \bibnamefont {{Schuster}}}, \bibinfo {author} {\bibfnamefont
  {L.}~\bibnamefont {{Shao}}}, \bibinfo {author} {\bibfnamefont
  {Z.}~\bibnamefont {{Shen}}}, \bibinfo {author} {\bibfnamefont
  {D.}~\bibnamefont {{Small}}}, \bibinfo {author} {\bibfnamefont {B.~W.}\
  \bibnamefont {{Sohn}}}, \bibinfo {author} {\bibfnamefont {J.}~\bibnamefont
  {{SooHoo}}}, \bibinfo {author} {\bibfnamefont {F.}~\bibnamefont {{Tazaki}}},
  \bibinfo {author} {\bibfnamefont {P.}~\bibnamefont {{Tiede}}}, \bibinfo
  {author} {\bibfnamefont {R.~P.~J.}\ \bibnamefont {{Tilanus}}}, \bibinfo
  {author} {\bibfnamefont {M.}~\bibnamefont {{Titus}}}, \bibinfo {author}
  {\bibfnamefont {K.}~\bibnamefont {{Toma}}}, \bibinfo {author} {\bibfnamefont
  {P.}~\bibnamefont {{Torne}}}, \bibinfo {author} {\bibfnamefont
  {T.}~\bibnamefont {{Trent}}}, \bibinfo {author} {\bibfnamefont
  {S.}~\bibnamefont {{Trippe}}}, \bibinfo {author} {\bibfnamefont
  {S.}~\bibnamefont {{Tsuda}}}, \bibinfo {author} {\bibfnamefont
  {I.}~\bibnamefont {{van Bemmel}}}, \bibinfo {author} {\bibfnamefont {H.~J.}\
  \bibnamefont {{van Langevelde}}}, \bibinfo {author} {\bibfnamefont {D.~R.}\
  \bibnamefont {{van Rossum}}}, \bibinfo {author} {\bibfnamefont
  {J.}~\bibnamefont {{Wagner}}}, \bibinfo {author} {\bibfnamefont
  {J.}~\bibnamefont {{Wardle}}}, \bibinfo {author} {\bibfnamefont
  {J.}~\bibnamefont {{Weintroub}}}, \bibinfo {author} {\bibfnamefont
  {N.}~\bibnamefont {{Wex}}}, \bibinfo {author} {\bibfnamefont
  {R.}~\bibnamefont {{Wharton}}}, \bibinfo {author} {\bibfnamefont
  {M.}~\bibnamefont {{Wielgus}}}, \bibinfo {author} {\bibfnamefont {G.~N.}\
  \bibnamefont {{Wong}}}, \bibinfo {author} {\bibfnamefont {Q.}~\bibnamefont
  {{Wu}}}, \bibinfo {author} {\bibfnamefont {K.}~\bibnamefont {{Young}}},\ and\
  \bibinfo {author} {\bibfnamefont {A.}~\bibnamefont {{Young}}},\ }\bibfield
  {title} {\bibinfo {title} {{First M87 Event Horizon Telescope Results. I. The
  Shadow of the Supermassive Black Hole}},\ }\href
  {https://doi.org/10.3847/2041-8213/ab0ec7} {\bibfield  {journal} {\bibinfo
  {journal} {\apjl}\ }\textbf {\bibinfo {volume} {875}},\ \bibinfo {eid} {L1}
  (\bibinfo {year} {2019})},\ \Eprint {https://arxiv.org/abs/1906.11238}
  {arXiv:1906.11238 [astro-ph.GA]} \BibitemShut {NoStop}%
\bibitem [{\citenamefont {{Romero}}\ \emph {et~al.}(2017)\citenamefont
  {{Romero}}, \citenamefont {{Boettcher}}, \citenamefont {{Markoff}},\ and\
  \citenamefont {{Tavecchio}}}]{2017SSRv..207....5R}%
  \BibitemOpen
  \bibfield  {author} {\bibinfo {author} {\bibfnamefont {G.~E.}\ \bibnamefont
  {{Romero}}}, \bibinfo {author} {\bibfnamefont {M.}~\bibnamefont
  {{Boettcher}}}, \bibinfo {author} {\bibfnamefont {S.}~\bibnamefont
  {{Markoff}}},\ and\ \bibinfo {author} {\bibfnamefont {F.}~\bibnamefont
  {{Tavecchio}}},\ }\bibfield  {title} {\bibinfo {title} {{Relativistic Jets in
  Active Galactic Nuclei and Microquasars}},\ }\href
  {https://doi.org/10.1007/s11214-016-0328-2} {\bibfield  {journal} {\bibinfo
  {journal} {\ssr}\ }\textbf {\bibinfo {volume} {207}},\ \bibinfo {pages} {5}
  (\bibinfo {year} {2017})},\ \Eprint {https://arxiv.org/abs/1611.09507}
  {arXiv:1611.09507 [astro-ph.HE]} \BibitemShut {NoStop}%
\bibitem [{\citenamefont {{Reig}}\ and\ \citenamefont
  {{Fabregat}}(2015)}]{2015A&A...574A..33R}%
  \BibitemOpen
  \bibfield  {author} {\bibinfo {author} {\bibfnamefont {P.}~\bibnamefont
  {{Reig}}}\ and\ \bibinfo {author} {\bibfnamefont {J.}~\bibnamefont
  {{Fabregat}}},\ }\bibfield  {title} {\bibinfo {title} {{Long-term variability
  of high-mass X-ray binaries. I. Photometry}},\ }\href
  {https://doi.org/10.1051/0004-6361/201425008} {\bibfield  {journal} {\bibinfo
   {journal} {\aap}\ }\textbf {\bibinfo {volume} {574}},\ \bibinfo {eid} {A33}
  (\bibinfo {year} {2015})},\ \Eprint {https://arxiv.org/abs/1411.7163}
  {arXiv:1411.7163 [astro-ph.HE]} \BibitemShut {NoStop}%
\bibitem [{\citenamefont {{Dai}}\ \emph {et~al.}(2015)\citenamefont {{Dai}},
  \citenamefont {{Zeng}}, \citenamefont {{Jiang}}, \citenamefont {{Fan}},
  \citenamefont {{Hu}}, \citenamefont {{Zhang}}, \citenamefont {{Yang}},
  \citenamefont {{Yan}}, \citenamefont {{Wang}},\ and\ \citenamefont
  {{Zhang}}}]{2015ApJS..218...18D}%
  \BibitemOpen
  \bibfield  {author} {\bibinfo {author} {\bibfnamefont {B.-z.}\ \bibnamefont
  {{Dai}}}, \bibinfo {author} {\bibfnamefont {W.}~\bibnamefont {{Zeng}}},
  \bibinfo {author} {\bibfnamefont {Z.-j.}\ \bibnamefont {{Jiang}}}, \bibinfo
  {author} {\bibfnamefont {Z.-h.}\ \bibnamefont {{Fan}}}, \bibinfo {author}
  {\bibfnamefont {W.}~\bibnamefont {{Hu}}}, \bibinfo {author} {\bibfnamefont
  {P.-f.}\ \bibnamefont {{Zhang}}}, \bibinfo {author} {\bibfnamefont {Q.-y.}\
  \bibnamefont {{Yang}}}, \bibinfo {author} {\bibfnamefont {D.-h.}\
  \bibnamefont {{Yan}}}, \bibinfo {author} {\bibfnamefont {D.}~\bibnamefont
  {{Wang}}},\ and\ \bibinfo {author} {\bibfnamefont {L.}~\bibnamefont
  {{Zhang}}},\ }\bibfield  {title} {\bibinfo {title} {{Long-term Multi-band
  Photometric Monitoring of Blazar S5 0716+714}},\ }\href
  {https://doi.org/10.1088/0067-0049/218/2/18} {\bibfield  {journal} {\bibinfo
  {journal} {\apjs}\ }\textbf {\bibinfo {volume} {218}},\ \bibinfo {eid} {18}
  (\bibinfo {year} {2015})}\BibitemShut {NoStop}%
\bibitem [{\citenamefont {{Padovani}}\ \emph {et~al.}(2017)\citenamefont
  {{Padovani}}, \citenamefont {{Alexander}}, \citenamefont {{Assef}},
  \citenamefont {{De Marco}}, \citenamefont {{Giommi}}, \citenamefont
  {{Hickox}}, \citenamefont {{Richards}}, \citenamefont {{Smol{\v{c}}i{\'c}}},
  \citenamefont {{Hatziminaoglou}}, \citenamefont {{Mainieri}},\ and\
  \citenamefont {{Salvato}}}]{2017A&ARv..25....2P}%
  \BibitemOpen
  \bibfield  {author} {\bibinfo {author} {\bibfnamefont {P.}~\bibnamefont
  {{Padovani}}}, \bibinfo {author} {\bibfnamefont {D.~M.}\ \bibnamefont
  {{Alexander}}}, \bibinfo {author} {\bibfnamefont {R.~J.}\ \bibnamefont
  {{Assef}}}, \bibinfo {author} {\bibfnamefont {B.}~\bibnamefont {{De Marco}}},
  \bibinfo {author} {\bibfnamefont {P.}~\bibnamefont {{Giommi}}}, \bibinfo
  {author} {\bibfnamefont {R.~C.}\ \bibnamefont {{Hickox}}}, \bibinfo {author}
  {\bibfnamefont {G.~T.}\ \bibnamefont {{Richards}}}, \bibinfo {author}
  {\bibfnamefont {V.}~\bibnamefont {{Smol{\v{c}}i{\'c}}}}, \bibinfo {author}
  {\bibfnamefont {E.}~\bibnamefont {{Hatziminaoglou}}}, \bibinfo {author}
  {\bibfnamefont {V.}~\bibnamefont {{Mainieri}}},\ and\ \bibinfo {author}
  {\bibfnamefont {M.}~\bibnamefont {{Salvato}}},\ }\bibfield  {title} {\bibinfo
  {title} {{Active galactic nuclei: what's in a name?}},\ }\href
  {https://doi.org/10.1007/s00159-017-0102-9} {\bibfield  {journal} {\bibinfo
  {journal} {\aapr}\ }\textbf {\bibinfo {volume} {25}},\ \bibinfo {eid} {2}
  (\bibinfo {year} {2017})},\ \Eprint {https://arxiv.org/abs/1707.07134}
  {arXiv:1707.07134 [astro-ph.GA]} \BibitemShut {NoStop}%
\bibitem [{\citenamefont {{Gierli{\'n}ski}}\ \emph {et~al.}(2008)\citenamefont
  {{Gierli{\'n}ski}}, \citenamefont {{Middleton}}, \citenamefont {{Ward}},\
  and\ \citenamefont {{Done}}}]{2008Natur.455..369G}%
  \BibitemOpen
  \bibfield  {author} {\bibinfo {author} {\bibfnamefont {M.}~\bibnamefont
  {{Gierli{\'n}ski}}}, \bibinfo {author} {\bibfnamefont {M.}~\bibnamefont
  {{Middleton}}}, \bibinfo {author} {\bibfnamefont {M.}~\bibnamefont
  {{Ward}}},\ and\ \bibinfo {author} {\bibfnamefont {C.}~\bibnamefont
  {{Done}}},\ }\bibfield  {title} {\bibinfo {title} {{A periodicity of
  \raisebox{-0.5ex}\textasciitilde1hour in X-ray emission from the active
  galaxy RE J1034+396}},\ }\href {https://doi.org/10.1038/nature07277}
  {\bibfield  {journal} {\bibinfo  {journal} {\nat}\ }\textbf {\bibinfo
  {volume} {455}},\ \bibinfo {pages} {369} (\bibinfo {year}
  {2008})}\BibitemShut {NoStop}%
\bibitem [{\citenamefont {{Mirabel}}\ and\ \citenamefont
  {{Rodr{\'\i}guez}}(1999)}]{1999ARA&A..37..409M}%
  \BibitemOpen
  \bibfield  {author} {\bibinfo {author} {\bibfnamefont {I.~F.}\ \bibnamefont
  {{Mirabel}}}\ and\ \bibinfo {author} {\bibfnamefont {L.~F.}\ \bibnamefont
  {{Rodr{\'\i}guez}}},\ }\bibfield  {title} {\bibinfo {title} {{Sources of
  Relativistic Jets in the Galaxy}},\ }\href
  {https://doi.org/10.1146/annurev.astro.37.1.409} {\bibfield  {journal}
  {\bibinfo  {journal} {\araa}\ }\textbf {\bibinfo {volume} {37}},\ \bibinfo
  {pages} {409} (\bibinfo {year} {1999})},\ \Eprint
  {https://arxiv.org/abs/astro-ph/9902062} {arXiv:astro-ph/9902062 [astro-ph]}
  \BibitemShut {NoStop}%
\bibitem [{\citenamefont {{Abeysekara}}\ \emph {et~al.}(2018)\citenamefont
  {{Abeysekara}}, \citenamefont {{Albert}}, \citenamefont {{Alfaro}},
  \citenamefont {{Alvarez}}, \citenamefont {{{\'A}lvarez}}, \citenamefont
  {{Arceo}}, \citenamefont {{Arteaga-Vel{\'a}zquez}}, \citenamefont {{Avila
  Rojas}}, \citenamefont {{Ayala Solares}}, \citenamefont {{Belmont-Moreno}},
  \citenamefont {{BenZvi}}, \citenamefont {{Brisbois}}, \citenamefont
  {{Caballero-Mora}}, \citenamefont {{Capistr{\'a}n}}, \citenamefont
  {{Carrami{\~n}ana}}, \citenamefont {{Casanova}}, \citenamefont {{Castillo}},
  \citenamefont {{Cotti}}, \citenamefont {{Cotzomi}}, \citenamefont
  {{Couti{\~n}o de Le{\'o}n}}, \citenamefont {{De Le{\'o}n}}, \citenamefont
  {{De la Fuente}}, \citenamefont {{D{\'\i}az-V{\'e}lez}}, \citenamefont
  {{Dichiara}}, \citenamefont {{Dingus}}, \citenamefont {{DuVernois}},
  \citenamefont {{Ellsworth}}, \citenamefont {{Engel}}, \citenamefont
  {{Espinoza}}, \citenamefont {{Fang}}, \citenamefont {{Fleischhack}},
  \citenamefont {{Fraija}}, \citenamefont {{Galv{\'a}n-G{\'a}mez}},
  \citenamefont {{Garc{\'\i}a-Gonz{\'a}lez}}, \citenamefont {{Garfias}},
  \citenamefont {{Gonz{\'a}lez-Mu{\~n}oz}}, \citenamefont {{Gonz{\'a}lez}},
  \citenamefont {{Goodman}}, \citenamefont {{Hampel-Arias}}, \citenamefont
  {{Harding}}, \citenamefont {{Hernandez}}, \citenamefont {{Hinton}},
  \citenamefont {{Hona}}, \citenamefont {{Hueyotl-Zahuantitla}}, \citenamefont
  {{Hui}}, \citenamefont {{H{\"u}ntemeyer}}, \citenamefont {{Iriarte}},
  \citenamefont {{Jardin-Blicq}}, \citenamefont {{Joshi}}, \citenamefont
  {{Kaufmann}}, \citenamefont {{Kar}}, \citenamefont {{Kunde}}, \citenamefont
  {{Lauer}}, \citenamefont {{Lee}}, \citenamefont {{Le{\'o}n Vargas}},
  \citenamefont {{Li}}, \citenamefont {{Linnemann}}, \citenamefont
  {{Longinotti}}, \citenamefont {{Luis-Raya}}, \citenamefont
  {{L{\'o}pez-Coto}}, \citenamefont {{Malone}}, \citenamefont {{Marinelli}},
  \citenamefont {{Martinez}}, \citenamefont {{Martinez-Castellanos}},
  \citenamefont {{Mart{\'\i}nez-Castro}}, \citenamefont {{Matthews}},
  \citenamefont {{Miranda-Romagnoli}}, \citenamefont {{Moreno}}, \citenamefont
  {{Mostaf{\'a}}}, \citenamefont {{Nayerhoda}}, \citenamefont {{Nellen}},
  \citenamefont {{Newbold}}, \citenamefont {{Nisa}}, \citenamefont
  {{Noriega-Papaqui}}, \citenamefont {{Pretz}}, \citenamefont
  {{P{\'e}rez-P{\'e}rez}}, \citenamefont {{Ren}}, \citenamefont {{Rho}},
  \citenamefont {{Rivi{\`e}re}}, \citenamefont {{Rosa-Gonz{\'a}lez}},
  \citenamefont {{Rosenberg}}, \citenamefont {{Ruiz-Velasco}}, \citenamefont
  {{Salesa Greus}}, \citenamefont {{Sandoval}}, \citenamefont {{Schneider}},
  \citenamefont {{Schoorlemmer}}, \citenamefont {{Seglar Arroyo}},
  \citenamefont {{Sinnis}}, \citenamefont {{Smith}}, \citenamefont
  {{Springer}}, \citenamefont {{Surajbali}}, \citenamefont {{Taboada}},
  \citenamefont {{Tibolla}}, \citenamefont {{Tollefson}}, \citenamefont
  {{Torres}}, \citenamefont {{Vianello}}, \citenamefont {{Villase{\~n}or}},
  \citenamefont {{Weisgarber}}, \citenamefont {{Werner}}, \citenamefont
  {{Westerhoff}}, \citenamefont {{Wood}}, \citenamefont {{Yapici}},
  \citenamefont {{Yodh}}, \citenamefont {{Zepeda}}, \citenamefont {{Zhang}},\
  and\ \citenamefont {{Zhou}}}]{2018Natur.562...82A}%
  \BibitemOpen
  \bibfield  {author} {\bibinfo {author} {\bibfnamefont {A.~U.}\ \bibnamefont
  {{Abeysekara}}}, \bibinfo {author} {\bibfnamefont {A.}~\bibnamefont
  {{Albert}}}, \bibinfo {author} {\bibfnamefont {R.}~\bibnamefont {{Alfaro}}},
  \bibinfo {author} {\bibfnamefont {C.}~\bibnamefont {{Alvarez}}}, \bibinfo
  {author} {\bibfnamefont {J.~D.}\ \bibnamefont {{{\'A}lvarez}}}, \bibinfo
  {author} {\bibfnamefont {R.}~\bibnamefont {{Arceo}}}, \bibinfo {author}
  {\bibfnamefont {J.~C.}\ \bibnamefont {{Arteaga-Vel{\'a}zquez}}}, \bibinfo
  {author} {\bibfnamefont {D.}~\bibnamefont {{Avila Rojas}}}, \bibinfo {author}
  {\bibfnamefont {H.~A.}\ \bibnamefont {{Ayala Solares}}}, \bibinfo {author}
  {\bibfnamefont {E.}~\bibnamefont {{Belmont-Moreno}}}, \bibinfo {author}
  {\bibfnamefont {S.~Y.}\ \bibnamefont {{BenZvi}}}, \bibinfo {author}
  {\bibfnamefont {C.}~\bibnamefont {{Brisbois}}}, \bibinfo {author}
  {\bibfnamefont {K.~S.}\ \bibnamefont {{Caballero-Mora}}}, \bibinfo {author}
  {\bibfnamefont {T.}~\bibnamefont {{Capistr{\'a}n}}}, \bibinfo {author}
  {\bibfnamefont {A.}~\bibnamefont {{Carrami{\~n}ana}}}, \bibinfo {author}
  {\bibfnamefont {S.}~\bibnamefont {{Casanova}}}, \bibinfo {author}
  {\bibfnamefont {M.}~\bibnamefont {{Castillo}}}, \bibinfo {author}
  {\bibfnamefont {U.}~\bibnamefont {{Cotti}}}, \bibinfo {author} {\bibfnamefont
  {J.}~\bibnamefont {{Cotzomi}}}, \bibinfo {author} {\bibfnamefont
  {S.}~\bibnamefont {{Couti{\~n}o de Le{\'o}n}}}, \bibinfo {author}
  {\bibfnamefont {C.}~\bibnamefont {{De Le{\'o}n}}}, \bibinfo {author}
  {\bibfnamefont {E.}~\bibnamefont {{De la Fuente}}}, \bibinfo {author}
  {\bibfnamefont {J.~C.}\ \bibnamefont {{D{\'\i}az-V{\'e}lez}}}, \bibinfo
  {author} {\bibfnamefont {S.}~\bibnamefont {{Dichiara}}}, \bibinfo {author}
  {\bibfnamefont {B.~L.}\ \bibnamefont {{Dingus}}}, \bibinfo {author}
  {\bibfnamefont {M.~A.}\ \bibnamefont {{DuVernois}}}, \bibinfo {author}
  {\bibfnamefont {R.~W.}\ \bibnamefont {{Ellsworth}}}, \bibinfo {author}
  {\bibfnamefont {K.}~\bibnamefont {{Engel}}}, \bibinfo {author} {\bibfnamefont
  {C.}~\bibnamefont {{Espinoza}}}, \bibinfo {author} {\bibfnamefont
  {K.}~\bibnamefont {{Fang}}}, \bibinfo {author} {\bibfnamefont
  {H.}~\bibnamefont {{Fleischhack}}}, \bibinfo {author} {\bibfnamefont
  {N.}~\bibnamefont {{Fraija}}}, \bibinfo {author} {\bibfnamefont
  {A.}~\bibnamefont {{Galv{\'a}n-G{\'a}mez}}}, \bibinfo {author} {\bibfnamefont
  {J.~A.}\ \bibnamefont {{Garc{\'\i}a-Gonz{\'a}lez}}}, \bibinfo {author}
  {\bibfnamefont {F.}~\bibnamefont {{Garfias}}}, \bibinfo {author}
  {\bibfnamefont {A.}~\bibnamefont {{Gonz{\'a}lez-Mu{\~n}oz}}}, \bibinfo
  {author} {\bibfnamefont {M.~M.}\ \bibnamefont {{Gonz{\'a}lez}}}, \bibinfo
  {author} {\bibfnamefont {J.~A.}\ \bibnamefont {{Goodman}}}, \bibinfo {author}
  {\bibfnamefont {Z.}~\bibnamefont {{Hampel-Arias}}}, \bibinfo {author}
  {\bibfnamefont {J.~P.}\ \bibnamefont {{Harding}}}, \bibinfo {author}
  {\bibfnamefont {S.}~\bibnamefont {{Hernandez}}}, \bibinfo {author}
  {\bibfnamefont {J.}~\bibnamefont {{Hinton}}}, \bibinfo {author}
  {\bibfnamefont {B.}~\bibnamefont {{Hona}}}, \bibinfo {author} {\bibfnamefont
  {F.}~\bibnamefont {{Hueyotl-Zahuantitla}}}, \bibinfo {author} {\bibfnamefont
  {C.~M.}\ \bibnamefont {{Hui}}}, \bibinfo {author} {\bibfnamefont
  {P.}~\bibnamefont {{H{\"u}ntemeyer}}}, \bibinfo {author} {\bibfnamefont
  {A.}~\bibnamefont {{Iriarte}}}, \bibinfo {author} {\bibfnamefont
  {A.}~\bibnamefont {{Jardin-Blicq}}}, \bibinfo {author} {\bibfnamefont
  {V.}~\bibnamefont {{Joshi}}}, \bibinfo {author} {\bibfnamefont
  {S.}~\bibnamefont {{Kaufmann}}}, \bibinfo {author} {\bibfnamefont
  {P.}~\bibnamefont {{Kar}}}, \bibinfo {author} {\bibfnamefont {G.~J.}\
  \bibnamefont {{Kunde}}}, \bibinfo {author} {\bibfnamefont {R.~J.}\
  \bibnamefont {{Lauer}}}, \bibinfo {author} {\bibfnamefont {W.~H.}\
  \bibnamefont {{Lee}}}, \bibinfo {author} {\bibfnamefont {H.}~\bibnamefont
  {{Le{\'o}n Vargas}}}, \bibinfo {author} {\bibfnamefont {H.}~\bibnamefont
  {{Li}}}, \bibinfo {author} {\bibfnamefont {J.~T.}\ \bibnamefont
  {{Linnemann}}}, \bibinfo {author} {\bibfnamefont {A.~L.}\ \bibnamefont
  {{Longinotti}}}, \bibinfo {author} {\bibfnamefont {G.}~\bibnamefont
  {{Luis-Raya}}}, \bibinfo {author} {\bibfnamefont {R.}~\bibnamefont
  {{L{\'o}pez-Coto}}}, \bibinfo {author} {\bibfnamefont {K.}~\bibnamefont
  {{Malone}}}, \bibinfo {author} {\bibfnamefont {S.~S.}\ \bibnamefont
  {{Marinelli}}}, \bibinfo {author} {\bibfnamefont {O.}~\bibnamefont
  {{Martinez}}}, \bibinfo {author} {\bibfnamefont {I.}~\bibnamefont
  {{Martinez-Castellanos}}}, \bibinfo {author} {\bibfnamefont {J.}~\bibnamefont
  {{Mart{\'\i}nez-Castro}}}, \bibinfo {author} {\bibfnamefont {J.~A.}\
  \bibnamefont {{Matthews}}}, \bibinfo {author} {\bibfnamefont
  {P.}~\bibnamefont {{Miranda-Romagnoli}}}, \bibinfo {author} {\bibfnamefont
  {E.}~\bibnamefont {{Moreno}}}, \bibinfo {author} {\bibfnamefont
  {M.}~\bibnamefont {{Mostaf{\'a}}}}, \bibinfo {author} {\bibfnamefont
  {A.}~\bibnamefont {{Nayerhoda}}}, \bibinfo {author} {\bibfnamefont
  {L.}~\bibnamefont {{Nellen}}}, \bibinfo {author} {\bibfnamefont
  {M.}~\bibnamefont {{Newbold}}}, \bibinfo {author} {\bibfnamefont {M.~U.}\
  \bibnamefont {{Nisa}}}, \bibinfo {author} {\bibfnamefont {R.}~\bibnamefont
  {{Noriega-Papaqui}}}, \bibinfo {author} {\bibfnamefont {J.}~\bibnamefont
  {{Pretz}}}, \bibinfo {author} {\bibfnamefont {E.~G.}\ \bibnamefont
  {{P{\'e}rez-P{\'e}rez}}}, \bibinfo {author} {\bibfnamefont {Z.}~\bibnamefont
  {{Ren}}}, \bibinfo {author} {\bibfnamefont {C.~D.}\ \bibnamefont {{Rho}}},
  \bibinfo {author} {\bibfnamefont {C.}~\bibnamefont {{Rivi{\`e}re}}}, \bibinfo
  {author} {\bibfnamefont {D.}~\bibnamefont {{Rosa-Gonz{\'a}lez}}}, \bibinfo
  {author} {\bibfnamefont {M.}~\bibnamefont {{Rosenberg}}}, \bibinfo {author}
  {\bibfnamefont {E.}~\bibnamefont {{Ruiz-Velasco}}}, \bibinfo {author}
  {\bibfnamefont {F.}~\bibnamefont {{Salesa Greus}}}, \bibinfo {author}
  {\bibfnamefont {A.}~\bibnamefont {{Sandoval}}}, \bibinfo {author}
  {\bibfnamefont {M.}~\bibnamefont {{Schneider}}}, \bibinfo {author}
  {\bibfnamefont {H.}~\bibnamefont {{Schoorlemmer}}}, \bibinfo {author}
  {\bibfnamefont {M.}~\bibnamefont {{Seglar Arroyo}}}, \bibinfo {author}
  {\bibfnamefont {G.}~\bibnamefont {{Sinnis}}}, \bibinfo {author}
  {\bibfnamefont {A.~J.}\ \bibnamefont {{Smith}}}, \bibinfo {author}
  {\bibfnamefont {R.~W.}\ \bibnamefont {{Springer}}}, \bibinfo {author}
  {\bibfnamefont {P.}~\bibnamefont {{Surajbali}}}, \bibinfo {author}
  {\bibfnamefont {I.}~\bibnamefont {{Taboada}}}, \bibinfo {author}
  {\bibfnamefont {O.}~\bibnamefont {{Tibolla}}}, \bibinfo {author}
  {\bibfnamefont {K.}~\bibnamefont {{Tollefson}}}, \bibinfo {author}
  {\bibfnamefont {I.}~\bibnamefont {{Torres}}}, \bibinfo {author}
  {\bibfnamefont {G.}~\bibnamefont {{Vianello}}}, \bibinfo {author}
  {\bibfnamefont {L.}~\bibnamefont {{Villase{\~n}or}}}, \bibinfo {author}
  {\bibfnamefont {T.}~\bibnamefont {{Weisgarber}}}, \bibinfo {author}
  {\bibfnamefont {F.}~\bibnamefont {{Werner}}}, \bibinfo {author}
  {\bibfnamefont {S.}~\bibnamefont {{Westerhoff}}}, \bibinfo {author}
  {\bibfnamefont {J.}~\bibnamefont {{Wood}}}, \bibinfo {author} {\bibfnamefont
  {T.}~\bibnamefont {{Yapici}}}, \bibinfo {author} {\bibfnamefont
  {G.}~\bibnamefont {{Yodh}}}, \bibinfo {author} {\bibfnamefont
  {A.}~\bibnamefont {{Zepeda}}}, \bibinfo {author} {\bibfnamefont
  {H.}~\bibnamefont {{Zhang}}},\ and\ \bibinfo {author} {\bibfnamefont
  {H.}~\bibnamefont {{Zhou}}},\ }\bibfield  {title} {\bibinfo {title}
  {{Very-high-energy particle acceleration powered by the jets of the
  microquasar SS 433}},\ }\href {https://doi.org/10.1038/s41586-018-0565-5}
  {\bibfield  {journal} {\bibinfo  {journal} {\nat}\ }\textbf {\bibinfo
  {volume} {562}},\ \bibinfo {pages} {82} (\bibinfo {year} {2018})},\ \Eprint
  {https://arxiv.org/abs/1810.01892} {arXiv:1810.01892 [astro-ph.HE]}
  \BibitemShut {NoStop}%
\bibitem [{\citenamefont {{Blandford}}\ \emph {et~al.}(2019)\citenamefont
  {{Blandford}}, \citenamefont {{Meier}},\ and\ \citenamefont
  {{Readhead}}}]{2019ARA&A..57..467B}%
  \BibitemOpen
  \bibfield  {author} {\bibinfo {author} {\bibfnamefont {R.}~\bibnamefont
  {{Blandford}}}, \bibinfo {author} {\bibfnamefont {D.}~\bibnamefont
  {{Meier}}},\ and\ \bibinfo {author} {\bibfnamefont {A.}~\bibnamefont
  {{Readhead}}},\ }\bibfield  {title} {\bibinfo {title} {{Relativistic Jets
  from Active Galactic Nuclei}},\ }\href
  {https://doi.org/10.1146/annurev-astro-081817-051948} {\bibfield  {journal}
  {\bibinfo  {journal} {\araa}\ }\textbf {\bibinfo {volume} {57}},\ \bibinfo
  {pages} {467} (\bibinfo {year} {2019})},\ \Eprint
  {https://arxiv.org/abs/1812.06025} {arXiv:1812.06025 [astro-ph.HE]}
  \BibitemShut {NoStop}%
\bibitem [{\citenamefont
  {Collaboration}(2024)}]{lhaasocollaboration2024ultrahighenergygammarayemissionassociated}%
  \BibitemOpen
  \bibfield  {author} {\bibinfo {author} {\bibfnamefont {L.}~\bibnamefont
  {Collaboration}},\ }\href {https://arxiv.org/abs/2410.08988} {\bibinfo
  {title} {Ultrahigh-energy gamma-ray emission associated with black hole-jet
  systems}} (\bibinfo {year} {2024}),\ \Eprint
  {https://arxiv.org/abs/2410.08988} {arXiv:2410.08988 [astro-ph.HE]}
  \BibitemShut {NoStop}%
\bibitem [{\citenamefont {{Gebhardt}}\ \emph {et~al.}(2000)\citenamefont
  {{Gebhardt}}, \citenamefont {{Bender}}, \citenamefont {{Bower}},
  \citenamefont {{Dressler}}, \citenamefont {{Faber}}, \citenamefont
  {{Filippenko}}, \citenamefont {{Green}}, \citenamefont {{Grillmair}},
  \citenamefont {{Ho}}, \citenamefont {{Kormendy}}, \citenamefont {{Lauer}},
  \citenamefont {{Magorrian}}, \citenamefont {{Pinkney}}, \citenamefont
  {{Richstone}},\ and\ \citenamefont {{Tremaine}}}]{2000ApJ...539L..13G}%
  \BibitemOpen
  \bibfield  {author} {\bibinfo {author} {\bibfnamefont {K.}~\bibnamefont
  {{Gebhardt}}}, \bibinfo {author} {\bibfnamefont {R.}~\bibnamefont
  {{Bender}}}, \bibinfo {author} {\bibfnamefont {G.}~\bibnamefont {{Bower}}},
  \bibinfo {author} {\bibfnamefont {A.}~\bibnamefont {{Dressler}}}, \bibinfo
  {author} {\bibfnamefont {S.~M.}\ \bibnamefont {{Faber}}}, \bibinfo {author}
  {\bibfnamefont {A.~V.}\ \bibnamefont {{Filippenko}}}, \bibinfo {author}
  {\bibfnamefont {R.}~\bibnamefont {{Green}}}, \bibinfo {author} {\bibfnamefont
  {C.}~\bibnamefont {{Grillmair}}}, \bibinfo {author} {\bibfnamefont {L.~C.}\
  \bibnamefont {{Ho}}}, \bibinfo {author} {\bibfnamefont {J.}~\bibnamefont
  {{Kormendy}}}, \bibinfo {author} {\bibfnamefont {T.~R.}\ \bibnamefont
  {{Lauer}}}, \bibinfo {author} {\bibfnamefont {J.}~\bibnamefont
  {{Magorrian}}}, \bibinfo {author} {\bibfnamefont {J.}~\bibnamefont
  {{Pinkney}}}, \bibinfo {author} {\bibfnamefont {D.}~\bibnamefont
  {{Richstone}}},\ and\ \bibinfo {author} {\bibfnamefont {S.}~\bibnamefont
  {{Tremaine}}},\ }\bibfield  {title} {\bibinfo {title} {{A Relationship
  between Nuclear Black Hole Mass and Galaxy Velocity Dispersion}},\ }\href
  {https://doi.org/10.1086/312840} {\bibfield  {journal} {\bibinfo  {journal}
  {\apjl}\ }\textbf {\bibinfo {volume} {539}},\ \bibinfo {pages} {L13}
  (\bibinfo {year} {2000})},\ \Eprint {https://arxiv.org/abs/astro-ph/0006289}
  {arXiv:astro-ph/0006289 [astro-ph]} \BibitemShut {NoStop}%
\bibitem [{\citenamefont {{Merloni}}\ \emph {et~al.}(2003)\citenamefont
  {{Merloni}}, \citenamefont {{Heinz}},\ and\ \citenamefont {{di
  Matteo}}}]{2003MNRAS.345.1057M}%
  \BibitemOpen
  \bibfield  {author} {\bibinfo {author} {\bibfnamefont {A.}~\bibnamefont
  {{Merloni}}}, \bibinfo {author} {\bibfnamefont {S.}~\bibnamefont {{Heinz}}},\
  and\ \bibinfo {author} {\bibfnamefont {T.}~\bibnamefont {{di Matteo}}},\
  }\bibfield  {title} {\bibinfo {title} {{A Fundamental Plane of black hole
  activity}},\ }\href {https://doi.org/10.1046/j.1365-2966.2003.07017.x}
  {\bibfield  {journal} {\bibinfo  {journal} {\mnras}\ }\textbf {\bibinfo
  {volume} {345}},\ \bibinfo {pages} {1057} (\bibinfo {year} {2003})},\ \Eprint
  {https://arxiv.org/abs/astro-ph/0305261} {arXiv:astro-ph/0305261 [astro-ph]}
  \BibitemShut {NoStop}%
\bibitem [{\citenamefont {{K{\"o}rding}}\ \emph {et~al.}(2006)\citenamefont
  {{K{\"o}rding}}, \citenamefont {{Falcke}},\ and\ \citenamefont
  {{Corbel}}}]{2006A&A...456..439K}%
  \BibitemOpen
  \bibfield  {author} {\bibinfo {author} {\bibfnamefont {E.}~\bibnamefont
  {{K{\"o}rding}}}, \bibinfo {author} {\bibfnamefont {H.}~\bibnamefont
  {{Falcke}}},\ and\ \bibinfo {author} {\bibfnamefont {S.}~\bibnamefont
  {{Corbel}}},\ }\bibfield  {title} {\bibinfo {title} {{Refining the
  fundamental plane of accreting black holes}},\ }\href
  {https://doi.org/10.1051/0004-6361:20054144} {\bibfield  {journal} {\bibinfo
  {journal} {\aap}\ }\textbf {\bibinfo {volume} {456}},\ \bibinfo {pages} {439}
  (\bibinfo {year} {2006})},\ \Eprint {https://arxiv.org/abs/astro-ph/0603117}
  {arXiv:astro-ph/0603117 [astro-ph]} \BibitemShut {NoStop}%
\bibitem [{\citenamefont {{Nisbet}}\ and\ \citenamefont
  {{Best}}(2016)}]{2016MNRAS.455.2551N}%
  \BibitemOpen
  \bibfield  {author} {\bibinfo {author} {\bibfnamefont {D.~M.}\ \bibnamefont
  {{Nisbet}}}\ and\ \bibinfo {author} {\bibfnamefont {P.~N.}\ \bibnamefont
  {{Best}}},\ }\bibfield  {title} {\bibinfo {title} {{The mass fraction of AGN
  and the Fundamental Plane of black hole activity from a large X-ray-selected
  sample of LINERs}},\ }\href {https://doi.org/10.1093/mnras/stv2450}
  {\bibfield  {journal} {\bibinfo  {journal} {\mnras}\ }\textbf {\bibinfo
  {volume} {455}},\ \bibinfo {pages} {2551} (\bibinfo {year} {2016})},\ \Eprint
  {https://arxiv.org/abs/1510.06746} {arXiv:1510.06746 [astro-ph.GA]}
  \BibitemShut {NoStop}%
\bibitem [{\citenamefont {{Emmanoulopoulos}}\ \emph {et~al.}(2013)\citenamefont
  {{Emmanoulopoulos}}, \citenamefont {{McHardy}},\ and\ \citenamefont
  {{Papadakis}}}]{2013MNRAS.433..907E}%
  \BibitemOpen
  \bibfield  {author} {\bibinfo {author} {\bibfnamefont {D.}~\bibnamefont
  {{Emmanoulopoulos}}}, \bibinfo {author} {\bibfnamefont {I.~M.}\ \bibnamefont
  {{McHardy}}},\ and\ \bibinfo {author} {\bibfnamefont {I.~E.}\ \bibnamefont
  {{Papadakis}}},\ }\bibfield  {title} {\bibinfo {title} {{Generating
  artificial light curves: revisited and updated}},\ }\href
  {https://doi.org/10.1093/mnras/stt764} {\bibfield  {journal} {\bibinfo
  {journal} {\mnras}\ }\textbf {\bibinfo {volume} {433}},\ \bibinfo {pages}
  {907} (\bibinfo {year} {2013})},\ \Eprint {https://arxiv.org/abs/1305.0304}
  {arXiv:1305.0304 [astro-ph.IM]} \BibitemShut {NoStop}%
\bibitem [{\citenamefont {{McHardy}}\ \emph {et~al.}(2006)\citenamefont
  {{McHardy}}, \citenamefont {{Koerding}}, \citenamefont {{Knigge}},
  \citenamefont {{Uttley}},\ and\ \citenamefont
  {{Fender}}}]{2006Natur.444..730M}%
  \BibitemOpen
  \bibfield  {author} {\bibinfo {author} {\bibfnamefont {I.~M.}\ \bibnamefont
  {{McHardy}}}, \bibinfo {author} {\bibfnamefont {E.}~\bibnamefont
  {{Koerding}}}, \bibinfo {author} {\bibfnamefont {C.}~\bibnamefont
  {{Knigge}}}, \bibinfo {author} {\bibfnamefont {P.}~\bibnamefont {{Uttley}}},\
  and\ \bibinfo {author} {\bibfnamefont {R.~P.}\ \bibnamefont {{Fender}}},\
  }\bibfield  {title} {\bibinfo {title} {{Active galactic nuclei as scaled-up
  Galactic black holes}},\ }\href {https://doi.org/10.1038/nature05389}
  {\bibfield  {journal} {\bibinfo  {journal} {\nat}\ }\textbf {\bibinfo
  {volume} {444}},\ \bibinfo {pages} {730} (\bibinfo {year} {2006})},\ \Eprint
  {https://arxiv.org/abs/astro-ph/0612273} {arXiv:astro-ph/0612273 [astro-ph]}
  \BibitemShut {NoStop}%
\bibitem [{\citenamefont {{Scaringi}}\ \emph {et~al.}(2015)\citenamefont
  {{Scaringi}}, \citenamefont {{Maccarone}}, \citenamefont {{Kording}},
  \citenamefont {{Knigge}}, \citenamefont {{Vaughan}}, \citenamefont {{Marsh}},
  \citenamefont {{Aranzana}}, \citenamefont {{Dhillon}},\ and\ \citenamefont
  {{Barros}}}]{2015SciA....1E0686S}%
  \BibitemOpen
  \bibfield  {author} {\bibinfo {author} {\bibfnamefont {S.}~\bibnamefont
  {{Scaringi}}}, \bibinfo {author} {\bibfnamefont {T.~J.}\ \bibnamefont
  {{Maccarone}}}, \bibinfo {author} {\bibfnamefont {E.}~\bibnamefont
  {{Kording}}}, \bibinfo {author} {\bibfnamefont {C.}~\bibnamefont {{Knigge}}},
  \bibinfo {author} {\bibfnamefont {S.}~\bibnamefont {{Vaughan}}}, \bibinfo
  {author} {\bibfnamefont {T.~R.}\ \bibnamefont {{Marsh}}}, \bibinfo {author}
  {\bibfnamefont {E.}~\bibnamefont {{Aranzana}}}, \bibinfo {author}
  {\bibfnamefont {V.~S.}\ \bibnamefont {{Dhillon}}},\ and\ \bibinfo {author}
  {\bibfnamefont {S.~C.~C.}\ \bibnamefont {{Barros}}},\ }\bibfield  {title}
  {\bibinfo {title} {{Accretion-induced variability links young stellar
  objects, white dwarfs, and black holes}},\ }\href
  {https://doi.org/10.1126/sciadv.1500686} {\bibfield  {journal} {\bibinfo
  {journal} {Science Advances}\ }\textbf {\bibinfo {volume} {1}},\ \bibinfo
  {pages} {e1500686} (\bibinfo {year} {2015})},\ \Eprint
  {https://arxiv.org/abs/1510.02471} {arXiv:1510.02471 [astro-ph.HE]}
  \BibitemShut {NoStop}%
\bibitem [{\citenamefont {{Burke}}\ \emph {et~al.}(2021)\citenamefont
  {{Burke}}, \citenamefont {{Shen}}, \citenamefont {{Blaes}}, \citenamefont
  {{Gammie}}, \citenamefont {{Horne}}, \citenamefont {{Jiang}}, \citenamefont
  {{Liu}}, \citenamefont {{McHardy}}, \citenamefont {{Morgan}}, \citenamefont
  {{Scaringi}},\ and\ \citenamefont {{Yang}}}]{2021Sci...373..789B}%
  \BibitemOpen
  \bibfield  {author} {\bibinfo {author} {\bibfnamefont {C.~J.}\ \bibnamefont
  {{Burke}}}, \bibinfo {author} {\bibfnamefont {Y.}~\bibnamefont {{Shen}}},
  \bibinfo {author} {\bibfnamefont {O.}~\bibnamefont {{Blaes}}}, \bibinfo
  {author} {\bibfnamefont {C.~F.}\ \bibnamefont {{Gammie}}}, \bibinfo {author}
  {\bibfnamefont {K.}~\bibnamefont {{Horne}}}, \bibinfo {author} {\bibfnamefont
  {Y.-F.}\ \bibnamefont {{Jiang}}}, \bibinfo {author} {\bibfnamefont
  {X.}~\bibnamefont {{Liu}}}, \bibinfo {author} {\bibfnamefont {I.~M.}\
  \bibnamefont {{McHardy}}}, \bibinfo {author} {\bibfnamefont {C.~W.}\
  \bibnamefont {{Morgan}}}, \bibinfo {author} {\bibfnamefont {S.}~\bibnamefont
  {{Scaringi}}},\ and\ \bibinfo {author} {\bibfnamefont {Q.}~\bibnamefont
  {{Yang}}},\ }\bibfield  {title} {\bibinfo {title} {{A characteristic optical
  variability time scale in astrophysical accretion disks}},\ }\href
  {https://doi.org/10.1126/science.abg9933} {\bibfield  {journal} {\bibinfo
  {journal} {Science}\ }\textbf {\bibinfo {volume} {373}},\ \bibinfo {pages}
  {789} (\bibinfo {year} {2021})},\ \Eprint {https://arxiv.org/abs/2108.05389}
  {arXiv:2108.05389 [astro-ph.GA]} \BibitemShut {NoStop}%
\bibitem [{\citenamefont {{Zhang}}\ \emph {et~al.}(2024)\citenamefont
  {{Zhang}}, \citenamefont {{Yang}},\ and\ \citenamefont
  {{Dai}}}]{2024ApJ...967L..18Z}%
  \BibitemOpen
  \bibfield  {author} {\bibinfo {author} {\bibfnamefont {H.}~\bibnamefont
  {{Zhang}}}, \bibinfo {author} {\bibfnamefont {S.}~\bibnamefont {{Yang}}},\
  and\ \bibinfo {author} {\bibfnamefont {B.}~\bibnamefont {{Dai}}},\ }\bibfield
   {title} {\bibinfo {title} {{Discovering the Mass-Scaled Damping Timescale
  from Microquasars to Blazars}},\ }\href
  {https://doi.org/10.3847/2041-8213/ad488d} {\bibfield  {journal} {\bibinfo
  {journal} {\apjl}\ }\textbf {\bibinfo {volume} {967}},\ \bibinfo {eid} {L18}
  (\bibinfo {year} {2024})},\ \Eprint {https://arxiv.org/abs/2405.05575}
  {arXiv:2405.05575 [astro-ph.HE]} \BibitemShut {NoStop}%
\bibitem [{\citenamefont {{Zhou}}\ \emph {et~al.}(2015)\citenamefont {{Zhou}},
  \citenamefont {{Yuan}}, \citenamefont {{Pan}},\ and\ \citenamefont
  {{Liu}}}]{2015ApJ...798L...5Z}%
  \BibitemOpen
  \bibfield  {author} {\bibinfo {author} {\bibfnamefont {X.-L.}\ \bibnamefont
  {{Zhou}}}, \bibinfo {author} {\bibfnamefont {W.}~\bibnamefont {{Yuan}}},
  \bibinfo {author} {\bibfnamefont {H.-W.}\ \bibnamefont {{Pan}}},\ and\
  \bibinfo {author} {\bibfnamefont {Z.}~\bibnamefont {{Liu}}},\ }\bibfield
  {title} {\bibinfo {title} {{Universal Scaling of the 3:2 Twin-peak
  Quasi-periodic Oscillation Frequencies With Black Hole Mass and Spin
  Revisited}},\ }\href {https://doi.org/10.1088/2041-8205/798/1/L5} {\bibfield
  {journal} {\bibinfo  {journal} {\apjl}\ }\textbf {\bibinfo {volume} {798}},\
  \bibinfo {eid} {L5} (\bibinfo {year} {2015})},\ \Eprint
  {https://arxiv.org/abs/1411.7731} {arXiv:1411.7731 [astro-ph.HE]}
  \BibitemShut {NoStop}%
\bibitem [{\citenamefont {{Smith}}\ \emph {et~al.}(2021)\citenamefont
  {{Smith}}, \citenamefont {{Tandon}},\ and\ \citenamefont
  {{Wagoner}}}]{2021ApJ...906...92S}%
  \BibitemOpen
  \bibfield  {author} {\bibinfo {author} {\bibfnamefont {K.~L.}\ \bibnamefont
  {{Smith}}}, \bibinfo {author} {\bibfnamefont {C.~R.}\ \bibnamefont
  {{Tandon}}},\ and\ \bibinfo {author} {\bibfnamefont {R.~V.}\ \bibnamefont
  {{Wagoner}}},\ }\bibfield  {title} {\bibinfo {title} {{Confrontation of
  Observation and Theory: High-frequency QPOs in X-Ray Binaries, Tidal
  Disruption Events, and Active Galactic Nuclei}},\ }\href
  {https://doi.org/10.3847/1538-4357/abc9b7} {\bibfield  {journal} {\bibinfo
  {journal} {\apj}\ }\textbf {\bibinfo {volume} {906}},\ \bibinfo {eid} {92}
  (\bibinfo {year} {2021})},\ \Eprint {https://arxiv.org/abs/2011.05346}
  {arXiv:2011.05346 [astro-ph.HE]} \BibitemShut {NoStop}%
\bibitem [{\citenamefont {{Zhang}}\ \emph {et~al.}(2025)\citenamefont
  {{Zhang}}, \citenamefont {{Meng}}, \citenamefont {{Zhang}},\ and\
  \citenamefont {{Dai}}}]{2025MNRAS.538.2161Z}%
  \BibitemOpen
  \bibfield  {author} {\bibinfo {author} {\bibfnamefont {H.}~\bibnamefont
  {{Zhang}}}, \bibinfo {author} {\bibfnamefont {L.}~\bibnamefont {{Meng}}},
  \bibinfo {author} {\bibfnamefont {L.}~\bibnamefont {{Zhang}}},\ and\ \bibinfo
  {author} {\bibfnamefont {B.}~\bibnamefont {{Dai}}},\ }\bibfield  {title}
  {\bibinfo {title} {{Revisiting the observation-theory confrontation in
  high-frequency QPOs: a new QPO in NGC 5506, intermediate-mass black holes,
  and the crucial role of accretion state}},\ }\href
  {https://doi.org/10.1093/mnras/staf391} {\bibfield  {journal} {\bibinfo
  {journal} {\mnras}\ }\textbf {\bibinfo {volume} {538}},\ \bibinfo {pages}
  {2161} (\bibinfo {year} {2025})},\ \Eprint {https://arxiv.org/abs/2503.03631}
  {arXiv:2503.03631 [astro-ph.HE]} \BibitemShut {NoStop}%
\bibitem [{\citenamefont {{Czerny}}(2006)}]{2006ASPC..360..265C}%
  \BibitemOpen
  \bibfield  {author} {\bibinfo {author} {\bibfnamefont {B.}~\bibnamefont
  {{Czerny}}},\ }\bibfield  {title} {\bibinfo {title} {{The Role of the
  Accretion Disk in AGN Variability}},\ }in\ \href@noop {} {\emph {\bibinfo
  {booktitle} {AGN Variability from X-Rays to Radio Waves}}},\ \bibinfo
  {series} {Astronomical Society of the Pacific Conference Series}, Vol.\
  \bibinfo {volume} {360},\ \bibinfo {editor} {edited by\ \bibinfo {editor}
  {\bibfnamefont {C.~M.}\ \bibnamefont {{Gaskell}}}, \bibinfo {editor}
  {\bibfnamefont {I.~M.}\ \bibnamefont {{McHardy}}}, \bibinfo {editor}
  {\bibfnamefont {B.~M.}\ \bibnamefont {{Peterson}}},\ and\ \bibinfo {editor}
  {\bibfnamefont {S.~G.}\ \bibnamefont {{Sergeev}}}}\ (\bibinfo {year} {2006})\
  p.\ \bibinfo {pages} {265}\BibitemShut {NoStop}%
\bibitem [{\citenamefont {{Ishibashi}}\ and\ \citenamefont
  {{Courvoisier}}(2012)}]{2012A&A...540L...2I}%
  \BibitemOpen
  \bibfield  {author} {\bibinfo {author} {\bibfnamefont {W.}~\bibnamefont
  {{Ishibashi}}}\ and\ \bibinfo {author} {\bibfnamefont {T.~J.~L.}\
  \bibnamefont {{Courvoisier}}},\ }\bibfield  {title} {\bibinfo {title} {{The
  physical origin of the X-ray power spectral density break timescale in
  accreting black holes}},\ }\href
  {https://doi.org/10.1051/0004-6361/201218889} {\bibfield  {journal} {\bibinfo
   {journal} {\aap}\ }\textbf {\bibinfo {volume} {540}},\ \bibinfo {eid} {L2}
  (\bibinfo {year} {2012})},\ \Eprint {https://arxiv.org/abs/1203.0553}
  {arXiv:1203.0553 [astro-ph.HE]} \BibitemShut {NoStop}%
\bibitem [{\citenamefont {{Suberlak}}\ \emph {et~al.}(2021)\citenamefont
  {{Suberlak}}, \citenamefont {{Ivezi{\'c}}},\ and\ \citenamefont
  {{MacLeod}}}]{2021ApJ...907...96S}%
  \BibitemOpen
  \bibfield  {author} {\bibinfo {author} {\bibfnamefont {K.~L.}\ \bibnamefont
  {{Suberlak}}}, \bibinfo {author} {\bibfnamefont {{\v{Z}}.}~\bibnamefont
  {{Ivezi{\'c}}}},\ and\ \bibinfo {author} {\bibfnamefont {C.}~\bibnamefont
  {{MacLeod}}},\ }\bibfield  {title} {\bibinfo {title} {{Improving Damped
  Random Walk Parameters for SDSS Stripe 82 Quasars with Pan-STARRS1}},\ }\href
  {https://doi.org/10.3847/1538-4357/abc698} {\bibfield  {journal} {\bibinfo
  {journal} {\apj}\ }\textbf {\bibinfo {volume} {907}},\ \bibinfo {eid} {96}
  (\bibinfo {year} {2021})},\ \Eprint {https://arxiv.org/abs/2012.12907}
  {arXiv:2012.12907 [astro-ph.GA]} \BibitemShut {NoStop}%
\bibitem [{\citenamefont {{Zhou}}\ \emph {et~al.}(2024)\citenamefont {{Zhou}},
  \citenamefont {{Sun}}, \citenamefont {{Cai}}, \citenamefont {{Ren}},
  \citenamefont {{Wang}},\ and\ \citenamefont {{Xue}}}]{2024ApJ...966....8Z}%
  \BibitemOpen
  \bibfield  {author} {\bibinfo {author} {\bibfnamefont {S.}~\bibnamefont
  {{Zhou}}}, \bibinfo {author} {\bibfnamefont {M.}~\bibnamefont {{Sun}}},
  \bibinfo {author} {\bibfnamefont {Z.-Y.}\ \bibnamefont {{Cai}}}, \bibinfo
  {author} {\bibfnamefont {G.}~\bibnamefont {{Ren}}}, \bibinfo {author}
  {\bibfnamefont {J.-X.}\ \bibnamefont {{Wang}}},\ and\ \bibinfo {author}
  {\bibfnamefont {Y.}~\bibnamefont {{Xue}}},\ }\bibfield  {title} {\bibinfo
  {title} {{How Long Will the Quasar UV/Optical Flickering Be Damped?}},\
  }\href {https://doi.org/10.3847/1538-4357/ad2fbc} {\bibfield  {journal}
  {\bibinfo  {journal} {\apj}\ }\textbf {\bibinfo {volume} {966}},\ \bibinfo
  {eid} {8} (\bibinfo {year} {2024})},\ \Eprint
  {https://arxiv.org/abs/2403.01691} {arXiv:2403.01691 [astro-ph.HE]}
  \BibitemShut {NoStop}%
\bibitem [{\citenamefont {{Blandford}}\ and\ \citenamefont
  {{Znajek}}(1977)}]{1977MNRAS.179..433B}%
  \BibitemOpen
  \bibfield  {author} {\bibinfo {author} {\bibfnamefont {R.~D.}\ \bibnamefont
  {{Blandford}}}\ and\ \bibinfo {author} {\bibfnamefont {R.~L.}\ \bibnamefont
  {{Znajek}}},\ }\bibfield  {title} {\bibinfo {title} {{Electromagnetic
  extraction of energy from Kerr black holes.}},\ }\href
  {https://doi.org/10.1093/mnras/179.3.433} {\bibfield  {journal} {\bibinfo
  {journal} {\mnras}\ }\textbf {\bibinfo {volume} {179}},\ \bibinfo {pages}
  {433} (\bibinfo {year} {1977})}\BibitemShut {NoStop}%
\bibitem [{\citenamefont {{Blandford}}\ and\ \citenamefont
  {{Payne}}(1982)}]{1982MNRAS.199..883B}%
  \BibitemOpen
  \bibfield  {author} {\bibinfo {author} {\bibfnamefont {R.~D.}\ \bibnamefont
  {{Blandford}}}\ and\ \bibinfo {author} {\bibfnamefont {D.~G.}\ \bibnamefont
  {{Payne}}},\ }\bibfield  {title} {\bibinfo {title} {{Hydromagnetic flows from
  accretion disks and the production of radio jets.}},\ }\href
  {https://doi.org/10.1093/mnras/199.4.883} {\bibfield  {journal} {\bibinfo
  {journal} {\mnras}\ }\textbf {\bibinfo {volume} {199}},\ \bibinfo {pages}
  {883} (\bibinfo {year} {1982})}\BibitemShut {NoStop}%
\bibitem [{\citenamefont {{Finke}}\ and\ \citenamefont
  {{Becker}}(2014)}]{2014ApJ...791...21F}%
  \BibitemOpen
  \bibfield  {author} {\bibinfo {author} {\bibfnamefont {J.~D.}\ \bibnamefont
  {{Finke}}}\ and\ \bibinfo {author} {\bibfnamefont {P.~A.}\ \bibnamefont
  {{Becker}}},\ }\bibfield  {title} {\bibinfo {title} {{Fourier Analysis of
  Blazar Variability}},\ }\href {https://doi.org/10.1088/0004-637X/791/1/21}
  {\bibfield  {journal} {\bibinfo  {journal} {\apj}\ }\textbf {\bibinfo
  {volume} {791}},\ \bibinfo {eid} {21} (\bibinfo {year} {2014})},\ \Eprint
  {https://arxiv.org/abs/1406.2333} {arXiv:1406.2333 [astro-ph.GA]}
  \BibitemShut {NoStop}%
\bibitem [{\citenamefont {{Sharma}}\ \emph {et~al.}(2025)\citenamefont
  {{Sharma}}, \citenamefont {{Prince}},\ and\ \citenamefont
  {{Bose}}}]{2025PhRvD.111h3049S}%
  \BibitemOpen
  \bibfield  {author} {\bibinfo {author} {\bibfnamefont {A.}~\bibnamefont
  {{Sharma}}}, \bibinfo {author} {\bibfnamefont {R.}~\bibnamefont {{Prince}}},\
  and\ \bibinfo {author} {\bibfnamefont {D.}~\bibnamefont {{Bose}}},\
  }\bibfield  {title} {\bibinfo {title} {{From microquasars to AGN: A uniform
  jet variability}},\ }\href {https://doi.org/10.1103/PhysRevD.111.083049}
  {\bibfield  {journal} {\bibinfo  {journal} {\prd}\ }\textbf {\bibinfo
  {volume} {111}},\ \bibinfo {eid} {083049} (\bibinfo {year} {2025})},\ \Eprint
  {https://arxiv.org/abs/2410.06653} {arXiv:2410.06653 [astro-ph.HE]}
  \BibitemShut {NoStop}%
\bibitem [{\citenamefont {{Zhang}}(2024)}]{2024ApJ...972...80Z}%
  \BibitemOpen
  \bibfield  {author} {\bibinfo {author} {\bibfnamefont {P.}~\bibnamefont
  {{Zhang}}},\ }\bibfield  {title} {\bibinfo {title} {{A New Puzzling Periodic
  Signal in GeV Energies of the {\ensuremath{\gamma}}-Ray Binary LS
  I+61{\textdegree}303}},\ }\href {https://doi.org/10.3847/1538-4357/ad6a16}
  {\bibfield  {journal} {\bibinfo  {journal} {\apj}\ }\textbf {\bibinfo
  {volume} {972}},\ \bibinfo {eid} {80} (\bibinfo {year} {2024})},\ \Eprint
  {https://arxiv.org/abs/2406.02042} {arXiv:2406.02042 [astro-ph.HE]}
  \BibitemShut {NoStop}%
\bibitem [{\citenamefont {{Ruan}}\ \emph {et~al.}(2012)\citenamefont {{Ruan}},
  \citenamefont {{Anderson}}, \citenamefont {{MacLeod}}, \citenamefont
  {{Becker}}, \citenamefont {{Burnett}}, \citenamefont {{Davenport}},
  \citenamefont {{Ivezi{\'c}}}, \citenamefont {{Kochanek}}, \citenamefont
  {{Plotkin}}, \citenamefont {{Sesar}},\ and\ \citenamefont
  {{Stuart}}}]{2012ApJ...760...51R}%
  \BibitemOpen
  \bibfield  {author} {\bibinfo {author} {\bibfnamefont {J.~J.}\ \bibnamefont
  {{Ruan}}}, \bibinfo {author} {\bibfnamefont {S.~F.}\ \bibnamefont
  {{Anderson}}}, \bibinfo {author} {\bibfnamefont {C.~L.}\ \bibnamefont
  {{MacLeod}}}, \bibinfo {author} {\bibfnamefont {A.~C.}\ \bibnamefont
  {{Becker}}}, \bibinfo {author} {\bibfnamefont {T.~H.}\ \bibnamefont
  {{Burnett}}}, \bibinfo {author} {\bibfnamefont {J.~R.~A.}\ \bibnamefont
  {{Davenport}}}, \bibinfo {author} {\bibfnamefont {{\v{Z}}.}~\bibnamefont
  {{Ivezi{\'c}}}}, \bibinfo {author} {\bibfnamefont {C.~S.}\ \bibnamefont
  {{Kochanek}}}, \bibinfo {author} {\bibfnamefont {R.~M.}\ \bibnamefont
  {{Plotkin}}}, \bibinfo {author} {\bibfnamefont {B.}~\bibnamefont {{Sesar}}},\
  and\ \bibinfo {author} {\bibfnamefont {J.~S.}\ \bibnamefont {{Stuart}}},\
  }\bibfield  {title} {\bibinfo {title} {{Characterizing the Optical
  Variability of Bright Blazars: Variability-based Selection of Fermi Active
  Galactic Nuclei}},\ }\href {https://doi.org/10.1088/0004-637X/760/1/51}
  {\bibfield  {journal} {\bibinfo  {journal} {\apj}\ }\textbf {\bibinfo
  {volume} {760}},\ \bibinfo {eid} {51} (\bibinfo {year} {2012})},\ \Eprint
  {https://arxiv.org/abs/1209.3770} {arXiv:1209.3770 [astro-ph.HE]}
  \BibitemShut {NoStop}%
\bibitem [{\citenamefont {{Sharma}}\ \emph {et~al.}(2024)\citenamefont
  {{Sharma}}, \citenamefont {{Kamaram}}, \citenamefont {{Prince}},
  \citenamefont {{Khatoon}},\ and\ \citenamefont
  {{Bose}}}]{2024MNRAS.527.2672S}%
  \BibitemOpen
  \bibfield  {author} {\bibinfo {author} {\bibfnamefont {A.}~\bibnamefont
  {{Sharma}}}, \bibinfo {author} {\bibfnamefont {S.~R.}\ \bibnamefont
  {{Kamaram}}}, \bibinfo {author} {\bibfnamefont {R.}~\bibnamefont {{Prince}}},
  \bibinfo {author} {\bibfnamefont {R.}~\bibnamefont {{Khatoon}}},\ and\
  \bibinfo {author} {\bibfnamefont {D.}~\bibnamefont {{Bose}}},\ }\bibfield
  {title} {\bibinfo {title} {{Probing the disc-jet coupling in S4 0954+65, PKS
  0903-57, and 4C +01.02 with {\ensuremath{\gamma}}-rays}},\ }\href
  {https://doi.org/10.1093/mnras/stad3399} {\bibfield  {journal} {\bibinfo
  {journal} {\mnras}\ }\textbf {\bibinfo {volume} {527}},\ \bibinfo {pages}
  {2672} (\bibinfo {year} {2024})},\ \Eprint {https://arxiv.org/abs/2311.01738}
  {arXiv:2311.01738 [astro-ph.HE]} \BibitemShut {NoStop}%
\bibitem [{\citenamefont {{Barthelmy}}\ \emph {et~al.}(2005)\citenamefont
  {{Barthelmy}}, \citenamefont {{Barbier}}, \citenamefont {{Cummings}},
  \citenamefont {{Fenimore}}, \citenamefont {{Gehrels}}, \citenamefont
  {{Hullinger}}, \citenamefont {{Krimm}}, \citenamefont {{Markwardt}},
  \citenamefont {{Palmer}}, \citenamefont {{Parsons}}, \citenamefont {{Sato}},
  \citenamefont {{Suzuki}}, \citenamefont {{Takahashi}}, \citenamefont
  {{Tashiro}},\ and\ \citenamefont {{Tueller}}}]{2005SSRv..120..143B}%
  \BibitemOpen
  \bibfield  {author} {\bibinfo {author} {\bibfnamefont {S.~D.}\ \bibnamefont
  {{Barthelmy}}}, \bibinfo {author} {\bibfnamefont {L.~M.}\ \bibnamefont
  {{Barbier}}}, \bibinfo {author} {\bibfnamefont {J.~R.}\ \bibnamefont
  {{Cummings}}}, \bibinfo {author} {\bibfnamefont {E.~E.}\ \bibnamefont
  {{Fenimore}}}, \bibinfo {author} {\bibfnamefont {N.}~\bibnamefont
  {{Gehrels}}}, \bibinfo {author} {\bibfnamefont {D.}~\bibnamefont
  {{Hullinger}}}, \bibinfo {author} {\bibfnamefont {H.~A.}\ \bibnamefont
  {{Krimm}}}, \bibinfo {author} {\bibfnamefont {C.~B.}\ \bibnamefont
  {{Markwardt}}}, \bibinfo {author} {\bibfnamefont {D.~M.}\ \bibnamefont
  {{Palmer}}}, \bibinfo {author} {\bibfnamefont {A.}~\bibnamefont {{Parsons}}},
  \bibinfo {author} {\bibfnamefont {G.}~\bibnamefont {{Sato}}}, \bibinfo
  {author} {\bibfnamefont {M.}~\bibnamefont {{Suzuki}}}, \bibinfo {author}
  {\bibfnamefont {T.}~\bibnamefont {{Takahashi}}}, \bibinfo {author}
  {\bibfnamefont {M.}~\bibnamefont {{Tashiro}}},\ and\ \bibinfo {author}
  {\bibfnamefont {J.}~\bibnamefont {{Tueller}}},\ }\bibfield  {title} {\bibinfo
  {title} {{The Burst Alert Telescope (BAT) on the SWIFT Midex Mission}},\
  }\href {https://doi.org/10.1007/s11214-005-5096-3} {\bibfield  {journal}
  {\bibinfo  {journal} {\ssr}\ }\textbf {\bibinfo {volume} {120}},\ \bibinfo
  {pages} {143} (\bibinfo {year} {2005})},\ \Eprint
  {https://arxiv.org/abs/astro-ph/0507410} {arXiv:astro-ph/0507410 [astro-ph]}
  \BibitemShut {NoStop}%
\bibitem [{\citenamefont {{Lien}}\ \emph {et~al.}(2025)\citenamefont {{Lien}},
  \citenamefont {{Krimm}}, \citenamefont {{Markwardt}}, \citenamefont {{Oh}},
  \citenamefont {{Marcotulli}}, \citenamefont {{Mushotzky}}, \citenamefont
  {{Collins}}, \citenamefont {{Barthelmy}}, \citenamefont {{Baumgartner}},
  \citenamefont {{Cenko}}, \citenamefont {{Koss}}, \citenamefont {{Laha}},
  \citenamefont {{Sakamoto}}, \citenamefont {{Palmer}},\ and\ \citenamefont
  {{Parsotan}}}]{2025ApJ...989..161L}%
  \BibitemOpen
  \bibfield  {author} {\bibinfo {author} {\bibfnamefont {A.~Y.}\ \bibnamefont
  {{Lien}}}, \bibinfo {author} {\bibfnamefont {H.~A.}\ \bibnamefont {{Krimm}}},
  \bibinfo {author} {\bibfnamefont {C.~B.}\ \bibnamefont {{Markwardt}}},
  \bibinfo {author} {\bibfnamefont {K.}~\bibnamefont {{Oh}}}, \bibinfo {author}
  {\bibfnamefont {L.}~\bibnamefont {{Marcotulli}}}, \bibinfo {author}
  {\bibfnamefont {R.}~\bibnamefont {{Mushotzky}}}, \bibinfo {author}
  {\bibfnamefont {N.~R.}\ \bibnamefont {{Collins}}}, \bibinfo {author}
  {\bibfnamefont {S.~D.}\ \bibnamefont {{Barthelmy}}}, \bibinfo {author}
  {\bibfnamefont {W.~H.}\ \bibnamefont {{Baumgartner}}}, \bibinfo {author}
  {\bibfnamefont {S.~B.}\ \bibnamefont {{Cenko}}}, \bibinfo {author}
  {\bibfnamefont {M.}~\bibnamefont {{Koss}}}, \bibinfo {author} {\bibfnamefont
  {S.}~\bibnamefont {{Laha}}}, \bibinfo {author} {\bibfnamefont
  {T.}~\bibnamefont {{Sakamoto}}}, \bibinfo {author} {\bibfnamefont {D.~M.}\
  \bibnamefont {{Palmer}}},\ and\ \bibinfo {author} {\bibfnamefont
  {T.}~\bibnamefont {{Parsotan}}},\ }\bibfield  {title} {\bibinfo {title} {{The
  157 Month Swift/BAT All-sky Hard X-Ray Survey}},\ }\href
  {https://doi.org/10.3847/1538-4357/ade676} {\bibfield  {journal} {\bibinfo
  {journal} {\apj}\ }\textbf {\bibinfo {volume} {989}},\ \bibinfo {eid} {161}
  (\bibinfo {year} {2025})},\ \Eprint {https://arxiv.org/abs/2506.04109}
  {arXiv:2506.04109 [astro-ph.HE]} \BibitemShut {NoStop}%
\bibitem [{\citenamefont {{Tauris}}\ and\ \citenamefont {{van den
  Heuvel}}(2006)}]{2006csxs.book..623T}%
  \BibitemOpen
  \bibfield  {author} {\bibinfo {author} {\bibfnamefont {T.~M.}\ \bibnamefont
  {{Tauris}}}\ and\ \bibinfo {author} {\bibfnamefont {E.~P.~J.}\ \bibnamefont
  {{van den Heuvel}}},\ }\bibfield  {title} {\bibinfo {title} {{Formation and
  evolution of compact stellar X-ray sources}},\ }in\ \href
  {https://doi.org/10.48550/arXiv.astro-ph/0303456} {\emph {\bibinfo
  {booktitle} {Compact stellar X-ray sources}}},\ Vol.~\bibinfo {volume} {39},\
  \bibinfo {editor} {edited by\ \bibinfo {editor} {\bibfnamefont {W.~H.~G.}\
  \bibnamefont {{Lewin}}}\ and\ \bibinfo {editor} {\bibfnamefont
  {M.}~\bibnamefont {{van der Klis}}}}\ (\bibinfo {year} {2006})\ pp.\ \bibinfo
  {pages} {623--665}\BibitemShut {NoStop}%
\bibitem [{\citenamefont {{Mundo}}\ and\ \citenamefont
  {{Mushotzky}}(2023)}]{2023MNRAS.526.4040M}%
  \BibitemOpen
  \bibfield  {author} {\bibinfo {author} {\bibfnamefont {S.~A.}\ \bibnamefont
  {{Mundo}}}\ and\ \bibinfo {author} {\bibfnamefont {R.}~\bibnamefont
  {{Mushotzky}}},\ }\bibfield  {title} {\bibinfo {title} {{Long-term hard X-ray
  variability properties of Swift-BAT blazars}},\ }\href
  {https://doi.org/10.1093/mnras/stad2991} {\bibfield  {journal} {\bibinfo
  {journal} {\mnras}\ }\textbf {\bibinfo {volume} {526}},\ \bibinfo {pages}
  {4040} (\bibinfo {year} {2023})},\ \Eprint {https://arxiv.org/abs/2310.04952}
  {arXiv:2310.04952 [astro-ph.HE]} \BibitemShut {NoStop}%
\bibitem [{\citenamefont {{Kelly}}\ \emph {et~al.}(2009)\citenamefont
  {{Kelly}}, \citenamefont {{Bechtold}},\ and\ \citenamefont
  {{Siemiginowska}}}]{2009ApJ...698..895K}%
  \BibitemOpen
  \bibfield  {author} {\bibinfo {author} {\bibfnamefont {B.~C.}\ \bibnamefont
  {{Kelly}}}, \bibinfo {author} {\bibfnamefont {J.}~\bibnamefont
  {{Bechtold}}},\ and\ \bibinfo {author} {\bibfnamefont {A.}~\bibnamefont
  {{Siemiginowska}}},\ }\bibfield  {title} {\bibinfo {title} {{Are the
  Variations in Quasar Optical Flux Driven by Thermal Fluctuations?}},\ }\href
  {https://doi.org/10.1088/0004-637X/698/1/895} {\bibfield  {journal} {\bibinfo
   {journal} {\apj}\ }\textbf {\bibinfo {volume} {698}},\ \bibinfo {pages}
  {895} (\bibinfo {year} {2009})},\ \Eprint {https://arxiv.org/abs/0903.5315}
  {arXiv:0903.5315 [astro-ph.CO]} \BibitemShut {NoStop}%
\bibitem [{\citenamefont {{MacLeod}}\ \emph {et~al.}(2010)\citenamefont
  {{MacLeod}}, \citenamefont {{Ivezi{\'c}}}, \citenamefont {{Kochanek}},
  \citenamefont {{Koz{\l}owski}}, \citenamefont {{Kelly}}, \citenamefont
  {{Bullock}}, \citenamefont {{Kimball}}, \citenamefont {{Sesar}},
  \citenamefont {{Westman}}, \citenamefont {{Brooks}}, \citenamefont
  {{Gibson}}, \citenamefont {{Becker}},\ and\ \citenamefont {{de
  Vries}}}]{2010ApJ...721.1014M}%
  \BibitemOpen
  \bibfield  {author} {\bibinfo {author} {\bibfnamefont {C.~L.}\ \bibnamefont
  {{MacLeod}}}, \bibinfo {author} {\bibfnamefont {{\v{Z}}.}~\bibnamefont
  {{Ivezi{\'c}}}}, \bibinfo {author} {\bibfnamefont {C.~S.}\ \bibnamefont
  {{Kochanek}}}, \bibinfo {author} {\bibfnamefont {S.}~\bibnamefont
  {{Koz{\l}owski}}}, \bibinfo {author} {\bibfnamefont {B.}~\bibnamefont
  {{Kelly}}}, \bibinfo {author} {\bibfnamefont {E.}~\bibnamefont {{Bullock}}},
  \bibinfo {author} {\bibfnamefont {A.}~\bibnamefont {{Kimball}}}, \bibinfo
  {author} {\bibfnamefont {B.}~\bibnamefont {{Sesar}}}, \bibinfo {author}
  {\bibfnamefont {D.}~\bibnamefont {{Westman}}}, \bibinfo {author}
  {\bibfnamefont {K.}~\bibnamefont {{Brooks}}}, \bibinfo {author}
  {\bibfnamefont {R.}~\bibnamefont {{Gibson}}}, \bibinfo {author}
  {\bibfnamefont {A.~C.}\ \bibnamefont {{Becker}}},\ and\ \bibinfo {author}
  {\bibfnamefont {W.~H.}\ \bibnamefont {{de Vries}}},\ }\bibfield  {title}
  {\bibinfo {title} {{Modeling the Time Variability of SDSS Stripe 82 Quasars
  as a Damped Random Walk}},\ }\href
  {https://doi.org/10.1088/0004-637X/721/2/1014} {\bibfield  {journal}
  {\bibinfo  {journal} {\apj}\ }\textbf {\bibinfo {volume} {721}},\ \bibinfo
  {pages} {1014} (\bibinfo {year} {2010})},\ \Eprint
  {https://arxiv.org/abs/1004.0276} {arXiv:1004.0276 [astro-ph.CO]}
  \BibitemShut {NoStop}%
\bibitem [{\citenamefont {{Zu}}\ \emph {et~al.}(2013)\citenamefont {{Zu}},
  \citenamefont {{Kochanek}}, \citenamefont {{Koz{\l}owski}},\ and\
  \citenamefont {{Udalski}}}]{2013ApJ...765..106Z}%
  \BibitemOpen
  \bibfield  {author} {\bibinfo {author} {\bibfnamefont {Y.}~\bibnamefont
  {{Zu}}}, \bibinfo {author} {\bibfnamefont {C.~S.}\ \bibnamefont
  {{Kochanek}}}, \bibinfo {author} {\bibfnamefont {S.}~\bibnamefont
  {{Koz{\l}owski}}},\ and\ \bibinfo {author} {\bibfnamefont {A.}~\bibnamefont
  {{Udalski}}},\ }\bibfield  {title} {\bibinfo {title} {{Is Quasar Optical
  Variability a Damped Random Walk?}},\ }\href
  {https://doi.org/10.1088/0004-637X/765/2/106} {\bibfield  {journal} {\bibinfo
   {journal} {\apj}\ }\textbf {\bibinfo {volume} {765}},\ \bibinfo {eid} {106}
  (\bibinfo {year} {2013})},\ \Eprint {https://arxiv.org/abs/1202.3783}
  {arXiv:1202.3783 [astro-ph.CO]} \BibitemShut {NoStop}%
\bibitem [{\citenamefont {{Koz{\l}owski}}(2017)}]{2017A&A...597A.128K}%
  \BibitemOpen
  \bibfield  {author} {\bibinfo {author} {\bibfnamefont {S.}~\bibnamefont
  {{Koz{\l}owski}}},\ }\bibfield  {title} {\bibinfo {title} {{Limitations on
  the recovery of the true AGN variability parameters using damped random walk
  modeling}},\ }\href {https://doi.org/10.1051/0004-6361/201629890} {\bibfield
  {journal} {\bibinfo  {journal} {\aap}\ }\textbf {\bibinfo {volume} {597}},\
  \bibinfo {eid} {A128} (\bibinfo {year} {2017})},\ \Eprint
  {https://arxiv.org/abs/1611.08248} {arXiv:1611.08248 [astro-ph.GA]}
  \BibitemShut {NoStop}%
\bibitem [{\citenamefont {{Stone}}\ \emph {et~al.}(2022)\citenamefont
  {{Stone}}, \citenamefont {{Shen}}, \citenamefont {{Burke}}, \citenamefont
  {{Chen}}, \citenamefont {{Yang}}, \citenamefont {{Liu}}, \citenamefont
  {{Gruendl}}, \citenamefont {{Adam{\'o}w}}, \citenamefont
  {{Andrade-Oliveira}}, \citenamefont {{Annis}}, \citenamefont {{Bacon}},
  \citenamefont {{Bertin}}, \citenamefont {{Bocquet}}, \citenamefont
  {{Brooks}}, \citenamefont {{Burke}}, \citenamefont {{Carnero Rosell}},
  \citenamefont {{Carrasco Kind}}, \citenamefont {{Carretero}}, \citenamefont
  {{da Costa}}, \citenamefont {{Pereira}}, \citenamefont {{De Vicente}},
  \citenamefont {{Desai}}, \citenamefont {{Diehl}}, \citenamefont {{Doel}},
  \citenamefont {{Ferrero}}, \citenamefont {{Friedel}}, \citenamefont
  {{Frieman}}, \citenamefont {{Garc{\'\i}a-Bellido}}, \citenamefont
  {{Gaztanaga}}, \citenamefont {{Gruen}}, \citenamefont {{Gutierrez}},
  \citenamefont {{Hinton}}, \citenamefont {{Hollowood}}, \citenamefont
  {{Honscheid}}, \citenamefont {{James}}, \citenamefont {{Kuehn}},
  \citenamefont {{Kuropatkin}}, \citenamefont {{Lidman}}, \citenamefont
  {{Maia}}, \citenamefont {{Menanteau}}, \citenamefont {{Miquel}},
  \citenamefont {{Morgan}}, \citenamefont {{Paz-Chinch{\'o}n}}, \citenamefont
  {{Pieres}}, \citenamefont {{Plazas Malag{\'o}n}}, \citenamefont
  {{Rodriguez-Monroy}}, \citenamefont {{Sanchez}}, \citenamefont {{Scarpine}},
  \citenamefont {{Serrano}}, \citenamefont {{Sevilla-Noarbe}}, \citenamefont
  {{Smith}}, \citenamefont {{Suchyta}}, \citenamefont {{Swanson}},
  \citenamefont {{Tarl{\'e}}}, \citenamefont {{To}},\ and\ \citenamefont {{DES
  Collaboration}}}]{2022MNRAS.514..164S}%
  \BibitemOpen
  \bibfield  {author} {\bibinfo {author} {\bibfnamefont {Z.}~\bibnamefont
  {{Stone}}}, \bibinfo {author} {\bibfnamefont {Y.}~\bibnamefont {{Shen}}},
  \bibinfo {author} {\bibfnamefont {C.~J.}\ \bibnamefont {{Burke}}}, \bibinfo
  {author} {\bibfnamefont {Y.-C.}\ \bibnamefont {{Chen}}}, \bibinfo {author}
  {\bibfnamefont {Q.}~\bibnamefont {{Yang}}}, \bibinfo {author} {\bibfnamefont
  {X.}~\bibnamefont {{Liu}}}, \bibinfo {author} {\bibfnamefont {R.~A.}\
  \bibnamefont {{Gruendl}}}, \bibinfo {author} {\bibfnamefont {M.}~\bibnamefont
  {{Adam{\'o}w}}}, \bibinfo {author} {\bibfnamefont {F.}~\bibnamefont
  {{Andrade-Oliveira}}}, \bibinfo {author} {\bibfnamefont {J.}~\bibnamefont
  {{Annis}}}, \bibinfo {author} {\bibfnamefont {D.}~\bibnamefont {{Bacon}}},
  \bibinfo {author} {\bibfnamefont {E.}~\bibnamefont {{Bertin}}}, \bibinfo
  {author} {\bibfnamefont {S.}~\bibnamefont {{Bocquet}}}, \bibinfo {author}
  {\bibfnamefont {D.}~\bibnamefont {{Brooks}}}, \bibinfo {author}
  {\bibfnamefont {D.~L.}\ \bibnamefont {{Burke}}}, \bibinfo {author}
  {\bibfnamefont {A.}~\bibnamefont {{Carnero Rosell}}}, \bibinfo {author}
  {\bibfnamefont {M.}~\bibnamefont {{Carrasco Kind}}}, \bibinfo {author}
  {\bibfnamefont {J.}~\bibnamefont {{Carretero}}}, \bibinfo {author}
  {\bibfnamefont {L.~N.}\ \bibnamefont {{da Costa}}}, \bibinfo {author}
  {\bibfnamefont {M.~E.~S.}\ \bibnamefont {{Pereira}}}, \bibinfo {author}
  {\bibfnamefont {J.}~\bibnamefont {{De Vicente}}}, \bibinfo {author}
  {\bibfnamefont {S.}~\bibnamefont {{Desai}}}, \bibinfo {author} {\bibfnamefont
  {H.~T.}\ \bibnamefont {{Diehl}}}, \bibinfo {author} {\bibfnamefont
  {P.}~\bibnamefont {{Doel}}}, \bibinfo {author} {\bibfnamefont
  {I.}~\bibnamefont {{Ferrero}}}, \bibinfo {author} {\bibfnamefont {D.~N.}\
  \bibnamefont {{Friedel}}}, \bibinfo {author} {\bibfnamefont {J.}~\bibnamefont
  {{Frieman}}}, \bibinfo {author} {\bibfnamefont {J.}~\bibnamefont
  {{Garc{\'\i}a-Bellido}}}, \bibinfo {author} {\bibfnamefont {E.}~\bibnamefont
  {{Gaztanaga}}}, \bibinfo {author} {\bibfnamefont {D.}~\bibnamefont
  {{Gruen}}}, \bibinfo {author} {\bibfnamefont {G.}~\bibnamefont
  {{Gutierrez}}}, \bibinfo {author} {\bibfnamefont {S.~R.}\ \bibnamefont
  {{Hinton}}}, \bibinfo {author} {\bibfnamefont {D.~L.}\ \bibnamefont
  {{Hollowood}}}, \bibinfo {author} {\bibfnamefont {K.}~\bibnamefont
  {{Honscheid}}}, \bibinfo {author} {\bibfnamefont {D.~J.}\ \bibnamefont
  {{James}}}, \bibinfo {author} {\bibfnamefont {K.}~\bibnamefont {{Kuehn}}},
  \bibinfo {author} {\bibfnamefont {N.}~\bibnamefont {{Kuropatkin}}}, \bibinfo
  {author} {\bibfnamefont {C.}~\bibnamefont {{Lidman}}}, \bibinfo {author}
  {\bibfnamefont {M.~A.~G.}\ \bibnamefont {{Maia}}}, \bibinfo {author}
  {\bibfnamefont {F.}~\bibnamefont {{Menanteau}}}, \bibinfo {author}
  {\bibfnamefont {R.}~\bibnamefont {{Miquel}}}, \bibinfo {author}
  {\bibfnamefont {R.}~\bibnamefont {{Morgan}}}, \bibinfo {author}
  {\bibfnamefont {F.}~\bibnamefont {{Paz-Chinch{\'o}n}}}, \bibinfo {author}
  {\bibfnamefont {A.}~\bibnamefont {{Pieres}}}, \bibinfo {author}
  {\bibfnamefont {A.~A.}\ \bibnamefont {{Plazas Malag{\'o}n}}}, \bibinfo
  {author} {\bibfnamefont {M.}~\bibnamefont {{Rodriguez-Monroy}}}, \bibinfo
  {author} {\bibfnamefont {E.}~\bibnamefont {{Sanchez}}}, \bibinfo {author}
  {\bibfnamefont {V.}~\bibnamefont {{Scarpine}}}, \bibinfo {author}
  {\bibfnamefont {S.}~\bibnamefont {{Serrano}}}, \bibinfo {author}
  {\bibfnamefont {I.}~\bibnamefont {{Sevilla-Noarbe}}}, \bibinfo {author}
  {\bibfnamefont {M.}~\bibnamefont {{Smith}}}, \bibinfo {author} {\bibfnamefont
  {E.}~\bibnamefont {{Suchyta}}}, \bibinfo {author} {\bibfnamefont {M.~E.~C.}\
  \bibnamefont {{Swanson}}}, \bibinfo {author} {\bibfnamefont {G.}~\bibnamefont
  {{Tarl{\'e}}}}, \bibinfo {author} {\bibfnamefont {C.}~\bibnamefont {{To}}},\
  and\ \bibinfo {author} {\bibnamefont {{DES Collaboration}}},\ }\bibfield
  {title} {\bibinfo {title} {{Optical variability of quasars with 20-yr
  photometric light curves}},\ }\href {https://doi.org/10.1093/mnras/stac1259}
  {\bibfield  {journal} {\bibinfo  {journal} {\mnras}\ }\textbf {\bibinfo
  {volume} {514}},\ \bibinfo {pages} {164} (\bibinfo {year} {2022})},\ \Eprint
  {https://arxiv.org/abs/2201.02762} {arXiv:2201.02762 [astro-ph.GA]}
  \BibitemShut {NoStop}%
\bibitem [{\citenamefont {{Zhang}}\ \emph {et~al.}(2022)\citenamefont
  {{Zhang}}, \citenamefont {{Yan}},\ and\ \citenamefont
  {{Zhang}}}]{2022ApJ...930..157Z}%
  \BibitemOpen
  \bibfield  {author} {\bibinfo {author} {\bibfnamefont {H.}~\bibnamefont
  {{Zhang}}}, \bibinfo {author} {\bibfnamefont {D.}~\bibnamefont {{Yan}}},\
  and\ \bibinfo {author} {\bibfnamefont {L.}~\bibnamefont {{Zhang}}},\
  }\bibfield  {title} {\bibinfo {title} {{Characterizing the
  {\ensuremath{\gamma}}-Ray Variability of Active Galactic Nuclei with the
  Stochastic Process Method}},\ }\href
  {https://doi.org/10.3847/1538-4357/ac679e} {\bibfield  {journal} {\bibinfo
  {journal} {\apj}\ }\textbf {\bibinfo {volume} {930}},\ \bibinfo {eid} {157}
  (\bibinfo {year} {2022})},\ \Eprint {https://arxiv.org/abs/2204.09987}
  {arXiv:2204.09987 [astro-ph.HE]} \BibitemShut {NoStop}%
\bibitem [{\citenamefont {{Zhang}}\ \emph
  {et~al.}(2023{\natexlab{a}})\citenamefont {{Zhang}}, \citenamefont {{Yan}},\
  and\ \citenamefont {{Zhang}}}]{2023ApJ...944..103Z}%
  \BibitemOpen
  \bibfield  {author} {\bibinfo {author} {\bibfnamefont {H.}~\bibnamefont
  {{Zhang}}}, \bibinfo {author} {\bibfnamefont {D.}~\bibnamefont {{Yan}}},\
  and\ \bibinfo {author} {\bibfnamefont {L.}~\bibnamefont {{Zhang}}},\
  }\bibfield  {title} {\bibinfo {title} {{Gaussian Process Modeling Blazar
  Multiwavelength Variability: Indirectly Resolving Jet Structure}},\ }\href
  {https://doi.org/10.3847/1538-4357/acafe5} {\bibfield  {journal} {\bibinfo
  {journal} {\apj}\ }\textbf {\bibinfo {volume} {944}},\ \bibinfo {eid} {103}
  (\bibinfo {year} {2023}{\natexlab{a}})},\ \Eprint
  {https://arxiv.org/abs/2301.01025} {arXiv:2301.01025 [astro-ph.HE]}
  \BibitemShut {NoStop}%
\bibitem [{\citenamefont {{Kelly}}\ \emph {et~al.}(2014)\citenamefont
  {{Kelly}}, \citenamefont {{Becker}}, \citenamefont {{Sobolewska}},
  \citenamefont {{Siemiginowska}},\ and\ \citenamefont
  {{Uttley}}}]{2014ApJ...788...33K}%
  \BibitemOpen
  \bibfield  {author} {\bibinfo {author} {\bibfnamefont {B.~C.}\ \bibnamefont
  {{Kelly}}}, \bibinfo {author} {\bibfnamefont {A.~C.}\ \bibnamefont
  {{Becker}}}, \bibinfo {author} {\bibfnamefont {M.}~\bibnamefont
  {{Sobolewska}}}, \bibinfo {author} {\bibfnamefont {A.}~\bibnamefont
  {{Siemiginowska}}},\ and\ \bibinfo {author} {\bibfnamefont {P.}~\bibnamefont
  {{Uttley}}},\ }\bibfield  {title} {\bibinfo {title} {{Flexible and Scalable
  Methods for Quantifying Stochastic Variability in the Era of Massive
  Time-domain Astronomical Data Sets}},\ }\href
  {https://doi.org/10.1088/0004-637X/788/1/33} {\bibfield  {journal} {\bibinfo
  {journal} {\apj}\ }\textbf {\bibinfo {volume} {788}},\ \bibinfo {eid} {33}
  (\bibinfo {year} {2014})},\ \Eprint {https://arxiv.org/abs/1402.5978}
  {arXiv:1402.5978 [astro-ph.IM]} \BibitemShut {NoStop}%
\bibitem [{\citenamefont {{Gillespie}}(1996)}]{1996AmJPh..64..225G}%
  \BibitemOpen
  \bibfield  {author} {\bibinfo {author} {\bibfnamefont {D.~T.}\ \bibnamefont
  {{Gillespie}}},\ }\bibfield  {title} {\bibinfo {title} {{The mathematics of
  Brownian motion and Johnson noise}},\ }\href
  {https://doi.org/10.1119/1.18210} {\bibfield  {journal} {\bibinfo  {journal}
  {American Journal of Physics}\ }\textbf {\bibinfo {volume} {64}},\ \bibinfo
  {pages} {225} (\bibinfo {year} {1996})}\BibitemShut {NoStop}%
\bibitem [{\citenamefont {{Yu}}\ \emph {et~al.}(2022)\citenamefont {{Yu}},
  \citenamefont {{Richards}}, \citenamefont {{Vogeley}}, \citenamefont
  {{Moreno}},\ and\ \citenamefont {{Graham}}}]{2022ApJ...936..132Y}%
  \BibitemOpen
  \bibfield  {author} {\bibinfo {author} {\bibfnamefont {W.}~\bibnamefont
  {{Yu}}}, \bibinfo {author} {\bibfnamefont {G.~T.}\ \bibnamefont
  {{Richards}}}, \bibinfo {author} {\bibfnamefont {M.~S.}\ \bibnamefont
  {{Vogeley}}}, \bibinfo {author} {\bibfnamefont {J.}~\bibnamefont
  {{Moreno}}},\ and\ \bibinfo {author} {\bibfnamefont {M.~J.}\ \bibnamefont
  {{Graham}}},\ }\bibfield  {title} {\bibinfo {title} {{Examining AGN
  UV/Optical Variability beyond the Simple Damped Random Walk}},\ }\href
  {https://doi.org/10.3847/1538-4357/ac8351} {\bibfield  {journal} {\bibinfo
  {journal} {\apj}\ }\textbf {\bibinfo {volume} {936}},\ \bibinfo {eid} {132}
  (\bibinfo {year} {2022})},\ \Eprint {https://arxiv.org/abs/2201.08943}
  {arXiv:2201.08943 [astro-ph.GA]} \BibitemShut {NoStop}%
\bibitem [{\citenamefont {{Xu}}\ \emph {et~al.}(2025)\citenamefont {{Xu}},
  \citenamefont {{Hu}}, \citenamefont {{Chen}},\ and\ \citenamefont
  {{Huang}}}]{2025ApJ...984...45X}%
  \BibitemOpen
  \bibfield  {author} {\bibinfo {author} {\bibfnamefont {J.}~\bibnamefont
  {{Xu}}}, \bibinfo {author} {\bibfnamefont {S.}~\bibnamefont {{Hu}}}, \bibinfo
  {author} {\bibfnamefont {X.}~\bibnamefont {{Chen}}},\ and\ \bibinfo {author}
  {\bibfnamefont {S.}~\bibnamefont {{Huang}}},\ }\bibfield  {title} {\bibinfo
  {title} {{Are Multiwavelength Variability Events in Blazars Driven by a
  Common Stochastic Process?}},\ }\href
  {https://doi.org/10.3847/1538-4357/adc399} {\bibfield  {journal} {\bibinfo
  {journal} {\apj}\ }\textbf {\bibinfo {volume} {984}},\ \bibinfo {eid} {45}
  (\bibinfo {year} {2025})}\BibitemShut {NoStop}%
\bibitem [{\citenamefont {{Yang}}\ \emph {et~al.}(2021)\citenamefont {{Yang}},
  \citenamefont {{Yan}}, \citenamefont {{Zhang}}, \citenamefont {{Dai}},\ and\
  \citenamefont {{Zhang}}}]{2021ApJ...907..105Y}%
  \BibitemOpen
  \bibfield  {author} {\bibinfo {author} {\bibfnamefont {S.}~\bibnamefont
  {{Yang}}}, \bibinfo {author} {\bibfnamefont {D.}~\bibnamefont {{Yan}}},
  \bibinfo {author} {\bibfnamefont {P.}~\bibnamefont {{Zhang}}}, \bibinfo
  {author} {\bibfnamefont {B.}~\bibnamefont {{Dai}}},\ and\ \bibinfo {author}
  {\bibfnamefont {L.}~\bibnamefont {{Zhang}}},\ }\bibfield  {title} {\bibinfo
  {title} {{Gaussian Process Modeling Fermi-LAT {\ensuremath{\gamma}}-Ray
  Blazar Variability: A Sample of Blazars with {\ensuremath{\gamma}}-Ray
  Quasi-periodicities}},\ }\href {https://doi.org/10.3847/1538-4357/abcbff}
  {\bibfield  {journal} {\bibinfo  {journal} {\apj}\ }\textbf {\bibinfo
  {volume} {907}},\ \bibinfo {eid} {105} (\bibinfo {year} {2021})},\ \Eprint
  {https://arxiv.org/abs/2011.10186} {arXiv:2011.10186 [astro-ph.HE]}
  \BibitemShut {NoStop}%
\bibitem [{\citenamefont {{Zhang}}\ \emph
  {et~al.}(2023{\natexlab{b}})\citenamefont {{Zhang}}, \citenamefont {{Yang}},\
  and\ \citenamefont {{Dai}}}]{2023ApJ...946...52Z}%
  \BibitemOpen
  \bibfield  {author} {\bibinfo {author} {\bibfnamefont {H.}~\bibnamefont
  {{Zhang}}}, \bibinfo {author} {\bibfnamefont {S.}~\bibnamefont {{Yang}}},\
  and\ \bibinfo {author} {\bibfnamefont {B.}~\bibnamefont {{Dai}}},\ }\bibfield
   {title} {\bibinfo {title} {{Search for X-Ray Quasiperiodicity of Six AGNs
  Using the Gaussian Process Method}},\ }\href
  {https://doi.org/10.3847/1538-4357/acbe37} {\bibfield  {journal} {\bibinfo
  {journal} {\apj}\ }\textbf {\bibinfo {volume} {946}},\ \bibinfo {eid} {52}
  (\bibinfo {year} {2023}{\natexlab{b}})},\ \Eprint
  {https://arxiv.org/abs/2304.08044} {arXiv:2304.08044 [astro-ph.HE]}
  \BibitemShut {NoStop}%
\bibitem [{\citenamefont {{Foreman-Mackey}}\ \emph {et~al.}(2017)\citenamefont
  {{Foreman-Mackey}}, \citenamefont {{Agol}}, \citenamefont {{Ambikasaran}},\
  and\ \citenamefont {{Angus}}}]{2017AJ....154..220F}%
  \BibitemOpen
  \bibfield  {author} {\bibinfo {author} {\bibfnamefont {D.}~\bibnamefont
  {{Foreman-Mackey}}}, \bibinfo {author} {\bibfnamefont {E.}~\bibnamefont
  {{Agol}}}, \bibinfo {author} {\bibfnamefont {S.}~\bibnamefont
  {{Ambikasaran}}},\ and\ \bibinfo {author} {\bibfnamefont {R.}~\bibnamefont
  {{Angus}}},\ }\bibfield  {title} {\bibinfo {title} {{Fast and Scalable
  Gaussian Process Modeling with Applications to Astronomical Time Series}},\
  }\href {https://doi.org/10.3847/1538-3881/aa9332} {\bibfield  {journal}
  {\bibinfo  {journal} {\aj}\ }\textbf {\bibinfo {volume} {154}},\ \bibinfo
  {eid} {220} (\bibinfo {year} {2017})},\ \Eprint
  {https://arxiv.org/abs/1703.09710} {arXiv:1703.09710 [astro-ph.IM]}
  \BibitemShut {NoStop}%
\bibitem [{\citenamefont {Zhang}\ \emph {et~al.}(2025)\citenamefont {Zhang},
  \citenamefont {Yang}, \citenamefont {Zhang},\ and\ \citenamefont
  {Dai}}]{zhang_2025_17830360}%
  \BibitemOpen
  \bibfield  {author} {\bibinfo {author} {\bibfnamefont {H.}~\bibnamefont
  {Zhang}}, \bibinfo {author} {\bibfnamefont {S.}~\bibnamefont {Yang}},
  \bibinfo {author} {\bibfnamefont {L.}~\bibnamefont {Zhang}},\ and\ \bibinfo
  {author} {\bibfnamefont {B.}~\bibnamefont {Dai}},\ }\bibfield  {title}
  {\bibinfo {title} {A mass-independent damping timescale in black hole
  accretion systems (supplementary materials)},\ }\href
  {https://doi.org/10.5281/zenodo.17830360} {10.5281/zenodo.17830360} (\bibinfo
  {year} {2025})\BibitemShut {NoStop}%
\bibitem [{\citenamefont {{Laha}}\ \emph {et~al.}(2025)\citenamefont {{Laha}},
  \citenamefont {{Ricci}}, \citenamefont {{Mather}}, \citenamefont {{Behar}},
  \citenamefont {{Gallo}}, \citenamefont {{Marin}}, \citenamefont {{Mbarek}},\
  and\ \citenamefont {{Hankla}}}]{2025FrASS..1130392L}%
  \BibitemOpen
  \bibfield  {author} {\bibinfo {author} {\bibfnamefont {S.}~\bibnamefont
  {{Laha}}}, \bibinfo {author} {\bibfnamefont {C.}~\bibnamefont {{Ricci}}},
  \bibinfo {author} {\bibfnamefont {J.~C.}\ \bibnamefont {{Mather}}}, \bibinfo
  {author} {\bibfnamefont {E.}~\bibnamefont {{Behar}}}, \bibinfo {author}
  {\bibfnamefont {L.}~\bibnamefont {{Gallo}}}, \bibinfo {author} {\bibfnamefont
  {F.}~\bibnamefont {{Marin}}}, \bibinfo {author} {\bibfnamefont
  {R.}~\bibnamefont {{Mbarek}}},\ and\ \bibinfo {author} {\bibfnamefont
  {A.}~\bibnamefont {{Hankla}}},\ }\bibfield  {title} {\bibinfo {title} {{X-ray
  properties of coronal emission in radio quiet active galactic nuclei}},\
  }\href {https://doi.org/10.3389/fspas.2024.1530392} {\bibfield  {journal}
  {\bibinfo  {journal} {Frontiers in Astronomy and Space Sciences}\ }\textbf
  {\bibinfo {volume} {11}},\ \bibinfo {eid} {1530392} (\bibinfo {year}
  {2025})},\ \Eprint {https://arxiv.org/abs/2412.11321} {arXiv:2412.11321
  [astro-ph.HE]} \BibitemShut {NoStop}%
\bibitem [{\citenamefont {{Harris}}\ and\ \citenamefont
  {{Krawczynski}}(2006)}]{2006ARA&A..44..463H}%
  \BibitemOpen
  \bibfield  {author} {\bibinfo {author} {\bibfnamefont {D.~E.}\ \bibnamefont
  {{Harris}}}\ and\ \bibinfo {author} {\bibfnamefont {H.}~\bibnamefont
  {{Krawczynski}}},\ }\bibfield  {title} {\bibinfo {title} {{X-Ray Emission
  from Extragalactic Jets}},\ }\href
  {https://doi.org/10.1146/annurev.astro.44.051905.092446} {\bibfield
  {journal} {\bibinfo  {journal} {\araa}\ }\textbf {\bibinfo {volume} {44}},\
  \bibinfo {pages} {463} (\bibinfo {year} {2006})},\ \Eprint
  {https://arxiv.org/abs/astro-ph/0607228} {arXiv:astro-ph/0607228 [astro-ph]}
  \BibitemShut {NoStop}%
\bibitem [{\citenamefont {{Liodakis}}\ \emph {et~al.}(2025)\citenamefont
  {{Liodakis}}, \citenamefont {{Zhang}}, \citenamefont {{Boula}}, \citenamefont
  {{Middei}}, \citenamefont {{Otero-Santos}}, \citenamefont {{Blinov}},
  \citenamefont {{Agudo}}, \citenamefont {{B{\"o}ttcher}}, \citenamefont
  {{Chen}}, \citenamefont {{Ehlert}}, \citenamefont {{Jorstad}}, \citenamefont
  {{Kaaret}}, \citenamefont {{Krawczynski}}, \citenamefont {{Peirson}},
  \citenamefont {{Romani}}, \citenamefont {{Tavecchio}}, \citenamefont
  {{Weisskopf}}, \citenamefont {{Kouch}}, \citenamefont {{Lindfors}},
  \citenamefont {{Nilsson}}, \citenamefont {{McCall}}, \citenamefont
  {{Jermak}}, \citenamefont {{Steele}}, \citenamefont {{Myserlis}},
  \citenamefont {{Gurwell}}, \citenamefont {{Keating}}, \citenamefont {{Rao}},
  \citenamefont {{Kang}}, \citenamefont {{Lee}}, \citenamefont {{Kim}},
  \citenamefont {{Yeon Cheong}}, \citenamefont {{Jeong}}, \citenamefont
  {{Angelakis}}, \citenamefont {{Kraus}}, \citenamefont {{Jos{\'e} Aceituno}},
  \citenamefont {{Bonnoli}}, \citenamefont {{Casanova}}, \citenamefont
  {{Escudero}}, \citenamefont {{Ag{\'\i}s-Gonz{\'a}lez}}, \citenamefont
  {{Morcuende}}, \citenamefont {{Sota}}, \citenamefont {{Bachev}},
  \citenamefont {{Grishina}}, \citenamefont {{Kopatskaya}}, \citenamefont
  {{Larionova}}, \citenamefont {{Morozova}}, \citenamefont {{Savchenko}},
  \citenamefont {{Shishkina}}, \citenamefont {{Troitskiy}}, \citenamefont
  {{Troitskaya}},\ and\ \citenamefont {{Vasilyev}}}]{2025A&A...698L..19L}%
  \BibitemOpen
  \bibfield  {author} {\bibinfo {author} {\bibfnamefont {I.}~\bibnamefont
  {{Liodakis}}}, \bibinfo {author} {\bibfnamefont {H.}~\bibnamefont {{Zhang}}},
  \bibinfo {author} {\bibfnamefont {S.}~\bibnamefont {{Boula}}}, \bibinfo
  {author} {\bibfnamefont {R.}~\bibnamefont {{Middei}}}, \bibinfo {author}
  {\bibfnamefont {J.}~\bibnamefont {{Otero-Santos}}}, \bibinfo {author}
  {\bibfnamefont {D.}~\bibnamefont {{Blinov}}}, \bibinfo {author}
  {\bibfnamefont {I.}~\bibnamefont {{Agudo}}}, \bibinfo {author} {\bibfnamefont
  {M.}~\bibnamefont {{B{\"o}ttcher}}}, \bibinfo {author} {\bibfnamefont
  {C.-T.}\ \bibnamefont {{Chen}}}, \bibinfo {author} {\bibfnamefont {S.~R.}\
  \bibnamefont {{Ehlert}}}, \bibinfo {author} {\bibfnamefont {S.~G.}\
  \bibnamefont {{Jorstad}}}, \bibinfo {author} {\bibfnamefont {P.}~\bibnamefont
  {{Kaaret}}}, \bibinfo {author} {\bibfnamefont {H.}~\bibnamefont
  {{Krawczynski}}}, \bibinfo {author} {\bibfnamefont {A.~L.}\ \bibnamefont
  {{Peirson}}}, \bibinfo {author} {\bibfnamefont {R.~W.}\ \bibnamefont
  {{Romani}}}, \bibinfo {author} {\bibfnamefont {F.}~\bibnamefont
  {{Tavecchio}}}, \bibinfo {author} {\bibfnamefont {M.~C.}\ \bibnamefont
  {{Weisskopf}}}, \bibinfo {author} {\bibfnamefont {P.~M.}\ \bibnamefont
  {{Kouch}}}, \bibinfo {author} {\bibfnamefont {E.}~\bibnamefont {{Lindfors}}},
  \bibinfo {author} {\bibfnamefont {K.}~\bibnamefont {{Nilsson}}}, \bibinfo
  {author} {\bibfnamefont {C.}~\bibnamefont {{McCall}}}, \bibinfo {author}
  {\bibfnamefont {H.~E.}\ \bibnamefont {{Jermak}}}, \bibinfo {author}
  {\bibfnamefont {I.~A.}\ \bibnamefont {{Steele}}}, \bibinfo {author}
  {\bibfnamefont {I.}~\bibnamefont {{Myserlis}}}, \bibinfo {author}
  {\bibfnamefont {M.}~\bibnamefont {{Gurwell}}}, \bibinfo {author}
  {\bibfnamefont {G.~K.}\ \bibnamefont {{Keating}}}, \bibinfo {author}
  {\bibfnamefont {R.}~\bibnamefont {{Rao}}}, \bibinfo {author} {\bibfnamefont
  {S.}~\bibnamefont {{Kang}}}, \bibinfo {author} {\bibfnamefont {S.-S.}\
  \bibnamefont {{Lee}}}, \bibinfo {author} {\bibfnamefont {S.}~\bibnamefont
  {{Kim}}}, \bibinfo {author} {\bibfnamefont {W.}~\bibnamefont {{Yeon
  Cheong}}}, \bibinfo {author} {\bibfnamefont {H.-W.}\ \bibnamefont {{Jeong}}},
  \bibinfo {author} {\bibfnamefont {E.}~\bibnamefont {{Angelakis}}}, \bibinfo
  {author} {\bibfnamefont {A.}~\bibnamefont {{Kraus}}}, \bibinfo {author}
  {\bibfnamefont {F.}~\bibnamefont {{Jos{\'e} Aceituno}}}, \bibinfo {author}
  {\bibfnamefont {G.}~\bibnamefont {{Bonnoli}}}, \bibinfo {author}
  {\bibfnamefont {V.}~\bibnamefont {{Casanova}}}, \bibinfo {author}
  {\bibfnamefont {J.}~\bibnamefont {{Escudero}}}, \bibinfo {author}
  {\bibfnamefont {B.}~\bibnamefont {{Ag{\'\i}s-Gonz{\'a}lez}}}, \bibinfo
  {author} {\bibfnamefont {D.}~\bibnamefont {{Morcuende}}}, \bibinfo {author}
  {\bibfnamefont {A.}~\bibnamefont {{Sota}}}, \bibinfo {author} {\bibfnamefont
  {R.}~\bibnamefont {{Bachev}}}, \bibinfo {author} {\bibfnamefont {T.~S.}\
  \bibnamefont {{Grishina}}}, \bibinfo {author} {\bibfnamefont {E.~N.}\
  \bibnamefont {{Kopatskaya}}}, \bibinfo {author} {\bibfnamefont {E.~G.}\
  \bibnamefont {{Larionova}}}, \bibinfo {author} {\bibfnamefont {D.~A.}\
  \bibnamefont {{Morozova}}}, \bibinfo {author} {\bibfnamefont {S.~S.}\
  \bibnamefont {{Savchenko}}}, \bibinfo {author} {\bibfnamefont {E.~V.}\
  \bibnamefont {{Shishkina}}}, \bibinfo {author} {\bibfnamefont {I.~S.}\
  \bibnamefont {{Troitskiy}}}, \bibinfo {author} {\bibfnamefont {Y.~V.}\
  \bibnamefont {{Troitskaya}}},\ and\ \bibinfo {author} {\bibfnamefont {A.~A.}\
  \bibnamefont {{Vasilyev}}},\ }\bibfield  {title} {\bibinfo {title}
  {{Determining the origin of the X-ray emission in blazars through
  multiwavelength polarization}},\ }\href
  {https://doi.org/10.1051/0004-6361/202554747} {\bibfield  {journal} {\bibinfo
   {journal} {\aap}\ }\textbf {\bibinfo {volume} {698}},\ \bibinfo {eid} {L19}
  (\bibinfo {year} {2025})},\ \Eprint {https://arxiv.org/abs/2505.13603}
  {arXiv:2505.13603 [astro-ph.HE]} \BibitemShut {NoStop}%
\bibitem [{\citenamefont {{Narayan}}\ \emph {et~al.}(1997)\citenamefont
  {{Narayan}}, \citenamefont {{Barret}},\ and\ \citenamefont
  {{McClintock}}}]{1997ApJ...482..448N}%
  \BibitemOpen
  \bibfield  {author} {\bibinfo {author} {\bibfnamefont {R.}~\bibnamefont
  {{Narayan}}}, \bibinfo {author} {\bibfnamefont {D.}~\bibnamefont
  {{Barret}}},\ and\ \bibinfo {author} {\bibfnamefont {J.~E.}\ \bibnamefont
  {{McClintock}}},\ }\bibfield  {title} {\bibinfo {title} {{Advection-dominated
  Accretion Model of the Black Hole V404 Cygni in Quiescence}},\ }\href
  {https://doi.org/10.1086/304134} {\bibfield  {journal} {\bibinfo  {journal}
  {\apj}\ }\textbf {\bibinfo {volume} {482}},\ \bibinfo {pages} {448} (\bibinfo
  {year} {1997})},\ \Eprint {https://arxiv.org/abs/astro-ph/9610014}
  {arXiv:astro-ph/9610014 [astro-ph]} \BibitemShut {NoStop}%
\bibitem [{\citenamefont {{Krawczynski}}\ \emph {et~al.}(2022)\citenamefont
  {{Krawczynski}}, \citenamefont {{Muleri}}, \citenamefont {{Dov{\v{c}}iak}},
  \citenamefont {{Veledina}}, \citenamefont {{Rodriguez Cavero}}, \citenamefont
  {{Svoboda}}, \citenamefont {{Ingram}}, \citenamefont {{Matt}}, \citenamefont
  {{Garcia}}, \citenamefont {{Loktev}}, \citenamefont {{Negro}}, \citenamefont
  {{Poutanen}}, \citenamefont {{Kitaguchi}}, \citenamefont {{Podgorn{\'y}}},
  \citenamefont {{Rankin}}, \citenamefont {{Zhang}}, \citenamefont
  {{Berdyugin}}, \citenamefont {{Berdyugina}}, \citenamefont {{Bianchi}},
  \citenamefont {{Blinov}}, \citenamefont {{Capitanio}}, \citenamefont {{Di
  Lalla}}, \citenamefont {{Draghis}}, \citenamefont {{Fabiani}}, \citenamefont
  {{Kagitani}}, \citenamefont {{Kravtsov}}, \citenamefont {{Kiehlmann}},
  \citenamefont {{Latronico}}, \citenamefont {{Lutovinov}}, \citenamefont
  {{Mandarakas}}, \citenamefont {{Marin}}, \citenamefont {{Marinucci}},
  \citenamefont {{Miller}}, \citenamefont {{Mizuno}}, \citenamefont {{Molkov}},
  \citenamefont {{Omodei}}, \citenamefont {{Petrucci}}, \citenamefont
  {{Ratheesh}}, \citenamefont {{Sakanoi}}, \citenamefont {{Semena}},
  \citenamefont {{Skalidis}}, \citenamefont {{Soffitta}}, \citenamefont
  {{Tennant}}, \citenamefont {{Thalhammer}}, \citenamefont {{Tombesi}},
  \citenamefont {{Weisskopf}}, \citenamefont {{Wilms}}, \citenamefont
  {{Zhang}}, \citenamefont {{Agudo}}, \citenamefont {{Antonelli}},
  \citenamefont {{Bachetti}}, \citenamefont {{Baldini}}, \citenamefont
  {{Baumgartner}}, \citenamefont {{Bellazzini}}, \citenamefont {{Bongiorno}},
  \citenamefont {{Bonino}}, \citenamefont {{Brez}}, \citenamefont
  {{Bucciantini}}, \citenamefont {{Castellano}}, \citenamefont {{Cavazzuti}},
  \citenamefont {{Ciprini}}, \citenamefont {{Costa}}, \citenamefont {{De
  Rosa}}, \citenamefont {{Del Monte}}, \citenamefont {{Di Gesu}}, \citenamefont
  {{Di Marco}}, \citenamefont {{Donnarumma}}, \citenamefont {{Doroshenko}},
  \citenamefont {{Ehlert}}, \citenamefont {{Enoto}}, \citenamefont
  {{Evangelista}}, \citenamefont {{Ferrazzoli}}, \citenamefont {{Gunji}},
  \citenamefont {{Hayashida}}, \citenamefont {{Heyl}}, \citenamefont
  {{Iwakiri}}, \citenamefont {{Jorstad}}, \citenamefont {{Karas}},
  \citenamefont {{Kolodziejczak}}, \citenamefont {{La Monaca}}, \citenamefont
  {{Liodakis}}, \citenamefont {{Maldera}}, \citenamefont {{Manfreda}},
  \citenamefont {{Marscher}}, \citenamefont {{Marshall}}, \citenamefont
  {{Mitsuishi}}, \citenamefont {{Ng}}, \citenamefont
  {{O{\textquoteright}Dell}}, \citenamefont {{Oppedisano}}, \citenamefont
  {{Papitto}}, \citenamefont {{Pavlov}}, \citenamefont {{Peirson}},
  \citenamefont {{Perri}}, \citenamefont {{Pesce-Rollins}}, \citenamefont
  {{Pilia}}, \citenamefont {{Possenti}}, \citenamefont {{Puccetti}},
  \citenamefont {{Ramsey}}, \citenamefont {{Romani}}, \citenamefont
  {{Sgr{\`o}}}, \citenamefont {{Slane}}, \citenamefont {{Spandre}},
  \citenamefont {{Tamagawa}}, \citenamefont {{Tavecchio}}, \citenamefont
  {{Taverna}}, \citenamefont {{Tawara}}, \citenamefont {{Thomas}},
  \citenamefont {{Trois}}, \citenamefont {{Tsygankov}}, \citenamefont
  {{Turolla}}, \citenamefont {{Vink}}, \citenamefont {{Wu}}, \citenamefont
  {{Xie}},\ and\ \citenamefont {{Zane}}}]{2022Sci...378..650K}%
  \BibitemOpen
  \bibfield  {author} {\bibinfo {author} {\bibfnamefont {H.}~\bibnamefont
  {{Krawczynski}}}, \bibinfo {author} {\bibfnamefont {F.}~\bibnamefont
  {{Muleri}}}, \bibinfo {author} {\bibfnamefont {M.}~\bibnamefont
  {{Dov{\v{c}}iak}}}, \bibinfo {author} {\bibfnamefont {A.}~\bibnamefont
  {{Veledina}}}, \bibinfo {author} {\bibfnamefont {N.}~\bibnamefont {{Rodriguez
  Cavero}}}, \bibinfo {author} {\bibfnamefont {J.}~\bibnamefont {{Svoboda}}},
  \bibinfo {author} {\bibfnamefont {A.}~\bibnamefont {{Ingram}}}, \bibinfo
  {author} {\bibfnamefont {G.}~\bibnamefont {{Matt}}}, \bibinfo {author}
  {\bibfnamefont {J.~A.}\ \bibnamefont {{Garcia}}}, \bibinfo {author}
  {\bibfnamefont {V.}~\bibnamefont {{Loktev}}}, \bibinfo {author}
  {\bibfnamefont {M.}~\bibnamefont {{Negro}}}, \bibinfo {author} {\bibfnamefont
  {J.}~\bibnamefont {{Poutanen}}}, \bibinfo {author} {\bibfnamefont
  {T.}~\bibnamefont {{Kitaguchi}}}, \bibinfo {author} {\bibfnamefont
  {J.}~\bibnamefont {{Podgorn{\'y}}}}, \bibinfo {author} {\bibfnamefont
  {J.}~\bibnamefont {{Rankin}}}, \bibinfo {author} {\bibfnamefont
  {W.}~\bibnamefont {{Zhang}}}, \bibinfo {author} {\bibfnamefont
  {A.}~\bibnamefont {{Berdyugin}}}, \bibinfo {author} {\bibfnamefont {S.~V.}\
  \bibnamefont {{Berdyugina}}}, \bibinfo {author} {\bibfnamefont
  {S.}~\bibnamefont {{Bianchi}}}, \bibinfo {author} {\bibfnamefont
  {D.}~\bibnamefont {{Blinov}}}, \bibinfo {author} {\bibfnamefont
  {F.}~\bibnamefont {{Capitanio}}}, \bibinfo {author} {\bibfnamefont
  {N.}~\bibnamefont {{Di Lalla}}}, \bibinfo {author} {\bibfnamefont
  {P.}~\bibnamefont {{Draghis}}}, \bibinfo {author} {\bibfnamefont
  {S.}~\bibnamefont {{Fabiani}}}, \bibinfo {author} {\bibfnamefont
  {M.}~\bibnamefont {{Kagitani}}}, \bibinfo {author} {\bibfnamefont
  {V.}~\bibnamefont {{Kravtsov}}}, \bibinfo {author} {\bibfnamefont
  {S.}~\bibnamefont {{Kiehlmann}}}, \bibinfo {author} {\bibfnamefont
  {L.}~\bibnamefont {{Latronico}}}, \bibinfo {author} {\bibfnamefont {A.~A.}\
  \bibnamefont {{Lutovinov}}}, \bibinfo {author} {\bibfnamefont
  {N.}~\bibnamefont {{Mandarakas}}}, \bibinfo {author} {\bibfnamefont
  {F.}~\bibnamefont {{Marin}}}, \bibinfo {author} {\bibfnamefont
  {A.}~\bibnamefont {{Marinucci}}}, \bibinfo {author} {\bibfnamefont {J.~M.}\
  \bibnamefont {{Miller}}}, \bibinfo {author} {\bibfnamefont {T.}~\bibnamefont
  {{Mizuno}}}, \bibinfo {author} {\bibfnamefont {S.~V.}\ \bibnamefont
  {{Molkov}}}, \bibinfo {author} {\bibfnamefont {N.}~\bibnamefont {{Omodei}}},
  \bibinfo {author} {\bibfnamefont {P.-O.}\ \bibnamefont {{Petrucci}}},
  \bibinfo {author} {\bibfnamefont {A.}~\bibnamefont {{Ratheesh}}}, \bibinfo
  {author} {\bibfnamefont {T.}~\bibnamefont {{Sakanoi}}}, \bibinfo {author}
  {\bibfnamefont {A.~N.}\ \bibnamefont {{Semena}}}, \bibinfo {author}
  {\bibfnamefont {R.}~\bibnamefont {{Skalidis}}}, \bibinfo {author}
  {\bibfnamefont {P.}~\bibnamefont {{Soffitta}}}, \bibinfo {author}
  {\bibfnamefont {A.~F.}\ \bibnamefont {{Tennant}}}, \bibinfo {author}
  {\bibfnamefont {P.}~\bibnamefont {{Thalhammer}}}, \bibinfo {author}
  {\bibfnamefont {F.}~\bibnamefont {{Tombesi}}}, \bibinfo {author}
  {\bibfnamefont {M.~C.}\ \bibnamefont {{Weisskopf}}}, \bibinfo {author}
  {\bibfnamefont {J.}~\bibnamefont {{Wilms}}}, \bibinfo {author} {\bibfnamefont
  {S.}~\bibnamefont {{Zhang}}}, \bibinfo {author} {\bibfnamefont
  {I.}~\bibnamefont {{Agudo}}}, \bibinfo {author} {\bibfnamefont {L.~A.}\
  \bibnamefont {{Antonelli}}}, \bibinfo {author} {\bibfnamefont
  {M.}~\bibnamefont {{Bachetti}}}, \bibinfo {author} {\bibfnamefont
  {L.}~\bibnamefont {{Baldini}}}, \bibinfo {author} {\bibfnamefont {W.~H.}\
  \bibnamefont {{Baumgartner}}}, \bibinfo {author} {\bibfnamefont
  {R.}~\bibnamefont {{Bellazzini}}}, \bibinfo {author} {\bibfnamefont {S.~D.}\
  \bibnamefont {{Bongiorno}}}, \bibinfo {author} {\bibfnamefont
  {R.}~\bibnamefont {{Bonino}}}, \bibinfo {author} {\bibfnamefont
  {A.}~\bibnamefont {{Brez}}}, \bibinfo {author} {\bibfnamefont
  {N.}~\bibnamefont {{Bucciantini}}}, \bibinfo {author} {\bibfnamefont
  {S.}~\bibnamefont {{Castellano}}}, \bibinfo {author} {\bibfnamefont
  {E.}~\bibnamefont {{Cavazzuti}}}, \bibinfo {author} {\bibfnamefont
  {S.}~\bibnamefont {{Ciprini}}}, \bibinfo {author} {\bibfnamefont
  {E.}~\bibnamefont {{Costa}}}, \bibinfo {author} {\bibfnamefont
  {A.}~\bibnamefont {{De Rosa}}}, \bibinfo {author} {\bibfnamefont
  {E.}~\bibnamefont {{Del Monte}}}, \bibinfo {author} {\bibfnamefont
  {L.}~\bibnamefont {{Di Gesu}}}, \bibinfo {author} {\bibfnamefont
  {A.}~\bibnamefont {{Di Marco}}}, \bibinfo {author} {\bibfnamefont
  {I.}~\bibnamefont {{Donnarumma}}}, \bibinfo {author} {\bibfnamefont
  {V.}~\bibnamefont {{Doroshenko}}}, \bibinfo {author} {\bibfnamefont {S.~R.}\
  \bibnamefont {{Ehlert}}}, \bibinfo {author} {\bibfnamefont {T.}~\bibnamefont
  {{Enoto}}}, \bibinfo {author} {\bibfnamefont {Y.}~\bibnamefont
  {{Evangelista}}}, \bibinfo {author} {\bibfnamefont {R.}~\bibnamefont
  {{Ferrazzoli}}}, \bibinfo {author} {\bibfnamefont {S.}~\bibnamefont
  {{Gunji}}}, \bibinfo {author} {\bibfnamefont {K.}~\bibnamefont
  {{Hayashida}}}, \bibinfo {author} {\bibfnamefont {J.}~\bibnamefont {{Heyl}}},
  \bibinfo {author} {\bibfnamefont {W.}~\bibnamefont {{Iwakiri}}}, \bibinfo
  {author} {\bibfnamefont {S.~G.}\ \bibnamefont {{Jorstad}}}, \bibinfo {author}
  {\bibfnamefont {V.}~\bibnamefont {{Karas}}}, \bibinfo {author} {\bibfnamefont
  {J.~J.}\ \bibnamefont {{Kolodziejczak}}}, \bibinfo {author} {\bibfnamefont
  {F.}~\bibnamefont {{La Monaca}}}, \bibinfo {author} {\bibfnamefont
  {I.}~\bibnamefont {{Liodakis}}}, \bibinfo {author} {\bibfnamefont
  {S.}~\bibnamefont {{Maldera}}}, \bibinfo {author} {\bibfnamefont
  {A.}~\bibnamefont {{Manfreda}}}, \bibinfo {author} {\bibfnamefont {A.~P.}\
  \bibnamefont {{Marscher}}}, \bibinfo {author} {\bibfnamefont {H.~L.}\
  \bibnamefont {{Marshall}}}, \bibinfo {author} {\bibfnamefont
  {I.}~\bibnamefont {{Mitsuishi}}}, \bibinfo {author} {\bibfnamefont {C.-Y.}\
  \bibnamefont {{Ng}}}, \bibinfo {author} {\bibfnamefont {S.~L.}\ \bibnamefont
  {{O{\textquoteright}Dell}}}, \bibinfo {author} {\bibfnamefont
  {C.}~\bibnamefont {{Oppedisano}}}, \bibinfo {author} {\bibfnamefont
  {A.}~\bibnamefont {{Papitto}}}, \bibinfo {author} {\bibfnamefont {G.~G.}\
  \bibnamefont {{Pavlov}}}, \bibinfo {author} {\bibfnamefont {A.~L.}\
  \bibnamefont {{Peirson}}}, \bibinfo {author} {\bibfnamefont {M.}~\bibnamefont
  {{Perri}}}, \bibinfo {author} {\bibfnamefont {M.}~\bibnamefont
  {{Pesce-Rollins}}}, \bibinfo {author} {\bibfnamefont {M.}~\bibnamefont
  {{Pilia}}}, \bibinfo {author} {\bibfnamefont {A.}~\bibnamefont {{Possenti}}},
  \bibinfo {author} {\bibfnamefont {S.}~\bibnamefont {{Puccetti}}}, \bibinfo
  {author} {\bibfnamefont {B.~D.}\ \bibnamefont {{Ramsey}}}, \bibinfo {author}
  {\bibfnamefont {R.~W.}\ \bibnamefont {{Romani}}}, \bibinfo {author}
  {\bibfnamefont {C.}~\bibnamefont {{Sgr{\`o}}}}, \bibinfo {author}
  {\bibfnamefont {P.}~\bibnamefont {{Slane}}}, \bibinfo {author} {\bibfnamefont
  {G.}~\bibnamefont {{Spandre}}}, \bibinfo {author} {\bibfnamefont
  {T.}~\bibnamefont {{Tamagawa}}}, \bibinfo {author} {\bibfnamefont
  {F.}~\bibnamefont {{Tavecchio}}}, \bibinfo {author} {\bibfnamefont
  {R.}~\bibnamefont {{Taverna}}}, \bibinfo {author} {\bibfnamefont
  {Y.}~\bibnamefont {{Tawara}}}, \bibinfo {author} {\bibfnamefont {N.~E.}\
  \bibnamefont {{Thomas}}}, \bibinfo {author} {\bibfnamefont {A.}~\bibnamefont
  {{Trois}}}, \bibinfo {author} {\bibfnamefont {S.}~\bibnamefont
  {{Tsygankov}}}, \bibinfo {author} {\bibfnamefont {R.}~\bibnamefont
  {{Turolla}}}, \bibinfo {author} {\bibfnamefont {J.}~\bibnamefont {{Vink}}},
  \bibinfo {author} {\bibfnamefont {K.}~\bibnamefont {{Wu}}}, \bibinfo {author}
  {\bibfnamefont {F.}~\bibnamefont {{Xie}}},\ and\ \bibinfo {author}
  {\bibfnamefont {S.}~\bibnamefont {{Zane}}},\ }\bibfield  {title} {\bibinfo
  {title} {{Polarized x-rays constrain the disk-jet geometry in the black hole
  x-ray binary Cygnus X-1}},\ }\href {https://doi.org/10.1126/science.add5399}
  {\bibfield  {journal} {\bibinfo  {journal} {Science}\ }\textbf {\bibinfo
  {volume} {378}},\ \bibinfo {pages} {650} (\bibinfo {year} {2022})},\ \Eprint
  {https://arxiv.org/abs/2206.09972} {arXiv:2206.09972 [astro-ph.HE]}
  \BibitemShut {NoStop}%
\bibitem [{\citenamefont {{Liodakis}}\ \emph {et~al.}(2017)\citenamefont
  {{Liodakis}}, \citenamefont {{Pavlidou}}, \citenamefont {{Papadakis}},
  \citenamefont {{Angelakis}}, \citenamefont {{Marchili}}, \citenamefont
  {{Zensus}}, \citenamefont {{Fuhrmann}}, \citenamefont {{Karamanavis}},
  \citenamefont {{Myserlis}}, \citenamefont {{Nestoras}}, \citenamefont
  {{Palaiologou}},\ and\ \citenamefont {{Readhead}}}]{2017ApJ...851..144L}%
  \BibitemOpen
  \bibfield  {author} {\bibinfo {author} {\bibfnamefont {I.}~\bibnamefont
  {{Liodakis}}}, \bibinfo {author} {\bibfnamefont {V.}~\bibnamefont
  {{Pavlidou}}}, \bibinfo {author} {\bibfnamefont {I.}~\bibnamefont
  {{Papadakis}}}, \bibinfo {author} {\bibfnamefont {E.}~\bibnamefont
  {{Angelakis}}}, \bibinfo {author} {\bibfnamefont {N.}~\bibnamefont
  {{Marchili}}}, \bibinfo {author} {\bibfnamefont {J.~A.}\ \bibnamefont
  {{Zensus}}}, \bibinfo {author} {\bibfnamefont {L.}~\bibnamefont
  {{Fuhrmann}}}, \bibinfo {author} {\bibfnamefont {V.}~\bibnamefont
  {{Karamanavis}}}, \bibinfo {author} {\bibfnamefont {I.}~\bibnamefont
  {{Myserlis}}}, \bibinfo {author} {\bibfnamefont {I.}~\bibnamefont
  {{Nestoras}}}, \bibinfo {author} {\bibfnamefont {E.}~\bibnamefont
  {{Palaiologou}}},\ and\ \bibinfo {author} {\bibfnamefont {A.~C.~S.}\
  \bibnamefont {{Readhead}}},\ }\bibfield  {title} {\bibinfo {title} {{Scale
  Invariant Jets: From Blazars to Microquasars}},\ }\href
  {https://doi.org/10.3847/1538-4357/aa9992} {\bibfield  {journal} {\bibinfo
  {journal} {\apj}\ }\textbf {\bibinfo {volume} {851}},\ \bibinfo {eid} {144}
  (\bibinfo {year} {2017})},\ \Eprint {https://arxiv.org/abs/1711.03979}
  {arXiv:1711.03979 [astro-ph.HE]} \BibitemShut {NoStop}%
\bibitem [{\citenamefont {{Narayan}}\ \emph {et~al.}(1998)\citenamefont
  {{Narayan}}, \citenamefont {{Mahadevan}},\ and\ \citenamefont
  {{Quataert}}}]{1998tbha.conf..148N}%
  \BibitemOpen
  \bibfield  {author} {\bibinfo {author} {\bibfnamefont {R.}~\bibnamefont
  {{Narayan}}}, \bibinfo {author} {\bibfnamefont {R.}~\bibnamefont
  {{Mahadevan}}},\ and\ \bibinfo {author} {\bibfnamefont {E.}~\bibnamefont
  {{Quataert}}},\ }\bibfield  {title} {\bibinfo {title} {{Advection-dominated
  accretion around black holes}},\ }in\ \href
  {https://doi.org/10.48550/arXiv.astro-ph/9803141} {\emph {\bibinfo
  {booktitle} {Theory of Black Hole Accretion Disks}}},\ \bibinfo {editor}
  {edited by\ \bibinfo {editor} {\bibfnamefont {M.~A.}\ \bibnamefont
  {{Abramowicz}}}, \bibinfo {editor} {\bibfnamefont {G.}~\bibnamefont
  {{Bj{\"o}rnsson}}},\ and\ \bibinfo {editor} {\bibfnamefont {J.~E.}\
  \bibnamefont {{Pringle}}}}\ (\bibinfo {year} {1998})\ pp.\ \bibinfo {pages}
  {148--182},\ \Eprint {https://arxiv.org/abs/astro-ph/9803141}
  {arXiv:astro-ph/9803141 [astro-ph]} \BibitemShut {NoStop}%
\bibitem [{\citenamefont {{Davis}}\ \emph {et~al.}(2010)\citenamefont
  {{Davis}}, \citenamefont {{Stone}},\ and\ \citenamefont
  {{Pessah}}}]{2010ApJ...713...52D}%
  \BibitemOpen
  \bibfield  {author} {\bibinfo {author} {\bibfnamefont {S.~W.}\ \bibnamefont
  {{Davis}}}, \bibinfo {author} {\bibfnamefont {J.~M.}\ \bibnamefont
  {{Stone}}},\ and\ \bibinfo {author} {\bibfnamefont {M.~E.}\ \bibnamefont
  {{Pessah}}},\ }\bibfield  {title} {\bibinfo {title} {{Sustained
  Magnetorotational Turbulence in Local Simulations of Stratified Disks with
  Zero Net Magnetic Flux}},\ }\href
  {https://doi.org/10.1088/0004-637X/713/1/52} {\bibfield  {journal} {\bibinfo
  {journal} {\apj}\ }\textbf {\bibinfo {volume} {713}},\ \bibinfo {pages} {52}
  (\bibinfo {year} {2010})},\ \Eprint {https://arxiv.org/abs/0909.1570}
  {arXiv:0909.1570 [astro-ph.HE]} \BibitemShut {NoStop}%
\bibitem [{\citenamefont {{Kotko}}\ and\ \citenamefont
  {{Lasota}}(2012)}]{2012A&A...545A.115K}%
  \BibitemOpen
  \bibfield  {author} {\bibinfo {author} {\bibfnamefont {I.}~\bibnamefont
  {{Kotko}}}\ and\ \bibinfo {author} {\bibfnamefont {J.~P.}\ \bibnamefont
  {{Lasota}}},\ }\bibfield  {title} {\bibinfo {title} {{The viscosity parameter
  {\ensuremath{\alpha}} and the properties of accretion disc outbursts in close
  binaries}},\ }\href {https://doi.org/10.1051/0004-6361/201219618} {\bibfield
  {journal} {\bibinfo  {journal} {\aap}\ }\textbf {\bibinfo {volume} {545}},\
  \bibinfo {eid} {A115} (\bibinfo {year} {2012})},\ \Eprint
  {https://arxiv.org/abs/1209.0017} {arXiv:1209.0017 [astro-ph.SR]}
  \BibitemShut {NoStop}%
\bibitem [{\citenamefont {{Martin}}\ \emph {et~al.}(2019)\citenamefont
  {{Martin}}, \citenamefont {{Nixon}}, \citenamefont {{Pringle}},\ and\
  \citenamefont {{Livio}}}]{2019NewA...70....7M}%
  \BibitemOpen
  \bibfield  {author} {\bibinfo {author} {\bibfnamefont {R.~G.}\ \bibnamefont
  {{Martin}}}, \bibinfo {author} {\bibfnamefont {C.~J.}\ \bibnamefont
  {{Nixon}}}, \bibinfo {author} {\bibfnamefont {J.~E.}\ \bibnamefont
  {{Pringle}}},\ and\ \bibinfo {author} {\bibfnamefont {M.}~\bibnamefont
  {{Livio}}},\ }\bibfield  {title} {\bibinfo {title} {{On the physical nature
  of accretion disc viscosity}},\ }\href
  {https://doi.org/10.1016/j.newast.2019.01.001} {\bibfield  {journal}
  {\bibinfo  {journal} {\na}\ }\textbf {\bibinfo {volume} {70}},\ \bibinfo
  {pages} {7} (\bibinfo {year} {2019})},\ \Eprint
  {https://arxiv.org/abs/1901.01580} {arXiv:1901.01580 [astro-ph.HE]}
  \BibitemShut {NoStop}%
\bibitem [{\citenamefont {{Maraschi}}\ and\ \citenamefont
  {{Tavecchio}}(2003)}]{2003ApJ...593..667M}%
  \BibitemOpen
  \bibfield  {author} {\bibinfo {author} {\bibfnamefont {L.}~\bibnamefont
  {{Maraschi}}}\ and\ \bibinfo {author} {\bibfnamefont {F.}~\bibnamefont
  {{Tavecchio}}},\ }\bibfield  {title} {\bibinfo {title} {{The Jet-Disk
  Connection and Blazar Unification}},\ }\href {https://doi.org/10.1086/342118}
  {\bibfield  {journal} {\bibinfo  {journal} {\apj}\ }\textbf {\bibinfo
  {volume} {593}},\ \bibinfo {pages} {667} (\bibinfo {year} {2003})},\ \Eprint
  {https://arxiv.org/abs/astro-ph/0205252} {arXiv:astro-ph/0205252 [astro-ph]}
  \BibitemShut {NoStop}%
\bibitem [{\citenamefont {{Livio}}\ \emph {et~al.}(2003)\citenamefont
  {{Livio}}, \citenamefont {{Pringle}},\ and\ \citenamefont
  {{King}}}]{2003ApJ...593..184L}%
  \BibitemOpen
  \bibfield  {author} {\bibinfo {author} {\bibfnamefont {M.}~\bibnamefont
  {{Livio}}}, \bibinfo {author} {\bibfnamefont {J.~E.}\ \bibnamefont
  {{Pringle}}},\ and\ \bibinfo {author} {\bibfnamefont {A.~R.}\ \bibnamefont
  {{King}}},\ }\bibfield  {title} {\bibinfo {title} {{The Disk-Jet Connection
  in Microquasars and Active Galactic Nuclei}},\ }\href
  {https://doi.org/10.1086/375872} {\bibfield  {journal} {\bibinfo  {journal}
  {\apj}\ }\textbf {\bibinfo {volume} {593}},\ \bibinfo {pages} {184} (\bibinfo
  {year} {2003})},\ \Eprint {https://arxiv.org/abs/astro-ph/0304367}
  {arXiv:astro-ph/0304367 [astro-ph]} \BibitemShut {NoStop}%
\bibitem [{\citenamefont {{Wang}}\ \emph {et~al.}(2004)\citenamefont {{Wang}},
  \citenamefont {{Luo}},\ and\ \citenamefont {{Ho}}}]{2004ApJ...615L...9W}%
  \BibitemOpen
  \bibfield  {author} {\bibinfo {author} {\bibfnamefont {J.-M.}\ \bibnamefont
  {{Wang}}}, \bibinfo {author} {\bibfnamefont {B.}~\bibnamefont {{Luo}}},\ and\
  \bibinfo {author} {\bibfnamefont {L.~C.}\ \bibnamefont {{Ho}}},\ }\bibfield
  {title} {\bibinfo {title} {{The Connection between Jets, Accretion Disks, and
  Black Hole Mass in Blazars}},\ }\href {https://doi.org/10.1086/426060}
  {\bibfield  {journal} {\bibinfo  {journal} {\apjl}\ }\textbf {\bibinfo
  {volume} {615}},\ \bibinfo {pages} {L9} (\bibinfo {year} {2004})},\ \Eprint
  {https://arxiv.org/abs/astro-ph/0412074} {arXiv:astro-ph/0412074 [astro-ph]}
  \BibitemShut {NoStop}%
\bibitem [{\citenamefont {{Inoue}}\ \emph {et~al.}(2017)\citenamefont
  {{Inoue}}, \citenamefont {{Doi}}, \citenamefont {{Tanaka}}, \citenamefont
  {{Sikora}},\ and\ \citenamefont {{Madejski}}}]{2017ApJ...840...46I}%
  \BibitemOpen
  \bibfield  {author} {\bibinfo {author} {\bibfnamefont {Y.}~\bibnamefont
  {{Inoue}}}, \bibinfo {author} {\bibfnamefont {A.}~\bibnamefont {{Doi}}},
  \bibinfo {author} {\bibfnamefont {Y.~T.}\ \bibnamefont {{Tanaka}}}, \bibinfo
  {author} {\bibfnamefont {M.}~\bibnamefont {{Sikora}}},\ and\ \bibinfo
  {author} {\bibfnamefont {G.~M.}\ \bibnamefont {{Madejski}}},\ }\bibfield
  {title} {\bibinfo {title} {{Disk-Jet Connection in Active Supermassive Black
  Holes in the Standard Accretion Disk Regime}},\ }\href
  {https://doi.org/10.3847/1538-4357/aa6b57} {\bibfield  {journal} {\bibinfo
  {journal} {\apj}\ }\textbf {\bibinfo {volume} {840}},\ \bibinfo {eid} {46}
  (\bibinfo {year} {2017})},\ \Eprint {https://arxiv.org/abs/1704.00123}
  {arXiv:1704.00123 [astro-ph.HE]} \BibitemShut {NoStop}%
\bibitem [{\citenamefont {{Wang}}\ \emph {et~al.}(2021)\citenamefont {{Wang}},
  \citenamefont {{Mastroserio}}, \citenamefont {{Kara}}, \citenamefont
  {{Garc{\'\i}a}}, \citenamefont {{Ingram}}, \citenamefont {{Connors}},
  \citenamefont {{van der Klis}}, \citenamefont {{Dauser}}, \citenamefont
  {{Steiner}}, \citenamefont {{Buisson}}, \citenamefont {{Homan}},
  \citenamefont {{Lucchini}}, \citenamefont {{Fabian}}, \citenamefont
  {{Bright}}, \citenamefont {{Fender}}, \citenamefont {{Cackett}},\ and\
  \citenamefont {{Remillard}}}]{2021ApJ...910L...3W}%
  \BibitemOpen
  \bibfield  {author} {\bibinfo {author} {\bibfnamefont {J.}~\bibnamefont
  {{Wang}}}, \bibinfo {author} {\bibfnamefont {G.}~\bibnamefont
  {{Mastroserio}}}, \bibinfo {author} {\bibfnamefont {E.}~\bibnamefont
  {{Kara}}}, \bibinfo {author} {\bibfnamefont {J.~A.}\ \bibnamefont
  {{Garc{\'\i}a}}}, \bibinfo {author} {\bibfnamefont {A.}~\bibnamefont
  {{Ingram}}}, \bibinfo {author} {\bibfnamefont {R.}~\bibnamefont {{Connors}}},
  \bibinfo {author} {\bibfnamefont {M.}~\bibnamefont {{van der Klis}}},
  \bibinfo {author} {\bibfnamefont {T.}~\bibnamefont {{Dauser}}}, \bibinfo
  {author} {\bibfnamefont {J.~F.}\ \bibnamefont {{Steiner}}}, \bibinfo {author}
  {\bibfnamefont {D.~J.~K.}\ \bibnamefont {{Buisson}}}, \bibinfo {author}
  {\bibfnamefont {J.}~\bibnamefont {{Homan}}}, \bibinfo {author} {\bibfnamefont
  {M.}~\bibnamefont {{Lucchini}}}, \bibinfo {author} {\bibfnamefont {A.~C.}\
  \bibnamefont {{Fabian}}}, \bibinfo {author} {\bibfnamefont {J.}~\bibnamefont
  {{Bright}}}, \bibinfo {author} {\bibfnamefont {R.}~\bibnamefont {{Fender}}},
  \bibinfo {author} {\bibfnamefont {E.~M.}\ \bibnamefont {{Cackett}}},\ and\
  \bibinfo {author} {\bibfnamefont {R.~A.}\ \bibnamefont {{Remillard}}},\
  }\bibfield  {title} {\bibinfo {title} {{Disk, Corona, Jet Connection in the
  Intermediate State of MAXI J1820+070 Revealed by NICER Spectral-timing
  Analysis}},\ }\href {https://doi.org/10.3847/2041-8213/abec79} {\bibfield
  {journal} {\bibinfo  {journal} {\apjl}\ }\textbf {\bibinfo {volume} {910}},\
  \bibinfo {eid} {L3} (\bibinfo {year} {2021})},\ \Eprint
  {https://arxiv.org/abs/2103.05616} {arXiv:2103.05616 [astro-ph.HE]}
  \BibitemShut {NoStop}%
\bibitem [{\citenamefont {{Ryan}}\ \emph {et~al.}(2019)\citenamefont {{Ryan}},
  \citenamefont {{Siemiginowska}}, \citenamefont {{Sobolewska}},\ and\
  \citenamefont {{Grindlay}}}]{2019ApJ...885...12R}%
  \BibitemOpen
  \bibfield  {author} {\bibinfo {author} {\bibfnamefont {J.~L.}\ \bibnamefont
  {{Ryan}}}, \bibinfo {author} {\bibfnamefont {A.}~\bibnamefont
  {{Siemiginowska}}}, \bibinfo {author} {\bibfnamefont {M.~A.}\ \bibnamefont
  {{Sobolewska}}},\ and\ \bibinfo {author} {\bibfnamefont {J.}~\bibnamefont
  {{Grindlay}}},\ }\bibfield  {title} {\bibinfo {title} {{Characteristic
  Variability Timescales in the Gamma-Ray Power Spectra of Blazars}},\ }\href
  {https://doi.org/10.3847/1538-4357/ab426a} {\bibfield  {journal} {\bibinfo
  {journal} {\apj}\ }\textbf {\bibinfo {volume} {885}},\ \bibinfo {eid} {12}
  (\bibinfo {year} {2019})},\ \Eprint {https://arxiv.org/abs/1909.04227}
  {arXiv:1909.04227 [astro-ph.HE]} \BibitemShut {NoStop}%
\bibitem [{\citenamefont {{Zhang}}\ \emph
  {et~al.}(2025{\natexlab{a}})\citenamefont {{Zhang}}, \citenamefont {{Yan}},
  \citenamefont {{Zhang}},\ and\ \citenamefont {{Tang}}}]{2025MNRAS.537.2380Z}%
  \BibitemOpen
  \bibfield  {author} {\bibinfo {author} {\bibfnamefont {H.}~\bibnamefont
  {{Zhang}}}, \bibinfo {author} {\bibfnamefont {D.}~\bibnamefont {{Yan}}},
  \bibinfo {author} {\bibfnamefont {L.}~\bibnamefont {{Zhang}}},\ and\ \bibinfo
  {author} {\bibfnamefont {N.}~\bibnamefont {{Tang}}},\ }\bibfield  {title}
  {\bibinfo {title} {{Evidence for magneto-gravitational processes in
  supermassive black hole binary PG 1553+113}},\ }\href
  {https://doi.org/10.1093/mnras/staf129} {\bibfield  {journal} {\bibinfo
  {journal} {\mnras}\ }\textbf {\bibinfo {volume} {537}},\ \bibinfo {pages}
  {2380} (\bibinfo {year} {2025}{\natexlab{a}})},\ \Eprint
  {https://arxiv.org/abs/2501.15169} {arXiv:2501.15169 [astro-ph.HE]}
  \BibitemShut {NoStop}%
\bibitem [{\citenamefont {{Yuan}}\ and\ \citenamefont
  {{Narayan}}(2014)}]{2014ARA&A..52..529Y}%
  \BibitemOpen
  \bibfield  {author} {\bibinfo {author} {\bibfnamefont {F.}~\bibnamefont
  {{Yuan}}}\ and\ \bibinfo {author} {\bibfnamefont {R.}~\bibnamefont
  {{Narayan}}},\ }\bibfield  {title} {\bibinfo {title} {{Hot Accretion Flows
  Around Black Holes}},\ }\href
  {https://doi.org/10.1146/annurev-astro-082812-141003} {\bibfield  {journal}
  {\bibinfo  {journal} {\araa}\ }\textbf {\bibinfo {volume} {52}},\ \bibinfo
  {pages} {529} (\bibinfo {year} {2014})},\ \Eprint
  {https://arxiv.org/abs/1401.0586} {arXiv:1401.0586 [astro-ph.HE]}
  \BibitemShut {NoStop}%
\bibitem [{\citenamefont {{R{\'o}{\.z}a{\'n}ska}}(1999)}]{1999MNRAS.308..751R}%
  \BibitemOpen
  \bibfield  {author} {\bibinfo {author} {\bibfnamefont {A.}~\bibnamefont
  {{R{\'o}{\.z}a{\'n}ska}}},\ }\bibfield  {title} {\bibinfo {title} {{Thermal
  conduction between an accretion disc and a corona in active galactic nuclei:
  vertical structure of the transition layer}},\ }\href
  {https://doi.org/10.1046/j.1365-8711.1999.02752.x} {\bibfield  {journal}
  {\bibinfo  {journal} {\mnras}\ }\textbf {\bibinfo {volume} {308}},\ \bibinfo
  {pages} {751} (\bibinfo {year} {1999})}\BibitemShut {NoStop}%
\bibitem [{\citenamefont {{Liu}}\ \emph {et~al.}(2012)\citenamefont {{Liu}},
  \citenamefont {{Liu}}, \citenamefont {{Qiao}},\ and\ \citenamefont
  {{Mineshige}}}]{2012ApJ...754...81L}%
  \BibitemOpen
  \bibfield  {author} {\bibinfo {author} {\bibfnamefont {J.~Y.}\ \bibnamefont
  {{Liu}}}, \bibinfo {author} {\bibfnamefont {B.~F.}\ \bibnamefont {{Liu}}},
  \bibinfo {author} {\bibfnamefont {E.~L.}\ \bibnamefont {{Qiao}}},\ and\
  \bibinfo {author} {\bibfnamefont {S.}~\bibnamefont {{Mineshige}}},\
  }\bibfield  {title} {\bibinfo {title} {{The Structure and Spectral Features
  of a Thin Disk and Evaporation-fed Corona in High-luminosity Active Galactic
  Nuclei}},\ }\href {https://doi.org/10.1088/0004-637X/754/2/81} {\bibfield
  {journal} {\bibinfo  {journal} {\apj}\ }\textbf {\bibinfo {volume} {754}},\
  \bibinfo {eid} {81} (\bibinfo {year} {2012})},\ \Eprint
  {https://arxiv.org/abs/1205.6958} {arXiv:1205.6958 [astro-ph.HE]}
  \BibitemShut {NoStop}%
\bibitem [{\citenamefont {{Zhang}}\ \emph
  {et~al.}(2025{\natexlab{b}})\citenamefont {{Zhang}}, \citenamefont {{Yan}},
  \citenamefont {{Zhang}},\ and\ \citenamefont {{Tang}}}]{2025ApJ...988..206Z}%
  \BibitemOpen
  \bibfield  {author} {\bibinfo {author} {\bibfnamefont {H.}~\bibnamefont
  {{Zhang}}}, \bibinfo {author} {\bibfnamefont {D.}~\bibnamefont {{Yan}}},
  \bibinfo {author} {\bibfnamefont {L.}~\bibnamefont {{Zhang}}},\ and\ \bibinfo
  {author} {\bibfnamefont {N.}~\bibnamefont {{Tang}}},\ }\bibfield  {title}
  {\bibinfo {title} {{Gaussian Process Modeling Coronal X-Ray Variability of
  Active Galactic Nuclei}},\ }\href {https://doi.org/10.3847/1538-4357/adec92}
  {\bibfield  {journal} {\bibinfo  {journal} {\apj}\ }\textbf {\bibinfo
  {volume} {988}},\ \bibinfo {eid} {206} (\bibinfo {year}
  {2025}{\natexlab{b}})},\ \Eprint {https://arxiv.org/abs/2507.04715}
  {arXiv:2507.04715 [astro-ph.HE]} \BibitemShut {NoStop}%
\bibitem [{\citenamefont {{Garcia}}\ \emph {et~al.}(2020)\citenamefont
  {{Garcia}}, \citenamefont {{Mendoza}}, \citenamefont {{Bautista}},
  \citenamefont {{Kallman}}, \citenamefont {{Deprince}}, \citenamefont
  {{Palmeri}},\ and\ \citenamefont {{Quinet}}}]{2020AAS...23534604G}%
  \BibitemOpen
  \bibfield  {author} {\bibinfo {author} {\bibfnamefont {J.~A.}\ \bibnamefont
  {{Garcia}}}, \bibinfo {author} {\bibfnamefont {C.}~\bibnamefont {{Mendoza}}},
  \bibinfo {author} {\bibfnamefont {M.}~\bibnamefont {{Bautista}}}, \bibinfo
  {author} {\bibfnamefont {T.}~\bibnamefont {{Kallman}}}, \bibinfo {author}
  {\bibfnamefont {J.}~\bibnamefont {{Deprince}}}, \bibinfo {author}
  {\bibfnamefont {P.}~\bibnamefont {{Palmeri}}},\ and\ \bibinfo {author}
  {\bibfnamefont {P.}~\bibnamefont {{Quinet}}},\ }\bibfield  {title} {\bibinfo
  {title} {{High-Density Plasma Effects in Accretion Disks around Black
  Holes}},\ }in\ \href@noop {} {\emph {\bibinfo {booktitle} {American
  Astronomical Society Meeting Abstracts \#235}}},\ \bibinfo {series} {American
  Astronomical Society Meeting Abstracts}, Vol.\ \bibinfo {volume} {235}\
  (\bibinfo {year} {2020})\ p.\ \bibinfo {pages} {346.04}\BibitemShut {NoStop}%
\bibitem [{\citenamefont {{Chainakun}}\ \emph {et~al.}(2022)\citenamefont
  {{Chainakun}}, \citenamefont {{Watcharangkool}},\ and\ \citenamefont
  {{Young}}}]{2022MNRAS.512..728C}%
  \BibitemOpen
  \bibfield  {author} {\bibinfo {author} {\bibfnamefont {P.}~\bibnamefont
  {{Chainakun}}}, \bibinfo {author} {\bibfnamefont {A.}~\bibnamefont
  {{Watcharangkool}}},\ and\ \bibinfo {author} {\bibfnamefont {A.~J.}\
  \bibnamefont {{Young}}},\ }\bibfield  {title} {\bibinfo {title} {{Effects of
  the refractive index of the X-ray corona on the emission lines in AGNs}},\
  }\href {https://doi.org/10.1093/mnras/stac362} {\bibfield  {journal}
  {\bibinfo  {journal} {\mnras}\ }\textbf {\bibinfo {volume} {512}},\ \bibinfo
  {pages} {728} (\bibinfo {year} {2022})},\ \Eprint
  {https://arxiv.org/abs/2202.03657} {arXiv:2202.03657 [astro-ph.HE]}
  \BibitemShut {NoStop}%
\bibitem [{\citenamefont {{Zhu}}\ \emph {et~al.}(2020)\citenamefont {{Zhu}},
  \citenamefont {{Brandt}}, \citenamefont {{Luo}}, \citenamefont {{Wu}},
  \citenamefont {{Xue}},\ and\ \citenamefont {{Yang}}}]{2020MNRAS.496..245Z}%
  \BibitemOpen
  \bibfield  {author} {\bibinfo {author} {\bibfnamefont {S.~F.}\ \bibnamefont
  {{Zhu}}}, \bibinfo {author} {\bibfnamefont {W.~N.}\ \bibnamefont {{Brandt}}},
  \bibinfo {author} {\bibfnamefont {B.}~\bibnamefont {{Luo}}}, \bibinfo
  {author} {\bibfnamefont {J.}~\bibnamefont {{Wu}}}, \bibinfo {author}
  {\bibfnamefont {Y.~Q.}\ \bibnamefont {{Xue}}},\ and\ \bibinfo {author}
  {\bibfnamefont {G.}~\bibnamefont {{Yang}}},\ }\bibfield  {title} {\bibinfo
  {title} {{The L$_{X}$-L$_{uv}$-L$_{radio}$ relation and corona-disc-jet
  connection in optically selected radio-loud quasars}},\ }\href
  {https://doi.org/10.1093/mnras/staa1411} {\bibfield  {journal} {\bibinfo
  {journal} {\mnras}\ }\textbf {\bibinfo {volume} {496}},\ \bibinfo {pages}
  {245} (\bibinfo {year} {2020})},\ \Eprint {https://arxiv.org/abs/2006.13226}
  {arXiv:2006.13226 [astro-ph.HE]} \BibitemShut {NoStop}%
\bibitem [{\citenamefont {{Schmidt}}(1963)}]{1963Natur.197.1040S}%
  \BibitemOpen
  \bibfield  {author} {\bibinfo {author} {\bibfnamefont {M.}~\bibnamefont
  {{Schmidt}}},\ }\bibfield  {title} {\bibinfo {title} {{3C 273 : A Star-Like
  Object with Large Red-Shift}},\ }\href {https://doi.org/10.1038/1971040a0}
  {\bibfield  {journal} {\bibinfo  {journal} {\nat}\ }\textbf {\bibinfo
  {volume} {197}},\ \bibinfo {pages} {1040} (\bibinfo {year}
  {1963})}\BibitemShut {NoStop}%
\bibitem [{\citenamefont {{Done}}\ \emph {et~al.}(2004)\citenamefont {{Done}},
  \citenamefont {{Wardzi{\'n}ski}},\ and\ \citenamefont
  {{Gierli{\'n}ski}}}]{2004MNRAS.349..393D}%
  \BibitemOpen
  \bibfield  {author} {\bibinfo {author} {\bibfnamefont {C.}~\bibnamefont
  {{Done}}}, \bibinfo {author} {\bibfnamefont {G.}~\bibnamefont
  {{Wardzi{\'n}ski}}},\ and\ \bibinfo {author} {\bibfnamefont {M.}~\bibnamefont
  {{Gierli{\'n}ski}}},\ }\bibfield  {title} {\bibinfo {title} {{GRS 1915+105:
  the brightest Galactic black hole}},\ }\href
  {https://doi.org/10.1111/j.1365-2966.2004.07545.x} {\bibfield  {journal}
  {\bibinfo  {journal} {\mnras}\ }\textbf {\bibinfo {volume} {349}},\ \bibinfo
  {pages} {393} (\bibinfo {year} {2004})},\ \Eprint
  {https://arxiv.org/abs/astro-ph/0308536} {arXiv:astro-ph/0308536 [astro-ph]}
  \BibitemShut {NoStop}%
\end{thebibliography}%
\appendix
\section{Two examples of DRW modeling}\label{App:A}
\begin{figure*}[ht]
	\includegraphics[width=0.49\textwidth]{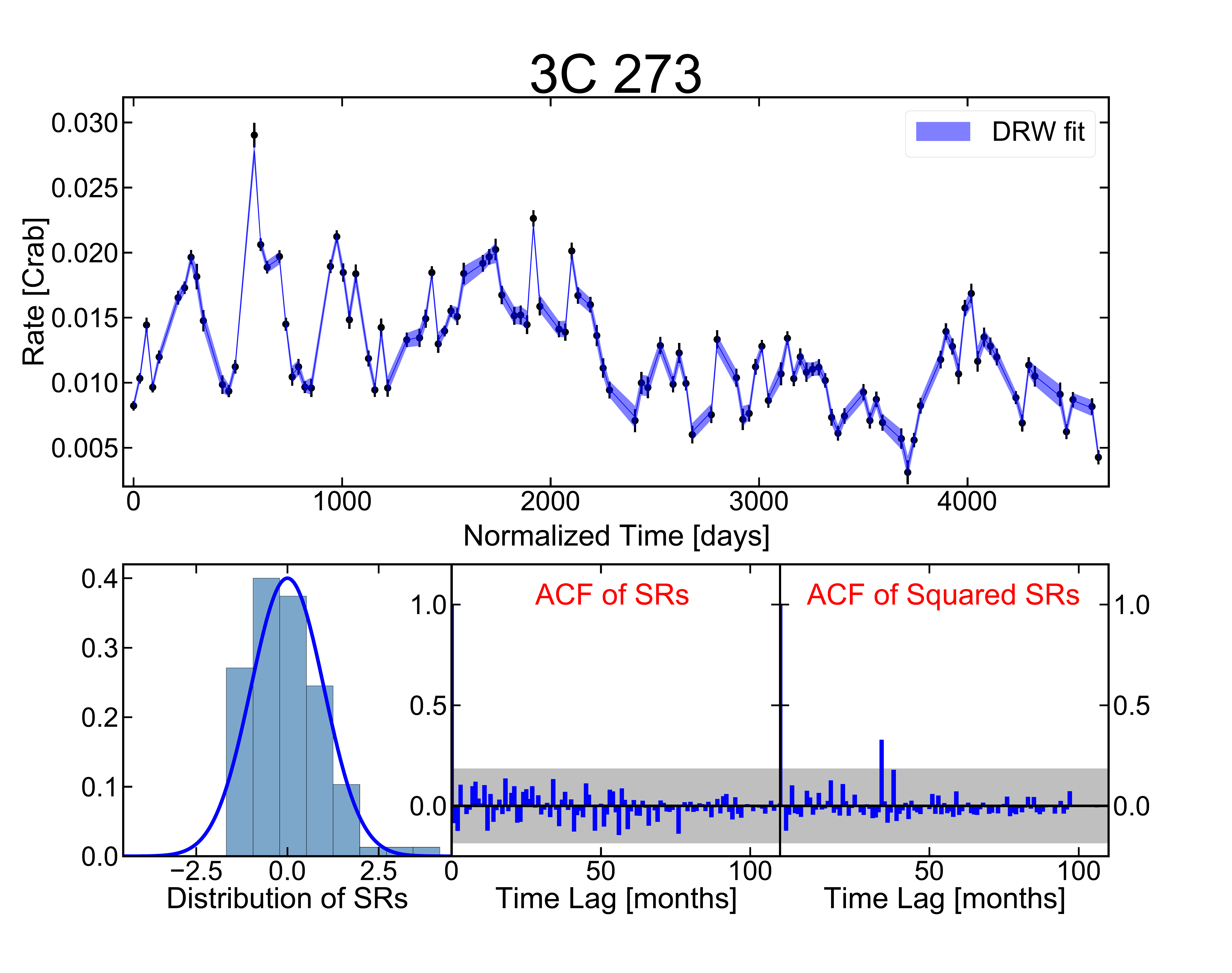}
	\includegraphics[width=0.49\textwidth]{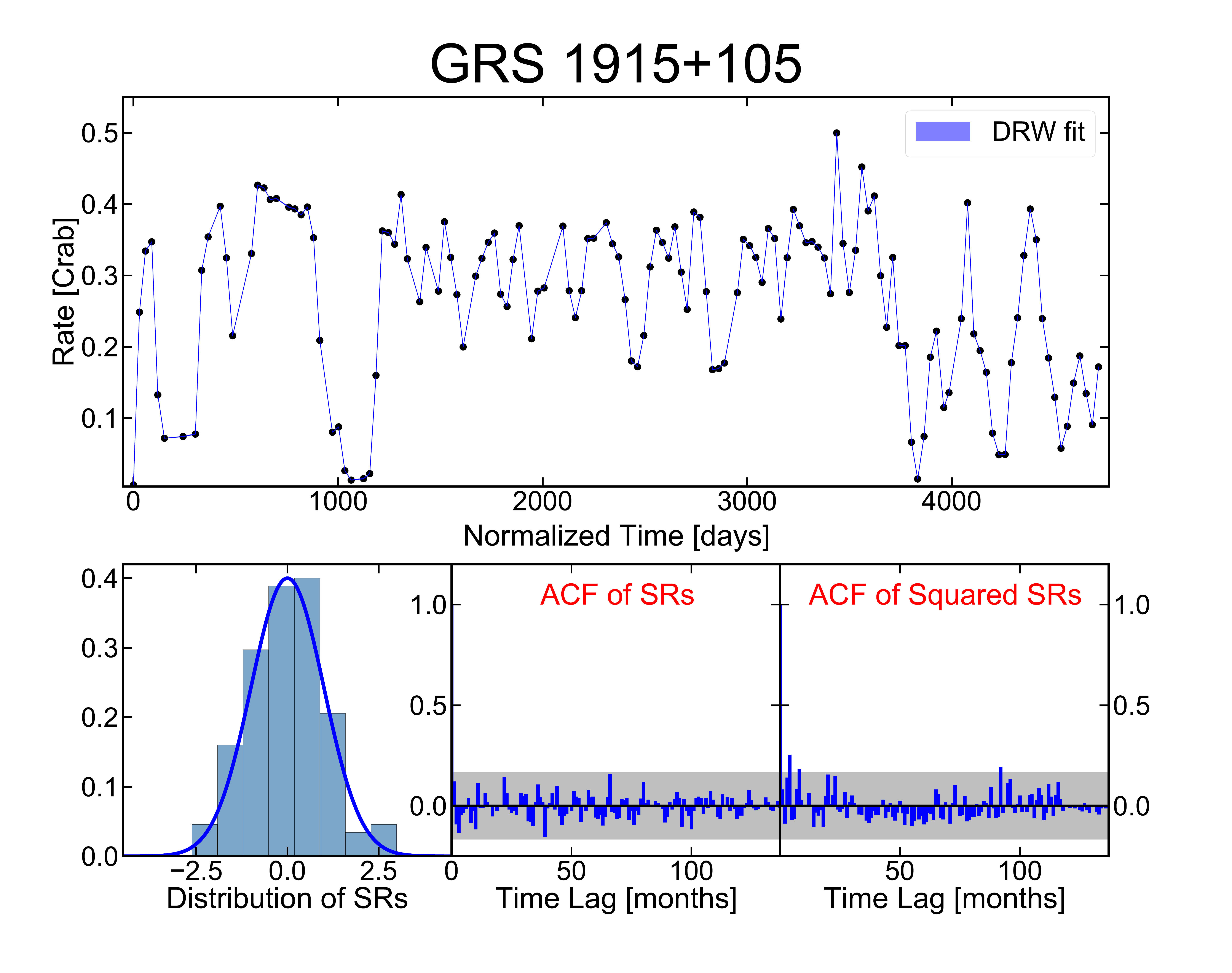}
	\caption{Modeling results for 3C~273 and GRS 1915+105. In the left and right panel, the top part indicates the best-fitting LC of 3C~273 and GRS 1915+105, respectively. The distribution of SRs, ACFs of SRs and squared SRs are shown in bottom part. The gray shaded region is the 95\% confidence interval of white noise. \label{fig:exam}}
\end{figure*}

\begin{figure*}[ht]
	\includegraphics[width=0.49\textwidth]{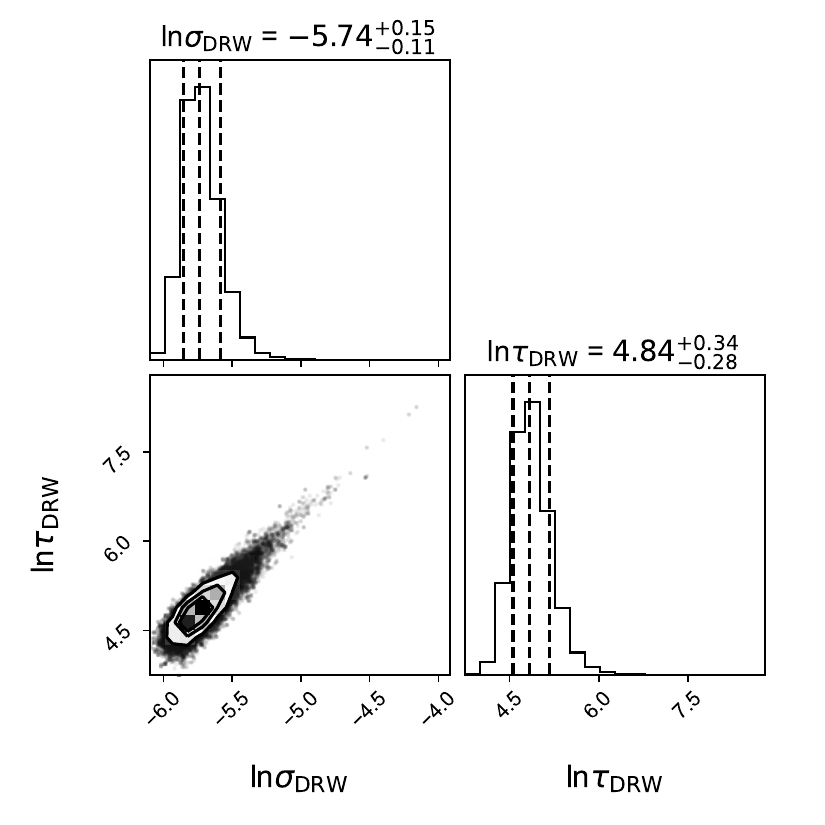}
	\includegraphics[width=0.49\textwidth]{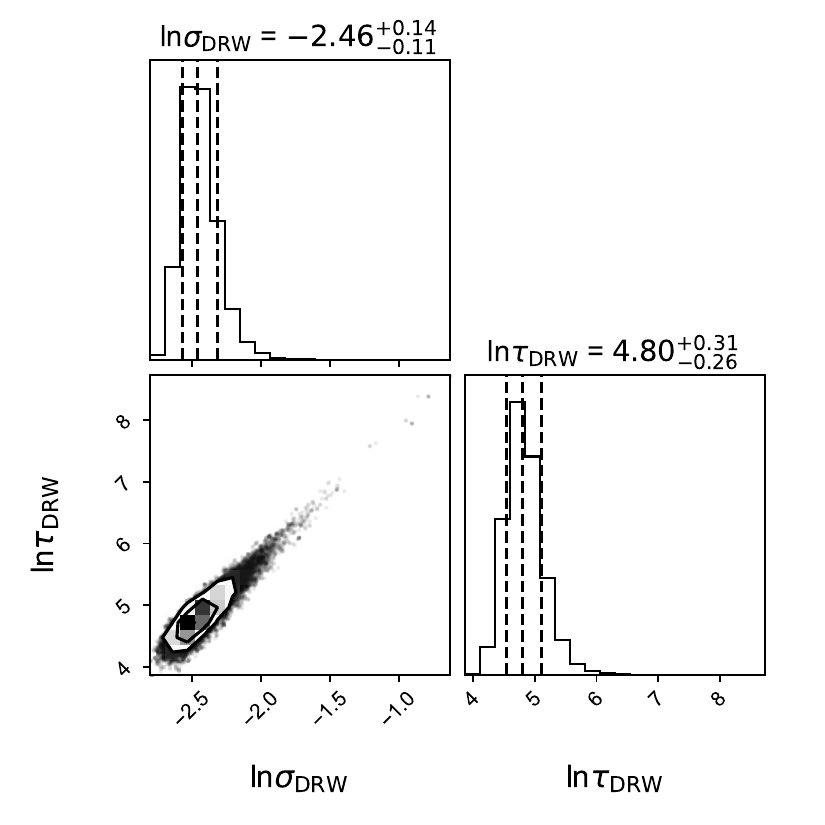}
	\caption{Posterior distributions of the model parameters for 3C~273 (left) and GRS 1915+105. \label{fig:pos}}
\end{figure*}

\begin{figure*}[ht]
	\includegraphics[width=0.49\textwidth]{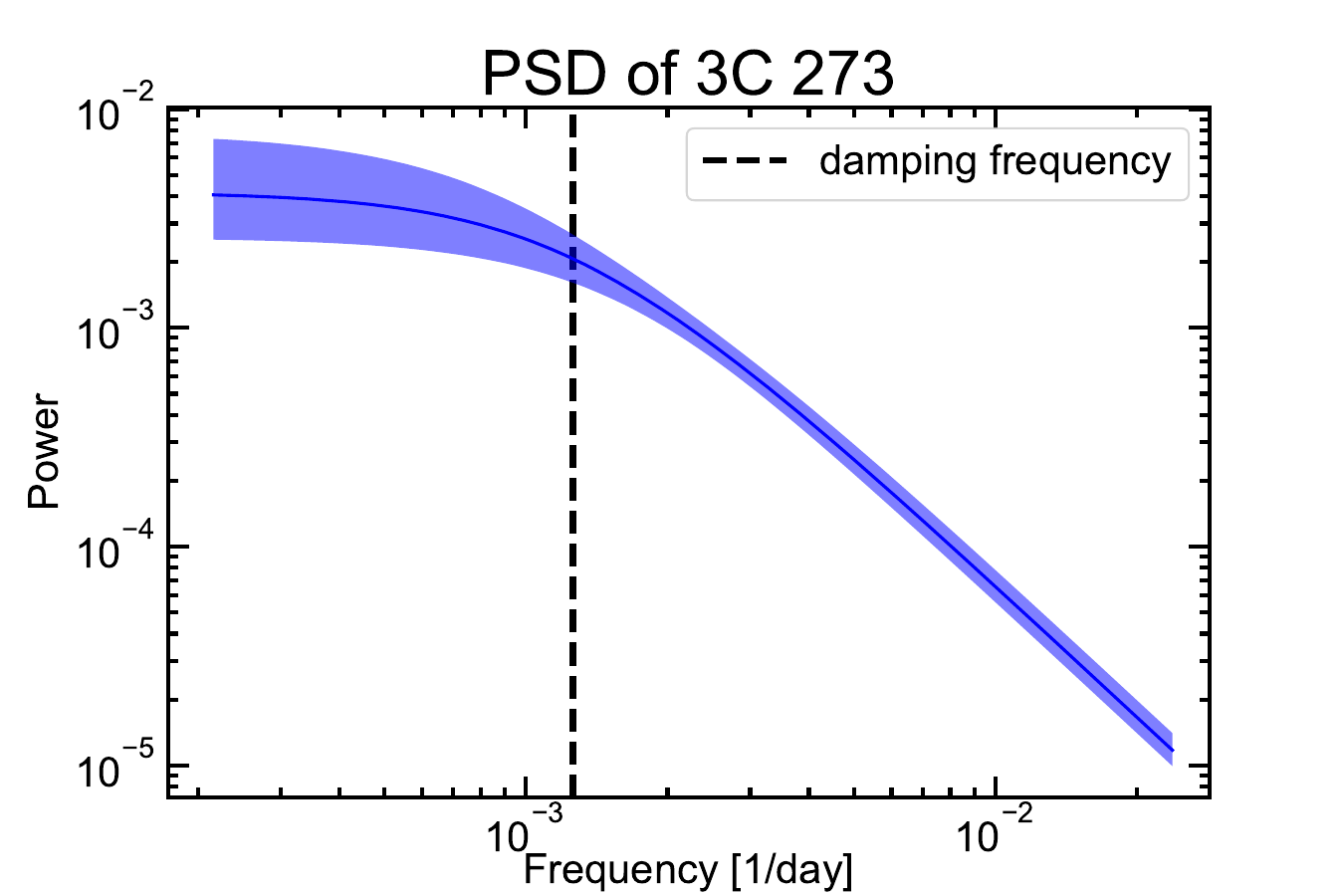}
	\includegraphics[width=0.49\textwidth]{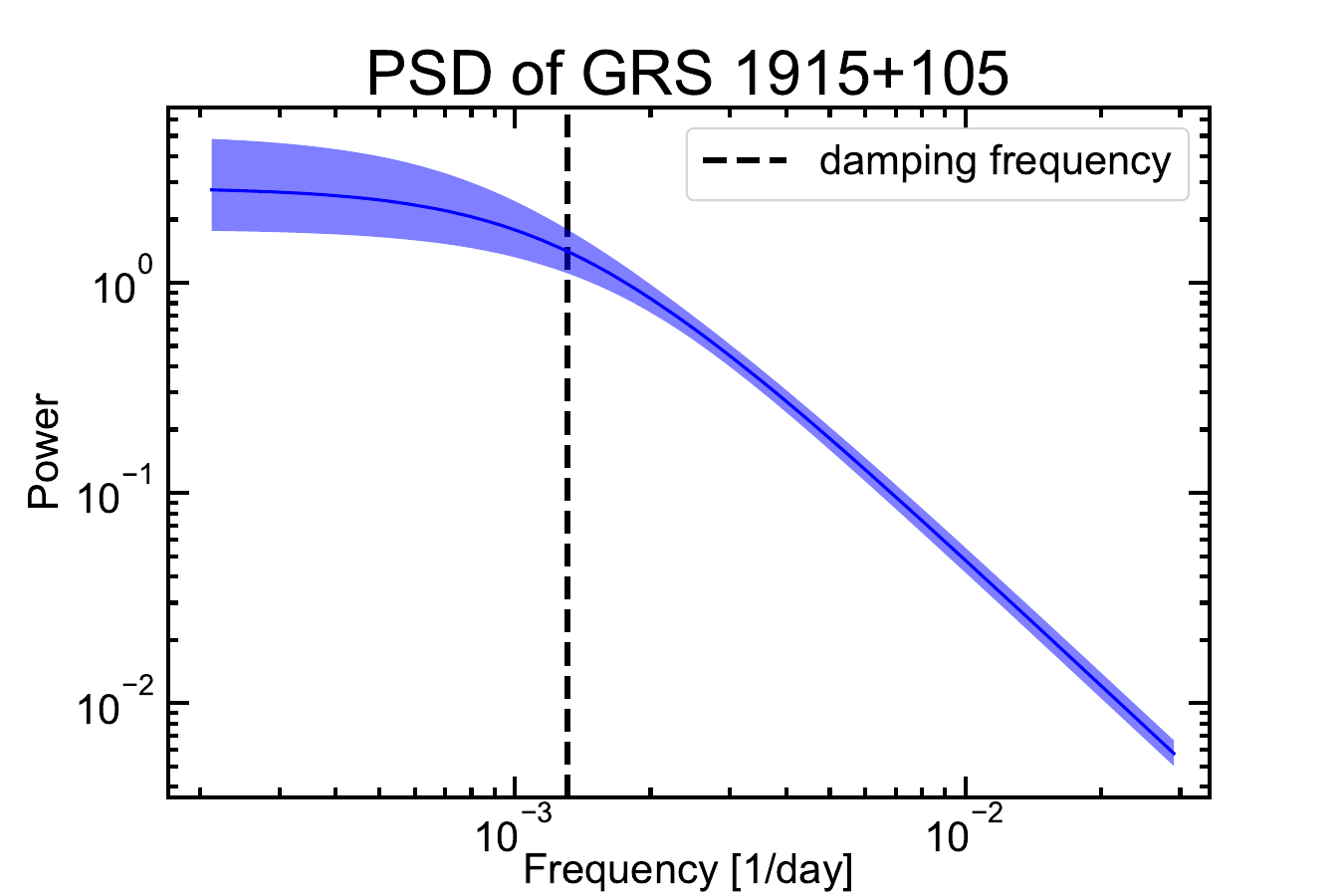}
	\caption{PSD of the fitted DRW model for 3C~273 (left) and GRS 1915+105. The blue lines and shaded region indicate the PSD of model and 1$\sigma$ confidence interval. The damping frequency of PSD is shown as black dashed lines. \label{fig:psd}}
\end{figure*}

We have plotted the modeling results of 3C~273 (SWIFT J1229.1+0202) and GRS 1915+105 (SWIFT J1915.3+1057) in Figure \ref{fig:exam}, representing the first discovered AGN \cite{1963Natur.197.1040S} and one of the brightest galactic black hole accretion system \cite{2004MNRAS.349..393D}, respectively. In the top parts of the two panels in Figure \ref{fig:exam}, the fitted model and 1$\sigma$ confidence interval are shown as the blue line and shaded region, respectively. In the bottom parts, the distribution of SRs, ACFs of the SRs and squared SRs are well represented by white noise. In Figure \ref{fig:pos}, the posterior distributions of the parameters show convergence, indicating good fits. In Figure \ref{fig:psd}, the black dashed line represents the damping frequency of the model PSD. 
\section{Fitting results of 106 valid sources}\label{App:B}
The fitting results of 106 valid sources are presented in Table \ref{tab:B2}--\ref{tab:B4}, the \textsc{XLSX} version of this table can see in Supplementary Materials \cite{zhang_2025_17830360}.

\begin{table*}[h]  
	\renewcommand{\arraystretch}{1.2}
	\setlength{\tabcolsep}{10pt}
	\centering
	\caption{}
	\begin{tabular}{cccccc}
		\hline
		SWIFT Source name & Type & Time length & Average cadence & ln $\sigma_{\text{DRW}}$ & ln $\tau_{\text{DRW}}^{\text{obs}}$ \\
		\hline
		J0036.0+5951     & BZB          & 4740.96 & 34.61 & $-6.20_{-0.09}^{+0.10}$ & $4.24_{-0.24}^{+0.27}$ \\
		J0138.6-4001     & Sy2          & 4717    & 35.73 & $-7.76_{-0.07}^{+0.28}$ & $5.75_{-1.78}^{+1.88}$ \\
		J0201.0-0648     & Sy2          & 4687.5  & 42.23 & $-7.19_{-0.11}^{+0.11}$ & $3.87_{-0.47}^{+0.43}$ \\
		J0234.1+3233     & Sy2          & 4740.96 & 41.23 & $-6.58_{-0.19}^{+0.31}$ & $5.71_{-0.46}^{+0.74}$ \\
		J0244.8-5829     & Beamed AGN   & 4717    & 35.73 & $-7.55_{-0.15}^{+0.21}$ & $5.38_{-0.44}^{+0.56}$ \\
		J0250.7+4142     & Sy2          & 4679.96 & 40.34 & $-7.77_{-0.06}^{+0.22}$ & $5.22_{-4.67}^{+3.37}$ \\
		J0336.6+3217     & BZQ          & 4656    & 43.92 & $-7.69_{-0.11}^{+0.17}$ & $4.66_{-0.72}^{+0.80}$ \\
		J0342.0-2115     & Sy1          & 4717    & 36.01 & $-7.53_{-0.18}^{+0.30}$ & $5.88_{-0.59}^{+0.86}$ \\
		J0407.4+0339     & Sy2          & 4717    & 44.92 & $-7.71_{-0.10}^{+0.21}$ & $4.93_{-1.10}^{+2.04}$ \\
		J0420.0-5457     & Sy1.5        & 4740.9  & 33.86 & $-7.61_{-0.16}^{+0.30}$ & $5.37_{-0.63}^{+0.99}$ \\
		J0444.1+2813     & Sy2          & 4686.5  & 47.34 & $-7.02_{-0.17}^{+0.23}$ & $5.34_{-0.46}^{+0.63}$ \\
		J0451.4-0346     & Sy1.5        & 4686.5  & 45.06 & $-7.75_{-0.07}^{+0.12}$ & $4.19_{-3.20}^{+0.86}$ \\
		J0452.2+4933     & Sy1.5        & 4656    & 41.95 & $-7.53_{-0.18}^{+0.31}$ & $5.82_{-0.63}^{+0.90}$ \\
		J0502.1+0332     & Sy1          & 4686.5  & 48.82 & $-7.67_{-0.12}^{+0.16}$ & $4.57_{-0.66}^{+0.95}$ \\
		J0505.8-2351     & Sy2          & 4686.5  & 36.61 & $-7.56_{-0.18}^{+0.33}$ & $5.91_{-0.69}^{+0.98}$ \\
		J0526.2-2118     & Sy2          & 4686.5  & 36.9  & $-7.70_{-0.11}^{+0.31}$ & $5.79_{-1.43}^{+2.03}$ \\
		J0552.2-0727     & Sy2          & 4686.5  & 43.8  & $-5.56_{-0.11}^{+0.15}$ & $4.87_{-0.28}^{+0.33}$ \\
		J0615.8+7101     & Sy1.9        & 4717    & 36.01 & $-6.69_{-0.09}^{+0.10}$ & $4.12_{-0.25}^{+0.28}$ \\
		J0748.8-6743     & LMXB         & 4740.96 & 34.86 & $-5.09_{-0.21}^{+0.38}$ & $6.13_{-0.44}^{+0.79}$ \\
		J0841.4+7052     & BZQ          & 4740.96 & 34.11 & $-7.36_{-0.16}^{+0.21}$ & $5.41_{-0.42}^{+0.57}$ \\
		J0920.5-5511     & LMXB         & 4740.96 & 36.75 & $-6.72_{-0.09}^{+0.10}$ & $4.21_{-0.26}^{+0.29}$ \\
		J0945.6-1420     & Sy1.9        & 4658    & 50.63 & $-6.38_{-0.11}^{+0.14}$ & $4.61_{-0.35}^{+0.39}$ \\
		J0959.5-2248     & Sy2          & 4717    & 46.7  & $-7.14_{-0.12}^{+0.12}$ & $3.94_{-0.45}^{+0.42}$ \\
		\hline
	\end{tabular}\\
    {Notes--- "LMXB" are low-mass XRBs. "Sy 1--2" are Seyferts. The rest were classified as blazars. All of measurements regard time in this table are given in units of days.} \label{tab:B2}

\end{table*}
\begin{table*} 
	\renewcommand{\arraystretch}{1.2}
	\setlength{\tabcolsep}{10pt}
	\centering
	\caption{(Continued).} \label{tab:B3}
	\begin{tabular}{cccccc}
		\hline
		SWIFT Source name & Type & Time length & Average cadence & ln $\sigma_{\text{DRW}}$ & ln $\tau_{\text{DRW}}^{\text{obs}}$ \\
		\hline
		J1104.4+3812     & BZB          & 4717    & 34.68 & $-5.40_{-0.10}^{+0.12}$ & $4.63_{-0.24}^{+0.30}$ \\
		J1106.5+7234     & Sy1.2        & 4740.96 & 33.15 & $-6.70_{-0.10}^{+0.12}$ & $4.46_{-0.26}^{+0.31}$ \\
		J1200.2-5350     & Sy2          & 4740.96 & 37.93 & $-7.75_{-0.08}^{+0.25}$ & $5.85_{-1.68}^{+2.92}$ \\
		J1210.5+3924     & Sy1.5        & 4717    & 34.18 & $-5.39_{-0.10}^{+0.12}$ & $4.55_{-0.24}^{+0.28}$ \\
		J1217.3+0714     & Sy1.2        & 4627.49 & 39.22 & $-7.33_{-0.12}^{+0.13}$ & $4.46_{-0.41}^{+0.45}$ \\
		J1221.3+3012     & BZB          & 4717    & 34.18 & $-7.74_{-0.08}^{+0.27}$ & $5.92_{-0.93}^{+1.26}$ \\
		J1225.8+1240     & Sy2          & 4717    & 38.04 & $-5.96_{-0.11}^{+0.13}$ & $4.74_{-0.27}^{+0.32}$ \\
		J1229.1+0202     & BZQ          & 4627.49 & 42.07 & $-5.74_{-0.11}^{+0.15}$ & $4.84_{-0.28}^{+0.34}$ \\
		J1235.6-3954     & Sy2          & 4717    & 43.28 & $-6.68_{-0.10}^{+0.10}$ & $3.93_{-0.32}^{+0.32}$ \\
		J1241.6-5748     & Sy1.5        & 4740.96 & 39.18 & $-7.47_{-0.22}^{+0.38}$ & $5.92_{-0.72}^{+1.20}$ \\
		J1249.3-5907     & LMXB         & 4740.96 & 38.86 & $-6.70_{-0.15}^{+0.23}$ & $5.37_{-0.41}^{+0.58}$ \\
		J1304.3-1022     & Sy2          & 4717    & 47.17 & $-7.49_{-0.15}^{+0.15}$ & $4.35_{-0.88}^{+0.72}$ \\
		J1305.4-4928     & Sy2          & 4717    & 41.74 & $-6.24_{-0.08}^{+0.09}$ & $3.90_{-0.26}^{+0.27}$ \\
		J1307.2-7049     & LMXB          & 4740.96 & 34.86 & $-6.13_{-0.08}^{+0.09}$ & $4.02_{-0.22}^{+0.24}$ \\
		J1313.6+3650B    & Sy1.9        & 4717    & 32.53 & $-7.78_{-0.05}^{+0.18}$ & $5.10_{-1.69}^{+1.72}$ \\
		J1325.4-4301     & Sy2          & 4717    & 43.28 & $-4.32_{-0.17}^{+0.24}$ & $5.64_{-0.37}^{+0.51}$ \\
		J1338.2+0433     & Sy2          & 4717    & 41.74 & $-6.56_{-0.17}^{+0.26}$ & $5.71_{-0.40}^{+0.58}$ \\
		J1345.5+4139     & Sy2          & 4717    & 32.53 & $-7.77_{-0.06}^{+0.11}$ & $4.64_{-0.59}^{+0.83}$ \\
		J1352.8+6917     & Sy1.5        & 4740.96 & 33.62 & $-7.31_{-0.10}^{+0.11}$ & $4.19_{-0.34}^{+0.35}$ \\
		J1417.9+2507     & Sy1.5        & 4717    & 33.69 & $-6.92_{-0.09}^{+0.10}$ & $4.04_{-0.28}^{+0.29}$ \\
		J1428.7+4234     & BZB          & 4717    & 32.99 & $-7.77_{-0.05}^{+0.10}$ & $4.68_{-0.50}^{+0.57}$ \\
		J1535.4-5714     & LMXB          & 4717    & 39.97 & $-2.42_{-0.08}^{+0.08}$ & $3.84_{-0.22}^{+0.23}$ \\
		J1539.2-6227     & LMXB         & 4717    & 38.98 & $-6.50_{-0.08}^{+0.09}$ & $3.81_{-0.27}^{+0.27}$ \\
		J1540.5+1416     & Sy1.9        & 4717    & 38.04 & $-7.74_{-0.08}^{+0.28}$ & $5.98_{-5.20}^{+3.05}$ \\
		J1612.9-5227     & LMXB         & 4656    & 39.79 & $-3.89_{-0.09}^{+0.11}$ & $4.39_{-0.24}^{+0.27}$ \\
		J1619.4-2808     & LMXB         & 4656    & 39.46 & $-7.35_{-0.12}^{+0.12}$ & $3.80_{-0.67}^{+0.51}$ \\
		J1626.5-2951     & BZQ          & 4656    & 39.46 & $-7.74_{-0.08}^{+0.24}$ & $5.55_{-1.84}^{+2.34}$ \\
		J1643.9+5739     & TDE          & 4740.96 & 35.38 & $-6.84_{-0.09}^{+0.09}$ & $3.98_{-0.23}^{+0.26}$ \\
		J1652.3-4520     & LMXB         & 4656    & 39.79 & $-5.92_{-0.10}^{+0.11}$ & $4.44_{-0.24}^{+0.28}$ \\
		J1654.0+3946     & BZB          & 4717    & 35.2  & $-5.98_{-0.10}^{+0.13}$ & $4.68_{-0.25}^{+0.30}$ \\
		J1656.3-3302     & BZQ          & 4656    & 39.13 & $-7.53_{-0.16}^{+0.18}$ & $4.88_{-0.51}^{+0.58}$ \\
		J1659.2-1515     & LMXB         & 4656    & 38.8  & $-4.54_{-0.10}^{+0.13}$ & $4.71_{-0.25}^{+0.31}$ \\
		J1700.6-4222     & LMXB         & 4656    & 39.79 & $-7.43_{-0.14}^{+0.14}$ & $4.41_{-0.64}^{+0.72}$ \\
		J1701.7-4050     & LMXB         & 4656    & 39.79 & $-5.78_{-0.12}^{+0.15}$ & $4.84_{-0.29}^{+0.36}$ \\
		J1702.8-4849     & LMXB         & 4656    & 39.79 & $-3.08_{-0.09}^{+0.11}$ & $4.28_{-0.24}^{+0.26}$ \\
		J1706.7+2401     & LMXB         & 4717    & 34.68 & $-5.76_{-0.12}^{+0.15}$ & $5.01_{-0.27}^{+0.33}$ \\
		J1708.9-4404     & LMXB         & 4656    & 39.46 & $-4.78_{-0.08}^{+0.08}$ & $3.73_{-0.23}^{+0.24}$ \\
		J1709.4-2639     & LMXB         & 4656    & 38.48 & $-6.21_{-0.08}^{+0.09}$ & $3.76_{-0.24}^{+0.25}$ \\
		J1709.8-3627A    & LMXB         & 4656    & 39.46 & $-4.85_{-0.08}^{+0.09}$ & $3.87_{-0.22}^{+0.23}$ \\
		J1709.8-3627B    & LMXB         & 4656    & 39.46 & $-5.37_{-0.08}^{+0.09}$ & $3.85_{-0.23}^{+0.25}$ \\
		J1727.4-3046     & LMXB         & 4656    & 37.85 & $-5.32_{-0.12}^{+0.15}$ & $4.96_{-0.27}^{+0.34}$ \\
		\hline
	\end{tabular}
\end{table*}
\begin{table*} 
	\renewcommand{\arraystretch}{1.2}
	\setlength{\tabcolsep}{10pt}
	\centering
	\caption{(Continued).} \label{tab:B4}
	\begin{tabular}{cccccc}
		\hline
		SWIFT Source name & Type & Time length & Average cadence & ln $\sigma_{\text{DRW}}$ & ln $\tau_{\text{DRW}}^{\text{obs}}$ \\
		\hline
		J1731.6-1657     & LMXB         & 4657    & 38.81 & $-5.98_{-0.11}^{+0.14}$ & $4.83_{-0.28}^{+0.35}$ \\
		J1731.9-2444     & LMXB         & 4656    & 37.85 & $-3.47_{-0.07}^{+0.08}$ & $3.69_{-0.24}^{+0.24}$ \\
		J1738.3-2657     & LMXB         & 4626.5  & 37.92 & $-5.87_{-0.09}^{+0.10}$ & $4.10_{-0.23}^{+0.27}$ \\
		J1740.6-2821A    & LMXB         & 4626.5  & 37.92 & $-6.01_{-0.08}^{+0.09}$ & $3.93_{-0.25}^{+0.26}$ \\
		J1745.1-2891     & LMXB         & 4626.5  & 38.24 & $-2.87_{-0.10}^{+0.12}$ & $4.56_{-0.24}^{+0.29}$ \\
		J174510.8-262411 & LMXB         & 4657    & 38.17 & $-3.46_{-0.10}^{+0.12}$ & $4.59_{-0.24}^{+0.28}$ \\
		J1746.3-2850C    & LMXB         & 4626.5  & 38.88 & $-5.43_{-0.08}^{+0.08}$ & $3.71_{-0.22}^{+0.23}$ \\
		J1746.3-2931B    & LMXB         & 4626.5  & 38.24 & $-5.01_{-0.11}^{+0.13}$ & $4.72_{-0.26}^{+0.30}$ \\
		J1747.4-2719     & LMXB         & 4657    & 38.81 & $-3.62_{-0.09}^{+0.11}$ & $4.41_{-0.22}^{+0.26}$ \\
		J1747.9-2631A    & LMXB         & 4657    & 38.81 & $-4.80_{-0.14}^{+0.18}$ & $5.24_{-0.30}^{+0.40}$ \\
		J1748.1-3605     & LMXB         & 4626.5  & 39.21 & $-6.07_{-0.09}^{+0.10}$ & $3.88_{-0.27}^{+0.27}$ \\
		J1749.4-2822     & LMXB         & 4657    & 39.13 & $-5.46_{-0.08}^{+0.09}$ & $3.89_{-0.24}^{+0.25}$ \\
		J1750.6-2903     & LMXB         & 4657    & 39.8  & $-5.48_{-0.11}^{+0.13}$ & $4.66_{-0.27}^{+0.32}$ \\
		J1752.1-2220     & LMXB         & 4657    & 39.47 & $-3.29_{-0.10}^{+0.11}$ & $4.46_{-0.24}^{+0.27}$ \\
		J1753.5-0130     & LMXB         & 4686.5  & 36.61 & $-3.59_{-0.09}^{+0.10}$ & $4.24_{-0.23}^{+0.26}$ \\
		J1801.5-2031     & LMXB         & 4657    & 40.15 & $-5.75_{-0.10}^{+0.12}$ & $4.29_{-0.28}^{+0.31}$ \\
		J1804.1-3421     & LMXB         & 4657    & 39.8  & $-4.46_{-0.10}^{+0.11}$ & $4.46_{-0.24}^{+0.28}$ \\
		J1806.5-2223     & LMXB         & 4657    & 40.5  & $-5.50_{-0.09}^{+0.11}$ & $4.39_{-0.25}^{+0.28}$ \\
		J1810.2-1903     & LMXB         & 4657    & 40.15 & $-5.05_{-0.09}^{+0.11}$ & $4.47_{-0.24}^{+0.28}$ \\
		J1810.8-2610     & LMXB         & 4657    & 40.85 & $-5.68_{-0.08}^{+0.08}$ & $3.80_{-0.22}^{+0.23}$ \\
		J1825.7-3705     & LMXB         & 4657    & 40.15 & $-6.20_{-0.10}^{+0.12}$ & $4.45_{-0.28}^{+0.32}$ \\
		J1829.4-2346     & LMXB         & 4657    & 40.85 & $-3.81_{-0.21}^{+0.36}$ & $6.06_{-0.44}^{+0.76}$ \\
		J1835.6-3259     & LMXB         & 4657    & 40.5  & $-5.68_{-0.19}^{+0.34}$ & $6.05_{-0.43}^{+0.72}$ \\
		J1835.7-1931     & LMXB         & 4657    & 40.85 & $-5.32_{-0.08}^{+0.10}$ & $4.13_{-0.23}^{+0.24}$ \\
		J1840.0+0501     & LMXB         & 4686.5  & 36.61 & $-6.06_{-0.09}^{+0.11}$ & $4.25_{-0.26}^{+0.28}$ \\
		J1856.8+0519     & LMXB         & 4686.5  & 36.61 & $-6.03_{-0.08}^{+0.09}$ & $3.73_{-0.28}^{+0.27}$ \\
		J1900.2-2455     & LMXB         & 4657    & 40.5  & $-4.57_{-0.14}^{+0.22}$ & $5.40_{-0.32}^{+0.47}$ \\
		J1910.2-0546     & LMXB         & 4657    & 38.81 & $-4.95_{-0.08}^{+0.08}$ & $3.85_{-0.21}^{+0.22}$ \\
		J1915.3+1057     & LMXB         & 4717    & 34.43 & $-2.46_{-0.11}^{+0.14}$ & $4.80_{-0.26}^{+0.31}$ \\
		J1922.7-1716     & LMXB         & 4657    & 41.21 & $-5.77_{-0.13}^{+0.17}$ & $5.08_{-0.31}^{+0.39}$ \\
		J1959.6+6507     & BZB          & 4740.96 & 33.86 & $-6.41_{-0.08}^{+0.09}$ & $3.88_{-0.23}^{+0.25}$ \\
		J2009.6-4851     & BZB          & 4657    & 41.95 & $-7.57_{-0.14}^{+0.16}$ & $4.38_{-0.57}^{+0.56}$ \\
		J2024.0-0246     & Sy1.9        & 4658    & 44.36 & $-7.73_{-0.08}^{+0.18}$ & $5.27_{-0.82}^{+1.12}$ \\
		J2034.4-3036     & Sy1          & 4627.5  & 45.37 & $-7.72_{-0.10}^{+0.27}$ & $5.69_{-1.25}^{+1.66}$ \\
		J2124.6+5057     & Sy1.9        & 4740.96 & 34.35 & $-6.67_{-0.08}^{+0.08}$ & $3.55_{-0.26}^{+0.26}$ \\
		J2152.0-3030     & BZQ          & 4658    & 58.23 & $-7.39_{-0.23}^{+0.35}$ & $6.06_{-0.66}^{+0.98}$ \\
		J2253.9+1608     & BZQ          & 4740.96 & 45.59 & $-6.27_{-0.10}^{+0.12}$ & $4.39_{-0.28}^{+0.31}$ \\
		J2254.1-1734     & Sy1.2        & 4627.49 & 49.23 & $-7.46_{-0.16}^{+0.16}$ & $4.41_{-0.61}^{+0.59}$ \\
		J2303.3+0852     & Sy1.5        & 4740.96 & 50.98 & $-7.71_{-0.10}^{+0.26}$ & $5.75_{-0.89}^{+1.21}$ \\
		J2304.8-0843     & Sy1.5        & 4627.49 & 49.76 & $-7.06_{-0.12}^{+0.13}$ & $4.32_{-0.37}^{+0.40}$ \\
		J2327.5+0938     & BZQ          & 4740.96 & 49.9  & $-7.74_{-0.08}^{+0.26}$ & $5.49_{-2.20}^{+2.54}$ \\
		J2346.8+5143     & BZB          & 4740.96 & 34.11 & $-7.78_{-0.05}^{+0.17}$ & $4.83_{-4.77}^{+2.41}$ \\
		\hline
	\end{tabular}
\end{table*}		
\end{document}